# Center-fixing of tropical cyclones using uncertainty-aware deep learning applied to high-temporal-resolution geostationary satellite imagery

**This is the third version (second revision) of an article under review in the AMS journal *Weather and Forecasting*.**

Ryan Lagerquist[a,b], Galina Chirokova[a], Robert DeMaria[a], Mark DeMaria[a], and Imme Ebert-Uphoff[a,c]

[a] *Cooperative Institute for Research in the Atmosphere, Colorado State University, Fort Collins, Colorado*

[b] *National Oceanic and Atmospheric Administration (NOAA) Global Systems Laboratory (GSL), Boulder, Colorado*

[c] *Department of Electrical and Computer Engineering, Colorado State University, Fort Collins, Colorado*

*Corresponding author*: Ryan Lagerquist, ralager@colostate.edu

ABSTRACT: Determining the location of a tropical cyclone's (TC) surface circulation center – "center-fixing" – is a critical first step in the TC-forecasting process, affecting current/future estimates of track, intensity, and structure. Despite a recent increase in automated center-fixing methods, only one such method (ARCHER-2) is operational, and its best performance is achieved when using microwave or scatterometer data, which are often unavailable. We develop a deep-learning algorithm called GeoCenter; besides a few scalars in the operational Automated Tropical Cyclone Forecasting System, it relies only on geostationary infrared (IR) satellite imagery, which is available for all TC basins at high frequency (10 min) and low latency (< 10 min) during both day and night. GeoCenter ingests an animation (time series) of IR images, including 9 channels at lag times up to 4 hours. The animation is centered at a "first guess" location, offset from the true TC-center location by 48 km on average and sometimes > 100 km; GeoCenter is tasked with correcting this offset. On an independent testing dataset, GeoCenter achieves a mean/median/RMS (root mean square) error of 26.6/22.2/32.4 km for all systems, 24.7/20.8/30.0 km for tropical systems, and 14.6/12.5/17.3 km for category-2–5 hurricanes. These values are similar to ARCHER-2 errors with microwave or scatterometer data, and better than ARCHER-2 errors when only IR data are available. GeoCenter also performs skillful uncertainty quantification, producing a well calibrated ensemble of 150 TC-center locations. Furthermore, all predictors used by GeoCenter are available in real time, which would make GeoCenter easy to implement operationally every 10 min.

SIGNIFICANCE STATEMENT: Estimating the location of a tropical cyclone's (TC) surface circulation center is a critical first step in the TC-forecasting process. Current and future estimates of several TC properties – including the TC track, intensity, and structure – are highly sensitive to this initial location estimate, called the "center fix". This paper describes a new deep-learning algorithm for center-fixing, called GeoCenter, whose main input is an animated time series of IR satellite imagery. GeoCenter performs competitively with existing methods for center-fixing, both operational and non-operational, and provides skillful estimates of uncertainty in the TC-center location. Furthermore, GeoCenter is designed so that it could be easily implemented in operations.

## 1. Introduction

Determining the location of a tropical cyclone's (TC) surface circulation center is a critical first step in the forecasting process. Even small errors in the center fix can lead to large errors in other TC properties, such as the current wind field (Figure 4 of Mayers and Ruf 2019) or intensity. For example, several articles emphasize the importance of an accurate center fix for Dvorak-based intensity estimation (Dvorak 1984; Velden et al. 2006; Olander and Velden 2007; Knaff et al. 2010), and although they do not provide a numerical estimate of this sensitivity, one can be derived from knowledge of the Dvorak technique. For example, Dvorak-based intensity estimation requires a scene type to be chosen from five categories. An incorrect center fix could lead to the wrong scene type – *e.g.*, shear instead of central dense overcast – which could lead to an error of 2-3 in the current intensity number (CI) or ~30 kt (Table 2 of Velden et al. 2006).

Small center-fixing errors can also lead to inaccurate forecasts of future TC processes, including rapid intensification (Kieper and Jiang 2012; Rozoff et al. 2015), and inaccurate forecasts of TC processes can lead to downstream errors in weather forecasts across the globe (Heming et al. 2019; Keller et al. 2019). Furthermore, accurate center-fixing is important for post-season analysis and research applications (Mayers and Ruf 2019) and initializing numerical weather prediction (NWP) models (Leslie and Holland 1995; Trabing and Bell 2020).

The most accurate data sources for center-fixing are aircraft reconnaissance and radar. However, aircraft reconnaissance is currently limited to storms in the North Atlantic and eastern North Pacific approaching landfall, while radar data are available only near land-based radar sites. Most TCs occur in the open ocean away from land, which makes satellite imagery the main data source for center-fixing.

Over the last half-century, many subjective and objective center-fixing methods have been developed, mostly based on satellite imagery. However, center-fixing remains a challenge, especially for weak systems and those undergoing extratropical transition. Currently, the only automated (objective) center-fixing method used operationally is Automated Rotational Center Hurricane Eye Retrieval (ARCHER-2; Wimmers and Velden 2016, henceforth WV16). ARCHER-2 has known limitations, including [1] large errors for weak or extratropical systems and [2] its reliance on microwave and scatterometer data, which are often unavailable, for best performance. For subjective center-fixing, operational forecasters rely primarily on the Dvorak (1975) technique, whose inputs are visible and IR satellite data. IR data from geostationary (GEO) satellites are the only satellite data available during both day and night, for all TC basins around the globe, with high temporal resolution (10 min) and low latency (< 10 min). Microwave and scatterometer data from low Earth orbit (LEO) satellites – which historically have provided the most accurate center fixes (*e.g.*, allowing ARCHER-2 to achieve a mean error of 24-31 km, versus 43-49 km for IR data, as shown in Table 3 of WV16) – are available in tropical regions only twice a day. Furthermore, due to the geometry of LEO configurations and TC tracks, it is not rare for a specific LEO satellite to have no observations of a specific TC for 2-3 days consecutively. Thus, it would be highly desirable to have an accurate, objective center-fixing method that relies only on GEO data.

The remainder of this section provides an in-depth review of existing center-fixing methods, ending with a short description of the method developed herein – a deep-learning algorithm called GeoCenter. GeoCenter uses only data available in real time – primarily IR data from GEO satellites – making it deployable in operations. Table 1 provides a list of acronyms for this entire article.

Table 1: List of acronyms.

| Acronym | Definition | Acronym | Definition |
|---|---|---|---|
| ABI | Advanced Baseline Imager onboard GOES-16/17/18 satellites | AHI | Advanced Himawari Imager onboard Himawari-8/9 satellites |
| AL | North Atlantic ocean basin | ARCHER-2 | Automated Rotational Center Hurricane Eye Retrieval |
| ATCF | Automated Tropical Cyclone Forecasting System | FBT | Final best track |
| CNN | Convolutional neural network | CPHC | Central Pacific Hurricane Center |
| CRPS | Continuous ranked probability score | DTMF | Monotonicity fraction from discard test |
| EESD | Euclidean ensemble standard deviation | EP | Eastern North Pacific ocean basin |
| GEO | Geostationary | IR | Infrared |
| JTWC | Joint Typhoon Warning Center | ML | Machine learning |
| MME | Multi-model ensemble | NHC | National Hurricane Center |
| REL | Reliability | RHD | Rank-histogram deviation |
| RMS | Root mean square | RMSD | Root mean squared distance (Euclidean) |
| RMSE | Root mean squared error | SSRAT | Spread-skill ratio |
| SSREL | Spread-skill reliability | TC | Tropical cyclone |
| UQ | Uncertainty quantification | WP | Western North Pacific ocean basin |

*a. Approaches based primarily on Dvorak technique*

The subjective Dvorak (1975, 1984) technique is the primary center-fixing method used by operational TC-forecasting centres worldwide, including the National Hurricane Center (NHC), Central Pacific Hurricane Center (CPHC), and Joint Typhoon Warning Center (JTWC). The main purpose of the Dvorak technique is to estimate the current TC intensity, which requires center-fixing as a first step. To perform center-fixing, the Dvorak technique categorizes the dominant cloud pattern in the TC (*e.g.*, curved band, central dense overcast, shear, etc.) and then fits a category-dependent cloud template to the satellite image. The Dvorak technique uses a single visible/IR image, rather than a time series (animation).

Several studies have taken steps to automate (*i.e.*, make objective) the Dvorak technique, leading to the objective Dvorak technique (ODT; Velden et al. 1998), advanced ODT (AODT; Olander et al. 2002), and advanced Dvorak technique (ADT; Olander and Velden 2007), which is now operational at the NOAA National Environmental Satellite, Data, and Information Service (NESDIS) (Olander and Velden 2019). The ADT uses ARCHER-2 for center-fixing, which incorporates parallax correction, spiral-fitting to curved cloud bands, ellipse-fitting to the inner eyewall (if present), thresholding of microwave imagery (if available), and fitting of ambiguity vectors from scatterometer retrievals (if available).

Other efforts to automate center-fixing are described below; note that none of these methods has been operationalized.

*b. Approaches based solely on fitting specific features*

Several methods focus on either spiral-fitting of curved cloud bands or ellipse-fitting of smaller features such as the inner eyewall or central dense overcast (CDO). In the spiral-fitting literature, Jaiswal and Kishtawal (2010) used image segmentation to isolate TC clouds from other systems, followed by image-processing techniques to reduce noise (*e.g.*, smoothing and filtering), and finally fitting a logarithmic spiral band to the denoised IR image. Lu et al. (2019) applied a similar approach to fused multi-band (visible, water vapour, and IR) imagery, while Shin et al. (2022) applied a similar approach to two-channel (longwave IR and water vapour) imagery, averaging center locations estimated from the two channels into an ensemble.

In the ellipse-fitting literature, Chaurasia et al. (2010) fit contours to the five coldest brightness temperatures ($T_b$) in the CDO region – starting with the minimum $T_b$ and incrementing by 2 K for each contour – then computed the center of each contour and took the mean of the five centers as the TC center. Wang et al. (2020) used edge detection to segment clusters of cold cloud tops (CCT) from two-channel IR imagery, then used a Hough circle transformation (Hough 1962; Duda and Hart 1972) to find the center of each CCT cluster, taking the mean center as the TC center. You et al. (2022) applied a similar approach to two-channel (one visible and one IR) imagery, but without the edge-detection step of Wang et al. (2020), which involves finding skeleton points and is sensitive to noise.

Shin et al. (2022) found that their method outperforms ARCHER-2 on a dataset of 190 TCs in the western North Pacific. However, a notable shortcoming of both spiral- and ellipse-fitting methods is that they apply only to a subset of TCs with certain structures (*e.g.*, clear spiral patterns or approximate elliptical symmetry near the TC center). For instance, Wang et al. (2020) note that their method can locate only "some specific types of TC centers," not including tropical depressions or TCs undergoing extratropical transition.

*c. Approaches using remotely sensed wind estimates*

Remotely sensed wind estimates can be at cloud level – *e.g.*, atmospheric motion vectors (AMV) derived from GEO satellite data – or at sea-surface level, derived from scatterometers or GPS-signal reflections.

Center-fixing methods based on cloud-level wind include Zheng et al. (2019), who computed AMVs from IR imagery provided by the Gaofen-4 satellite[1], converted the AMVs to a field called magnitude of the mean of the direction vectors (MMDV), and then took the minimum of this field as the center of concentric motion – *i.e.*, the TC center. However, the computation of high-quality AMVs was made possible by the extremely high spatial and temporal resolution of the Gaofen-4 satellite. It is unclear how well the approach of Zheng et al. (2019) would generalize to typical GEO satellites, which have much coarser resolution (10 min and 0.5-2.0 km, compared to 20 s and 50-400 m for Gaofen).

Methods based on sea-surface wind include Mayers and Ruf (2019), who developed a center-fixing system called MTrack, based on GPS-signal reflections from the Cyclone Global Navigation Satellite System (CYGNSS). They showed that MTrack reduces uncertainty in center fixes from the final best-track dataset (Section 2a), but it is unclear how well this system works for asymmetric storms. Also, for weak storms without a well defined structure, scatterometers do not allow for precise localization of features, due to directional ambiguity (Lin et al. 2013). Thus, center-fixing methods based on remotely sensed wind seem to work best for complementing other methods.

*d. Deep-learning approaches*

The foregoing automated methods are considered expert systems (Reiss and Hofmann 1988; Kumar et al. 1994), *i.e.*, algorithms with preset rules defined by humans. While human expertise is crucial, expert systems are often outperformed by statistical models that can learn autonomously from large datasets, in particular deep learning (Goodfellow et al. 2016). In the last few years, deep learning, especially convolutional neural networks (CNN), have become popular tools for center-fixing. Unlike traditional NNs and other statistical models, CNNs can directly ingest image data, including animations and multispectral imagery (*e.g.*, multiple satellite channels). For example, Wang et al. (2019) trained a CNN on 4-channel IR imagery, with the image center deviating from the true TC center in both the *x*- and *y*-directions, and tasked the CNN with finding the true TC center. Yang et al. (2019) used a more complex architecture, based on the U-net (Ronneberger et al. 2015), to perform semantic segmentation, producing a probability-of-TC-center-presence at every pixel in the satellite image. Wang and Li (2023) applied a ResNet (He et al. 2016) architecture with transfer learning[2] to 3-channel imagery, with the image center deviating from the true TC center in only one direction. Smith and Tuomi (2021) trained three CNNs – each on animations from one IR channel, containing lag times of 0-5 hours – with the image center being the true TC center from 5 hours ago. They averaged center locations estimated from the three CNNs into an ensemble. Ho et al. (2024) trained an ensemble of CNNs and convolutional long-short-term memory (ConvLSTM) models – using animations from multiple IR channels and the visible channel during the day, containing lag times of 0-50 min – with the image center being a 6- or 12-hour forecast from the ECMWF. We compare the performance of GeoCenter with several of these deep-learning approaches and ARCHER-2 (Section 5a).

---

[1] A GEO satellite launched in 2015, stationed at 105.5°E, with very high spatial (50 m for visible, 400 m for IR) and temporal (20 s) resolution. See Xin et al. (2022) for details.

*e. Our deep-learning approach*

To overcome the limitations of existing center-fixing methods, particularly those that hinder operational deployment, we present GeoCenter: a deep-learning method powered by an ensemble of CNNs. Below, we highlight the key features of GeoCenter that enhance its suitability for operational use.

1. All input data – including IR satellite images and nine scalars from the Automated Tropical Cyclone Forecasting System (ATCF; Sampson and Schrader 2000) – are available in real time.

2. The only satellite data used are geostationary IR images, available for all TC basins during both day and night with 10–15-min resolution and < 10-min latency.

---

[2]The model was pre-trained on photographs from the ImageNet dataset (Deng et al. 2009) – where the possible classes are "cat," "dog," "envelope," etc. – and then fine-tuned to detect TC centers instead.

3. Preliminary results suggest that the performance of GeoCenter is similar to the best ARCHER-2 estimates, which are obtained from microwave and scatterometer data.

4. GeoCenter is trained and evaluated on 377 TCs from three ocean basins – the North Atlantic, eastern North Pacific, and western North Pacific – using the full lifetime of every TC, including when classified as a disturbance/low, depression, subtropical storm, or extratropical storm. This makes GeoCenter nearly global and applicable to a wide variety of systems.

5. GeoCenter provides uncertainty quantification (UQ), which can help forecasters gauge the uncertainty in each center fix. While ARCHER-2 provides UQ as well, in the form of a confidence score, GeoCenter provides a full set of possible solutions via ensemble prediction, and we verify that these ensembles are well calibrated.

6. GeoCenter leverages multispectral animations, containing data from 9 IR channels at lag times extending up to 4 hours in the past. The basic idea behind this design choice is that subjective center-fixing by human forecasters is aided by considering multiple channels and time steps. Only one existing automated method (Smith and Tuomi 2021) uses multispectral animations.

## 2. Input data

This section discusses the labels and predictors used to train CNNs. Labels (or "ground truth") are the correct answers used to evaluate CNNs.

### a. Labels

The ATCF is an operational software package and database with information on multiple TC properties, including the intensity, structure, and center location. This information includes current analyses, official forecasts from TC-forecasting centres in the U.S., and forecasts from several models. The ATCF is also used as a comprehensive database for historical TCs. After every hurricane season, operational centres use this database to generate final "best track" data, containing the best estimates for 6-hourly center locations (Landsea and Franklin 2013). Since these final best tracks (FBT) are created by reanalyzing all available post-event data, they are sometimes different from preliminary best tracks generated during the season. However, FBTs are not the ultimate truth. Since FBTs are generated from multiple data sources, each with inherent uncertainty, they represent a smoothed storm track that may not account for high-frequency changes. For details on FBT accuracy, see Landsea and Franklin (2013). Despite these caveats, FBT is the best available archive of TC locations, so we use these data to create labels for CNN-training. For convenience, we will refer to the FBT center as the "true" center.

FBTs are available for every global TC throughout its lifetime, which may include many storm types: tropical depression/storm, hurricane, disturbance, low, tropical wave, or remnant system (extra-/post-/sub-tropical). We train and evaluate GeoCenter on all 6-hourly time points, including all these storm types.

Due to computing limitations, we train models with IR data at 30-min, rather than 10-min, time intervals. To match the IR satellite data (Section 2b) and increase the size of our training dataset, we interpolate FBT center fixes to 30-min time steps. Specifically, we use a cubic B-spline with no smoothing to interpolate both the latitude and longitude coordinates. To assess the accuracy of this interpolation, we conduct an experiment: we interpolate using *only* synoptic-time FBT centers as anchor points, reserving the special (off-synoptic-time)[3] FBT centers for evaluation. Specifically, at every special time in the FBT dataset, we compare the interpolated center fix with the FBT center fix, taking the latter as truth. The error distribution (Figure 1a) has a mean of 9.6 km and median of 7.3 km. Since FBT center fixes are rounded to the nearest $0.1°$ in latitude and longitude – yielding a representation error of 7.9 km at the equator and 6.8 km at $45°$N – we consider these interpolation errors acceptable. Thus, we train CNNs with data at 30-min time steps, most of which have an interpolated label (Figure 1b), with both synoptic and special times as anchor points for the interpolation. However, for validation and testing we evaluate only at synoptic times, which have more accurate labels. Also, note that for an eventual operational implementation of GeoCenter, the trained model could be applied with higher cadence than once every 30 min (*e.g.*, to new satellite data every 10 min).

---

[3]Synoptic times – 0000, 0600, 1200, and 1800 UTC every day – comprise the vast majority of the FBT data. Special times often correspond to landfall.

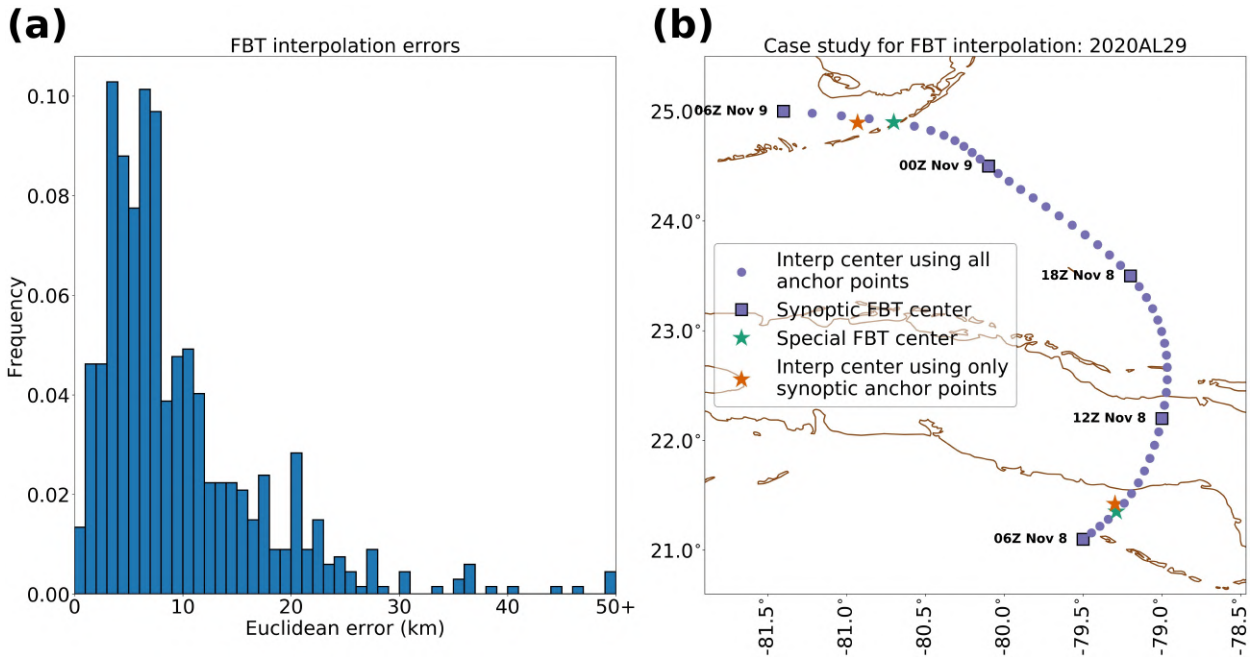

Figure 1: Interpolation of FBT centers, using the B-spline method discussed in the main text. [a] Histogram of interpolation errors. [b] Case study for a 24-hour period (0600 UTC 8 Nov – 0600 UTC 9 Nov 2020) of Tropical Storm Eta (AL292020). A "special FBT center" (see legend) is one at an off-synoptic time. For each special FBT center (green star), there is an interpolated center (orange star) valid at the same time, the interpolation using only synoptic FBT centers as anchor points. The purple dots are from a second, more accurate, interpolation, which uses all FBT centers (synoptic and special) as anchor points. The interpolation error at the first special time (0800 UTC 8 Nov) is 7.9 km; the error at the second special time (0400 UTC 9 Nov) is 23.1 km.

*b. Predictors: IR satellite data*

Our first source of predictor data is IR imagery from geostationary sensors: the GOES-16 Advanced Baseline Imager (ABI) for the AL basin, GOES-17/18 ABI for the EP basin, and Himawari-8/9 Advanced Himawari Imager (AHI) for the WP basin. We use a CIRA satellite archive to generate TC-centered images at 30-min time steps; although geostationary data are available every 10 min, we subsample due to computing limitations. We generate images throughout the full lifetime of every TC in the AL from 2017-2021, in the EP from 2018-2022, and in the WP from 2016-2021. For every TC sample (definition: one TC at one time step), we reproject imagery from the CIRA database to a 2500-by-2500-km grid with 2-km spacing centered on the FBT center fix, using the *plate carrée* projection as specified in the `pyresample` library. The reprojected imagery contains all ten IR channels available on the GOES ABI and Himawari AHI; see Figure 2 and Table 2. Although the ABI and AHI channels differ slightly in wavelength, we train GeoCenter with data from both sensors, making no adjustment to account for inexact correspondence between channels. An alternative approach would be training a separate version of GeoCenter for each sensor, but this would result in less training data for each of the two versions. A benefit of our approach – with a unified GeoCenter model for ABI and AHI – is that the unified GeoCenter model should generalize well to other geostationary sensors with slight wavelength differences, such as the Geo-Kompsat-2A Advanced Meteorological Imager (AMI) or Meteosat Third Generation (MTG) Flexible Combined Imager (FCI).

We have developed our own quality-control algorithm for the IR data, described in Appendix A, which correctly rejects samples affected by the loop heat pipe (NOAA STAR 2019; Van Naarden and Lindsey 2019; Section 3.3 of Goodman et al. 2019) issue on the GOES-17 satellite.

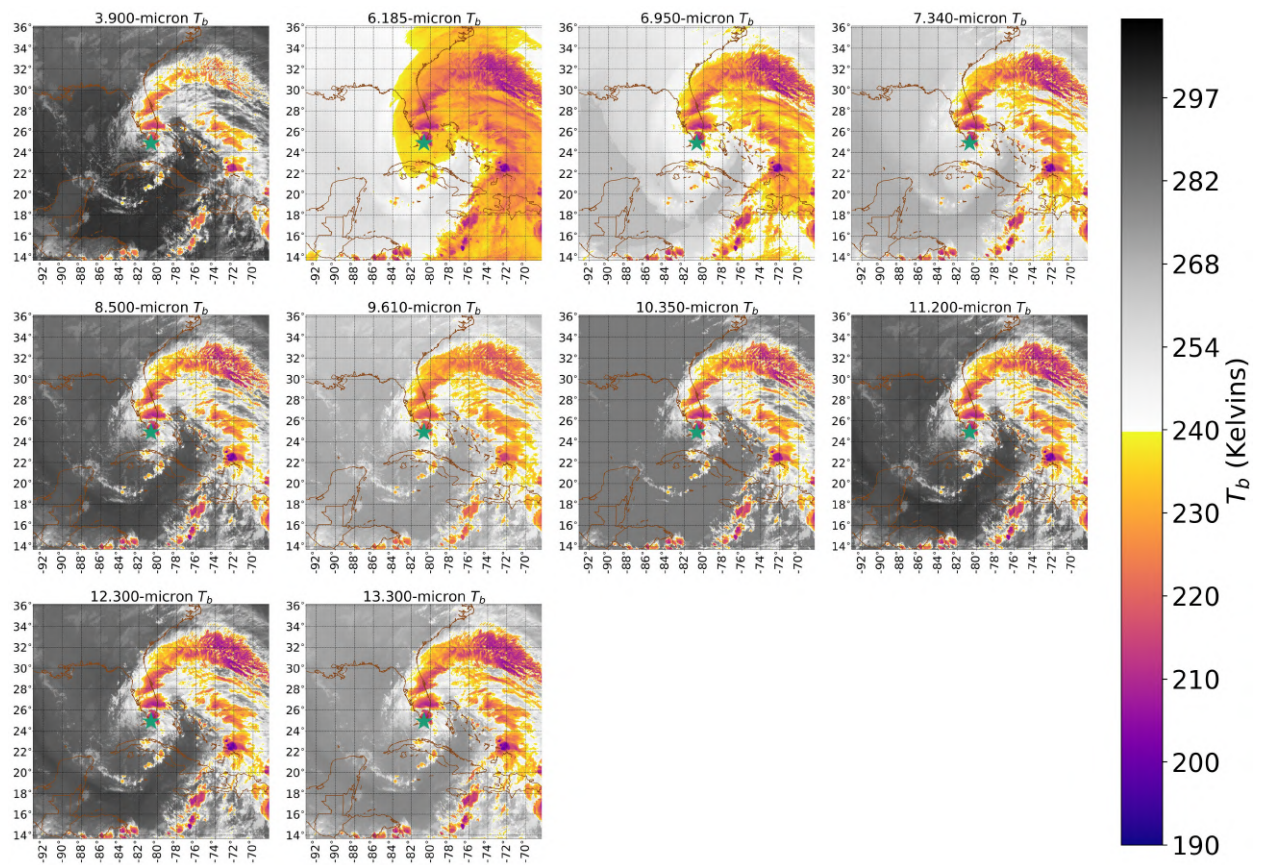

Figure 2: IR satellite imagery valid at 0400 UTC 9 Nov 2020 for Tropical Storm Eta (AL292020). Each panel shows brightness temperature at one spectral channel (colour fill) and the TC center from FBT (green star).

Table 2: IR channels used in this study. Channel *k* from the GOES ABI and channel *k* from the Himawari AHI are treated as equivalent, even in cases where the central wavelength or spectral response function is slightly different.

| Channel | Central wavelength for GOES-16/17/18 ABI ($\mu$m) | Central wavelength for Himawari-8/9 AHI ($\mu$m) |
|---|---|---|
| 7 | 3.9 | 3.9 |
| 8 | 6.185 | 6.2 |
| 9 | 6.95 | 6.9 |
| 10 | 7.34 | 7.3 |
| 11 | 8.5 | 8.6 |
| 12 | 9.61 | 9.6 |
| 13 | 10.35 | 10.4 |
| 14 | 11.2 | 11.2 |
| 15 | 12.3 | 12.4 |
| 16 | 13.3 | 13.3 |

*c. Predictors: ATCF scalars*

Our second source of predictor data is scalar TC properties from the ATCF. We use the CARQ (combined automated response to query) line in the A-deck file, which provides real-time estimates of TC properties. Specifically, for every TC sample, we extract the nine scalars listed in Table 3 from the second-most recent synoptic time step, 6 to 12 hours ago. Thus, all GeoCenter predictors – both IR satellite images and ATCF scalars – include only data available in real-time operational settings.

Table 3: Scalar ATCF variables used as predictors in this study. Variables below the horizontal line are related to storm type. "Spatial maximum (minimum)" is a maximum (minimum) over all points inside the TC.

| Variable | Explanation |
| --- | --- |
| Latitude | Latitude of TC center. |
| Sine of longitude | Longitude of TC center. Because longitude is a circular variable (*e.g.*, the difference between -179° and +179° is 2°, not 358°), we use the sine and cosine. |
| Cosine of longitude | See above. |
| Intensity | Spatial-maximum sustained wind speed (1-minute average) at 10 m above ground |
| Minimum pressure | Spatial-minimum pressure at mean sea level |
| **Storm type** | |
| Tropical flag | Binary flag. If the ATCF storm type is TD (tropical depression), TS (tropical storm), TY (typhoon), ST (super typhoon), TC (tropical cyclone), or HU (hurricane), the flag is 1. Otherwise, the flag is 0. |
| Subtropical flag | Binary flag. If the ATCF storm type is SD (subtropical depression) or SS (subtropical storm), the flag is 1. Otherwise, the flag is 0. |
| Extratropical flag | Binary flag. If the ATCF storm type is EX (extratropical) or PT (post-tropical), the flag is 1. Otherwise, the flag is 0. |
| Disturbance flag | Binary flag. If the ATCF storm type is DB (disturbance) or LO (low), the flag is 1. Otherwise, the flag is 0. |

*d. Pre-processing: Splitting and normalization*

Following common practice, we split our data into three subsets: training, validation, and testing. The training data are used to optimize CNN parameters (weights and biases); the validation data, to optimize hyperparameters (settings that remain static during training, like number of neurons/layers, learning rate, etc.); and the testing data, to evaluate the selected CNN model on independent data. We split the data by basin-year, as shown in Table 4.

After splitting, we normalize the predictor variables. Specifically, for each TC sample, the following steps are applied independently to 15 predictors (all 10 IR channels and the 5 ATCF scalars that are not binary flags).

1. Quantile normalization. Convert the variable $x$ to its quantile over the distribution of $x$ in the training data. Let the result be $x'$, ranging from $[0, 1]$.

2. $z$-score normalization. Convert $x'$ to a $z$-score, using the inverse cumulative distribution function (CDF) for the standard normal distribution. Let the result be $x''$, which ranges from $[-4.75, +4.75]$.[4]

Table 4: Splitting of data into independent subsets. One "TC sample" is one TC at one time step. In the training data these time steps are every 30 min; in the validation and testing data these time steps are every 6 h (synoptic times only).

| Data subset | Basin-years | Number of unique TCs | Number of TC samples |
|---|---|---|---|
| Training | AL 2017-2019, EP 2018-2020, WP 2016-2019 | 234 | 62 740 |
| Validation | AL 2020, EP 2021, WP 2020 | 76 | 1160 |
| Testing | AL 2021, EP 2022, WP 2021 | 67 | 1073 |

---

[4]In general, $z$-scores range from $(-\infty, +\infty)$. For example, the inverse CDF of 0 is $-\infty$, and the inverse CDF of 1 is $+\infty$. To avoid infinities and extremely large values – which both hamper CNN performance – we clip $x'$-values to the range $[0.000001, 0.999999]$, which results in $x''$-values ranging from $[-4.75, +4.75]$.

## 3. Methods

Table 5 provides an overview of the challenges involved in machine learning (ML)-based center-fixing and how we overcome those challenges. For discussions involving hardware, note that each CNN is trained with eight Tesla P100 graphics-processing units (GPU) and 16 GB of memory per GPU on NOAA's Hera supercomputer.

Table 5: Overview of challenges involved in ML center-fixing. Note that this table is an overview of the paper and therefore contains many yet-to-be-defined concepts.

| Challenge | Solution | Solution described in |
|---|---|---|
| • **Features at many scales are important.** Features used for operational center-fixing range from 10s of km (inner eyewall, convective cells) to 100s of km (spiral rain bands).<br>• Thus, the NN must "look at" both very small and very large areas. In ML jargon: convolutional filters need receptive fields of many different sizes.<br>• NNs have small receptive fields by default; large receptive fields require a "wide" or "deep" architecture (Nguyen et al. 2020).<br>    1. Wide NN: large filters ($9 \times 9$ pixels, $11 \times 11$, or even greater).<br>    2. Deep NN: small filters ($3 \times 3$) but many pooling layers, which serially coarsen image resolution so that for the deepest convolutional layers, a 3-by-3-pixel filter covers a large physical area. | • **We opt for the deep architecture**, which is less memory-intensive. | Section 3a |
| • **Small sample size.** Our training dataset has a *nominal* sample size of 62 740, but these samples come from 234 unique TCs (see Table 4). Temporal autocorrelation within each TC makes the *effective* sample size $\ll$ 62 740.<br>• Furthermore, the original images are already TC-centered, leaving nothing for the NN to do. | • **Data augmentation** (DA). For each original image, use random translations ("jittering") to create many new images.<br>• Each new image contains a different **first-guess error** (offset between image center and true TC center) to be corrected by the NN. **This allows the NN to learn to correct different kinds of errors.**<br>• DA increases the *nominal* sample size more than the *effective* sample size, because all images created by translating one original image are closely related. Nonetheless, DA increases effective sample size by a little. | Section 3b |
| • **High dimensionality.** One TC sample has dimensions of 1250 (rows) $\times$ 1250 (columns) $\times$ 10 (IR channels).<br>• If we also train the NN with a video containing $K$ lag times, dimensions grow to $1250 \times 1250 \times 10 \times K$ *for one data sample*.<br>• *A priori*, **we do not know what information in this hypercube is most relevant.** How much spatial context (grid size) is needed for accurate center-fixing? How much temporal context (lag times)? How much spectral info (channels)?<br>• Furthermore, using all available data leads to slow training and out-of-memory errors. | • We perform a **hyperparameter experiment** to determine what information is most relevant. | Sections 3e and 4 |
| • **Ensemble adds complexity.** We provide an ensemble of TC-center locations, not a single estimate. Evaluating an ensemble – or any prediction that includes uncertainty quantification (UQ) – requires **additional evaluation tools**. | • We use the continuous ranked probability score (CRPS), spread-skill plot & associated metrics, discard test & assoc. metrics, rank histogram & assoc. metrics. | Section 3f |

*a. CNN architectures*

Our CNN architecture is similar to the temporal U-net introduced by Chiu et al. (2020), which is specially designed for predicting a future image based on an animation (time series) of images from the recent past. An example[5] is shown in Figure 3. Our architecture is narrow and deep, which is a common choice in the ML literature (Nguyen et al. 2020). "Narrow" means that we use small convolutional filters (here, $3 \times 3$ pixels); "deep" means that we use many pooling layers to detect features at many spatial resolutions (here, doubling successively from the original 2-km grid spacing to 128 km). Thus, for the shallowest convolutional layers (left column of figure), the physical size of the receptive field is $6 \times 6$ km – while for the deepest convolutional layers (right column), the receptive field is $384 \times 384$ km. This allows the CNN to detect important features at many scales, ranging from the inner eyewall and convective cells (10s of km) to spiral rain bands (100s of km). The CNN outputs estimates of the offset between the image center and TC center, in row-column coordinates. More details about our architecture can be found in Appendix B; a full list of hyperparameter settings can be found in Supplemental Section 1.

---

[5]With predictors including 300-by-300-pixel IR images at 9 lag times and 3 channels. Note that different predictor sets lead to slightly different architectures.

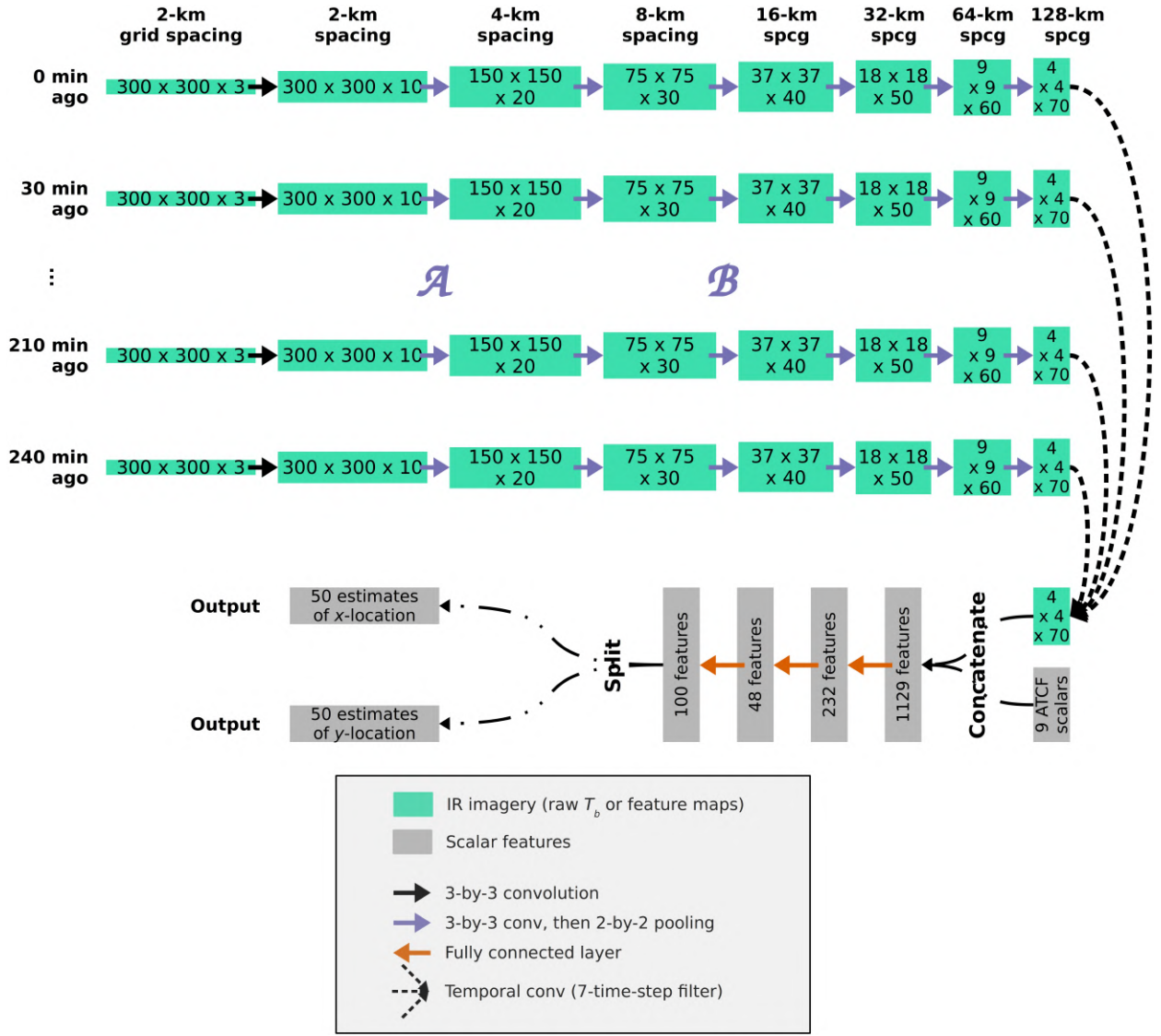

Figure 3: One of our specific CNN architectures – with IR predictors (green boxes at left) including 9 lag times and a 300-by-300-pixel domain. The numbers in each green box are tensor dimensions. For the raw IR imagery (leftmost column), the dimensions are $N_{\text{rows}} \times N_{\text{columns}} \times N_{\text{wavelengths}}$; for all other green tensors (which have passed through at least one CNN layer, so can no longer be interpreted as raw IR imagery), the dimensions are $N_{\text{rows}} \times N_{\text{columns}} \times N_{\text{feature maps}}$. Each column of convolutional layers (*e.g.*, those marked "$\mathcal{A}$" and "$\mathcal{B}$") is time-distributed, meaning that weights are shared across the layers. The dashed black arrows at the right are the "forecasting module" of Chiu et al. (2020), which uses temporal convolution to combine information across the nine lag times of IR imagery. Besides the IR imagery, CNN predictors include the ATCF scalars (grey box at bottom-right). These scalars are concatenated with flattened feature maps from the forecasting module, yielding a feature vector with length $4 \times 4 \times 70 + 9 = 1129$. This feature vector is passed through three fully connected layers, terminating with a length-100 vector. The first (last) 50 elements are taken as an ensemble of estimates for the *x*- (*y*-)location of the TC center.

## b. First-guess centers for CNN-training

During pre-processing, we center every IR image on the TC center from FBT, which is the best available representation of the truth. Thus, in image-center-relative coordinates, the true TC center is always at $x = y = 0$ km. However, in real time we do not know the true TC center. Instead, we have access to a forecast, which could be an official NHC/CPHC/JTWC forecast or an extrapolated track from the previous forecast cycle. This forecast is often called the "first guess". To mimic an operational setting for CNN-training, we create our own first guesses. Specifically, we randomly translate each IR image, creating an offset between the image center and true TC center, and task the CNN with correcting this offset. In the ML literature, applying such random perturbations to the input data is often called data augmentation (Section 5.2.5 of Chollet 2018).

The data presented to the CNN include a time series of recent IR imagery, representing the TC's recent track. In an operational setting, errors occur in the track's absolute geographic position (whole-track error) and its shape (track-shape error). Thus, our procedure for creating first-guess centers includes both types of error (Figure 4).

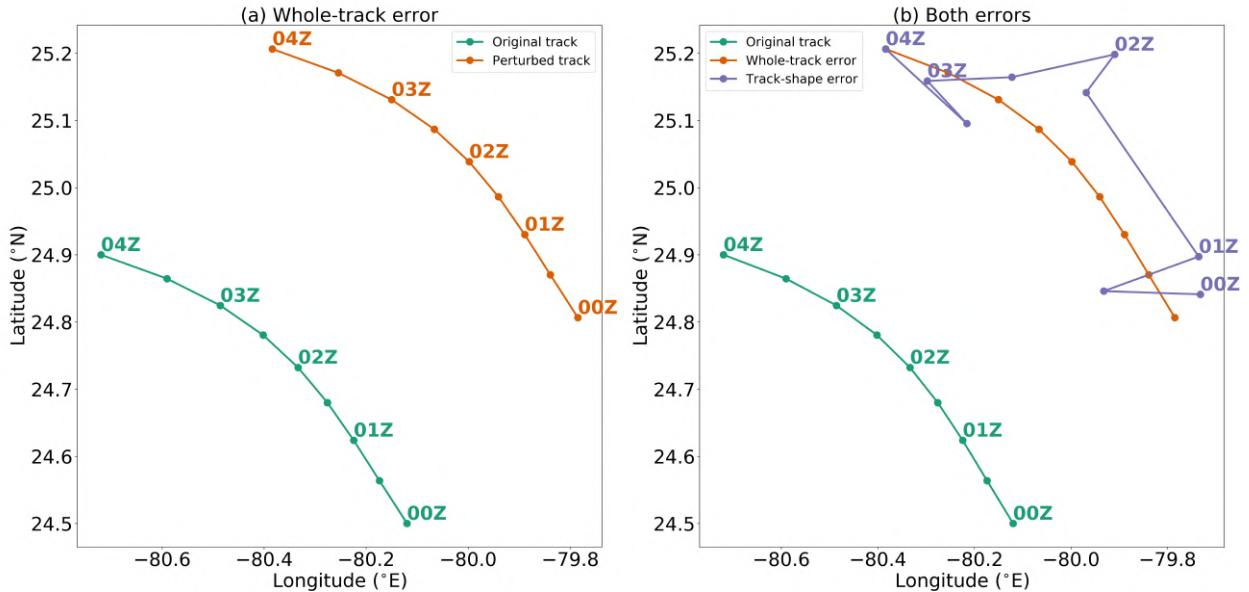

Figure 4: The two types of error applied to create first-guess TC centers for CNN-training. [a] Whole-track error (WTE) only, wherein the shape of the track does not change. This does not realistically mimic an operational setting, where there would be a different error at each lag time. [b] Both WTE and track-shape error (TSE). Note that TSE is applied independently to every non-zero lag time, *i.e.*, every time step except the last. However, since WTE ≫ TSE, total error (= WTE + TSE) is highly autocorrelated between time steps.

Our procedure for whole-track error is as follows. The same translation vector is applied to every 2-D slice of the CNN input, *i.e.*, every pair of lag time and IR channel. The distribution in step 1 is similar to first-guess errors seen in operations. For example, based on short-track data (described in Section 5b) obtained for the AL/EP/WP basins in 2024, the mean Euclidean error is 42.4 km. This is based on short-track forecasts valid at synoptic times and initialized three hours earlier, *i.e.*, halfway between synoptic times.

1. Draw the translation vector from a random distribution. Specifically, draw the length from $\max\{0 \text{ km}, \mathcal{N}(48 \text{ km}, 24 \text{ km})\}$ – which is a zero-truncated Gaussian distribution with mean of 48 km and standard deviation of 24 km – and the direction from $\mathcal{U}(0, 2\pi)$, which is a uniform distribution over the unit circle.

2. The distribution of translation vectors (first-guess errors) is shown in Figure 5a. If the CNN did nothing, this would be the error distribution for GeoCenter.

3. Convert the translation vector to Cartesian coordinates ($\Delta x$ and $\Delta y$ in km), then to row-column coordinates ($\Delta r$ and $\Delta c$ in number of pixels). Since the IR imagery has 2-km grid spacing, we use the equations $\Delta r = g(\frac{\Delta y}{2 \text{ km}})$ and $\Delta c = g(\frac{\Delta x}{2 \text{ km}})$, where $g()$ rounds to the nearest integer.

4. Translate the IR image by the given $\Delta r$ and $\Delta c$, avoiding edge effects. For example, if the domain size used by the CNN is 300 × 300, the pre-translation image must be larger than 300 × 300. Otherwise, the post-translation image will have an obvious border, which the CNN can easily use to "cheat". For example, if the border in Figure 5c is 30 rows thick on the south edge of the image and 40 columns thick on the east edge, the CNN can trivially determine that $\Delta r = +30$ and $\Delta c = -40$.

5. The CNN is tasked with estimating $\Delta r$ and $\Delta c$, providing a 50-member ensemble for each coordinate (Figure 3).

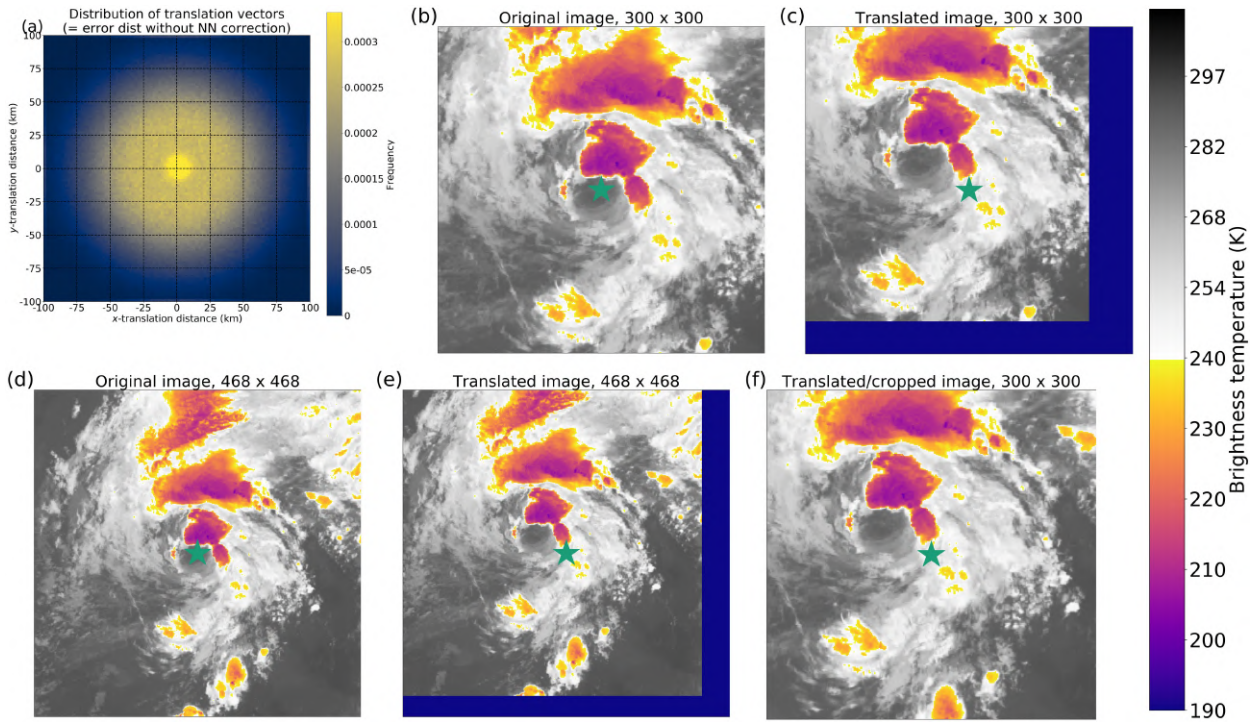

Figure 5: CNN-training with random first guesses. [a] Distribution of translation vectors for whole-track error, *i.e.*, offsets between the first-guess center and true TC center at lag time = 0 min. [b-c] First-guess procedure without accounting for edge effects – not ideal, because the CNN can use the size of the image border in panel c to "cheat". [d-f] First-guess procedure with a buffer zone to eliminate edge effects. The *translated and cropped* image (panel f) is the one used for CNN-training.

The image center in panels b-f is marked with a green star. The procedure illustrated in panels d-f is used for both whole-track error and track-shape error. For each lag time, the two errors are added and the procedure is applied once (keeping in mind that track-shape error is 0.0 km for the 0-min lag time).

Our procedure for track-shape error is identical but with two exceptions. First, the length of the translation vector is drawn from $\max\{0 \text{ km}, \mathcal{N}(10 \text{ km}, 5 \text{ km})\}$, which makes track-shape error $\ll$ whole-track error. Second, a different translation vector is drawn for every non-zero lag time, as shown in Figure 4b.

We train the CNN with batches of 20 TC samples each, applying the whole first-guess procedure (whole-track and track-shape errors) to each sample. One batch never contains more than one sample from the same TC, which reduces in-batch autocorrelation and increases in-batch diversity. Because the CNN is trained over many epochs, every TC sample (*e.g.*, Hurricane X at time *t*) is used more than once, but with a different first-guess error every time. This allows the CNN to learn to correct different kinds of errors.

*c. Uncertainty quantification*

Uncertainty can be decomposed into two parts: aleatory uncertainty, arising from deficiencies of the training data, and epistemic uncertainty, arising from deficiencies of the ML model itself. Most methods for machine-learned uncertainty quantification (ML-UQ) resolve only one type of uncertainty, so we combine two methods. To resolve aleatory uncertainty, we use the continuous ranked probability score (CRPS) method, where the CNN is trained to produce a 50-member ensemble that minimizes the CRPS loss function. To resolve epistemic uncertainty, we use the multi-model ensemble (MME) method, which involves training several CNNs. In a typical MME, all models have the same hyperparameters, differing only in the random seed used to initialize weights before training. But in our MME, one hyperparameter varies among the models, namely the set of IR channels used in the predictors. To combine the CRPS and MME methods, we train four CNN models – each with its own set of IR channels (Table 6) – with the CRPS loss function and an ensemble size of 50.

By default, we would train each CNN in the ensemble with all 10 channels. However, in an earlier channel-selection experiment (not shown), we found that 10-channel CNNs perform poorly. This is because 10-channel CNNs are very expensive – in terms of both memory, limiting the complexity of CNNs we can train, and time, limiting the number of training epochs that can be completed. In the same experiment we found that 1-channel CNNs perform poorly due to a lack of spectral information, while 3-channel CNNs provide a good compromise between the two extremes. We then performed a second experiment to determine the best 3-channel sets out of $\binom{10}{3} = 120$ possibilities. For that experiment we used the same model-selection methods described in Section 3f, but with one extra constraint. Namely, we chose the best four 3-channel sets such that all 10 channels are used in at least one CNN. The selected channel sets are shown in Table 6; note that one of the four channel sets is omitted from the final GeoCenter ensemble, for reasons discussed in Section 5.

For more background on ML-UQ techniques, see Haynes et al. (2023); for more details on our methodology, see Appendix C.

Table 6: Channel sets used to train CNNs. By experimental design, the ensemble should contain four CNNs, each using a different channel set in the predictors. However, as discussed in Section 5, the final GeoCenter ensemble contains three CNNs, omitting the third channel set listed below. Ranks in this table are based on an earlier channel-selection experiment, discussed briefly in Section 3c.

| Rank in earlier channel-selection experiment | Channels in set (band number) | Channels in set (GOES ABI wavelength in $\mu$m) |
|---|---|---|
| Best | 11, 12, 15 | 8.5, 9.61, 12.3 |
| Second-best | 7, 10, 16 | 3.9, 7.34, 13.3 |
| Third-best | 7, 8, 9 | 3.9, 6.185, 6.95 |
| Fourth-best | 9, 13, 14 | 6.95, 10.35, 11.2 |

*d. Bias correction*

For every CNN, we bias-correct the ensemble mean via isotonic regression (Barlow and Brunk 1972). Isotonic regression is a post-processing method, trained separately from the CNN. See Appendix D for details.

*e. Hyperparameter experiment*

This experiment aims to find the optimal dimensions for IR predictors. Namely, we optimize two hyperparameters – animation length (number of lag times) and domain size (number of pixels), which control the amount of temporal and spatial context available to the CNN. Also, these hyperparameters largely control the memory required for CNN-training, which is important because our hardware is memory-limited. We experiment with one additional hyperparameter: whether to train with first-guess errors drawn from a Gaussian distribution, as in Section 3b, or from a uniform distribution. Specifically, we may draw whole-track error from the distribution $\mathcal{U}(0 \text{ km}, 120 \text{ km})$, instead of $\max\{0 \text{ km}, \mathcal{N}(48 \text{ km}, 24 \text{ km})\}$. The uniform distribution leads to a higher average first-guess error and a higher frequency of very large first-guess errors (*e.g.*, > 100 km). We speculate that training with this wider distribution could make the CNNs more resilient to large first-guess errors. However, the trade-off is that training with a wider error distribution makes the center-fixing task inherently more difficult, which could hinder successful learning. At inference time (for validation and testing), we draw whole-track error from the Gaussian distribution, which is closer to the first-guess errors seen in operations. Note that CNNs have many additional hyperparameters, besides the three optimized in this experiment. See Supplemental Section 1 for a list of fixed hyperparameters and the values we choose.

We run the experiment as a grid search, trying all $6 \times 4 \times 2 = 48$ combinations of the values listed in Table 7. In this experiment we use the CRPS-only method for UQ, which requires one CNN per hyperparameter combination, hence 48 CNNs in total. The hybrid CRPS/MME method – used for our final model (Section 3c) – would require an ensemble of four CNNs per hyperparameter combination, hence $48 \times 4 = 192$ CNNs in total. The former approach takes about one week of wall-clock time on our hardware, while the latter approach would take a prohibitive four weeks. We train all CRPS-only models in this experiment with the same set of IR channels: the first listed in Table 6. After determining the best CRPS-only model (based on validation data), we train four CRPS-only models with the same hyperparameters, except using a different set of IR channels for each. These four models comprise the final CRPS/MME model, which we evaluate on testing data, using the evaluation metrics outlined below.

Table 7: Experimental hyperparameters. Exact times in the animation are spaced at 30-min intervals. For example, when the animation length is 3 time steps, the time steps included are $t_0, t_0 - 30$ min, and $t_0 - 60$ min, where $t_0$ is both the initialization time and the valid time. In other words, the task is to find the TC's current center – at $t_0$ – using IR imagery from $t_0$ and the recent past.

| Hyperparameter | Values attempted |
|---|---|
| Number of lag times | 1, 3, 5, 7, 9, 11 |
| Domain size (pixels per side) | 300, 400, 500, 600 |
| Domain size (km per side) | 600, 800, 1000, 1200 |
| Distribution for first-guess errors | Gaussian, uniform |

*f. Evaluation metrics*

To determine the best CRPS-only model, we use the evaluation metrics listed in Table 8. The first three metrics are based on Euclidean distance between the true TC center and the mean estimated TC center (henceforth, just "ensemble mean"). Mean absolute bias is also based on the ensemble mean but averaged over both coordinates. For example, if the bias in the *x*-coordinate is -4.2 km and the bias in the *y*-coordinate is +2.8 km, the mean absolute bias is $\frac{1}{2}(4.2 + 2.8) = 3.5$ km. Reliability – also based on the ensemble mean and coordinate-averaged – is based on the reliability/attributes diagram. This diagram is used to diagnose conditional bias, *i.e.*, bias as a function of the model's prediction (Hsu and Murphy 1986; Lagerquist et al. 2021).

To take full advantage of the uncertainty estimates provided by our CNNs, we need additional evaluation tools. This is the motivation for the last five metrics, explained briefly here and thoroughly in Section 4 of Haynes et al. (2023). All five metrics are coordinate-averaged. The CRPS is MAE $- \frac{1}{2}$MAPD, where MAE is the mean absolute error achieved by the ensemble average and MAPD is the mean absolute pairwise difference among ensemble members. The spread-skill reliability (SSREL) is the mean deviation of the spread-skill curve (Delle Monache et al. 2013) from the 1:1 line. A perfect SSREL of 0.0 means that, given any value of model spread (ensemble standard deviation; $\sigma_{\mathrm{ens}}$), the expected error (RMSE of ensemble mean; RMSE($\mu_{\mathrm{ens}}$)) equals the spread. The spread-skill ratio (SSRAT) is the mean ratio of $\sigma_{\mathrm{ens}}$ to RMSE($\mu_{\mathrm{ens}}$) over the entire dataset, regardless of spread. The ideal SSRAT is 1.0, with smaller values meaning that the model is overconfident and larger values meaning that it is underconfident.

The monotonicity fraction is based on the discard test, hence the acronym "DTMF". In general, the discard test involves incrementing a discard fraction (*e.g.*, from 0% to 5% to 10% ... to 95%); for every discard fraction $k$, the $k$% of highest-uncertainty samples are discarded, and the model error is recomputed. For a model with highly calibrated uncertainty estimates (such that the model's uncertainty is highly correlated with its own error), model error should decrease monotonically with discard fraction. For the discard tests in this study, we define "model error" as the mean Euclidean distance between the ensemble mean and true TC center, and we define "uncertainty" as the Euclidean ensemble standard deviation (EESD). Mathematically, EESD is

$$\mathrm{EESD}^2 = \frac{1}{N-1} \sum_{i=1}^{N} \left[ (x_{\mathrm{est}}^i - \overline{x_{\mathrm{est}}})^2 + (y_{\mathrm{est}}^i - \overline{y_{\mathrm{est}}})^2 \right], \qquad (1)$$

where $N$ is the ensemble size; $x_{\mathrm{est}}^i$ and $y_{\mathrm{est}}^i$ are the estimated TC-center coordinates from the $i^{\mathrm{th}}$ ensemble member; and $\overline{x_{\mathrm{est}}}$ and $\overline{y_{\mathrm{est}}}$ are the ensemble-mean TC-center coordinates. Thus, the bracketed term on the right-hand side is the squared Euclidean distance between the $i^{\mathrm{th}}$ member and the ensemble mean. A perfect DTMF of 1.0 means that model error always decreases when more high-uncertainty samples are discarded, *i.e.*, the model's estimate of its own uncertainty is perfectly correlated with its own error.

Finally, the rank-histogram deviation (RHD) is the mean deviation of the model's rank histogram (RH; Hamill 2001) from the perfect RH, which is uniform. A uniform RH means that the correct answer achieves every rank in the ensemble of predictions with equal frequency. For a perfect CNN, the *y*-coordinate of the true TC center achieves every rank in the ensemble – $1^{\mathrm{st}}$, $2^{\mathrm{nd}}$, ..., $51^{\mathrm{st}}$ – exactly $\frac{1}{51} = 1.96$% of the time. The same applies to the *x*-coordinate.

We combine the ten metrics in Table 8 subjectively, trading off several desires: a model with small Euclidean errors, small bias, good reliability, and good uncertainty estimates.

Table 8: Evaluation metrics used for model selection. RMS = root mean square; CRPS = continuous ranked probability score; SSREL = spread-skill reliability; SSRAT = spread-skill ratio; DTMF = monotonicity fraction in discard test; RHD = rank-histogram deviation. "Coord-averaged?" indicates whether the metric is averaged over the *x*- and *y*-coordinates of the TC's position vector. "Full ensemble?" indicates whether the metric considers the full ensemble or *just* the ensemble mean.

| Metric | Coord-averaged? | Full ensemble? | Range and optimal value |
| --- | --- | --- | --- |
| Mean Euclidean distance | | | $[0, \infty)$ km; 0 km |
| Median Euclidean distance | | | $[0, \infty)$ km; 0 km |
| RMS Euclidean distance | | | $[0, \infty)$ km; 0 km |
| Mean absolute bias | ✓ | | $[0, \infty)$ km; 0 km |
| Reliability | ✓ | | $[0, \infty)$ km$^2$; 0 km$^2$ |
| CRPS | ✓ | ✓ | $[0, \infty)$ km; 0 km |
| SSREL | ✓ | ✓ | $[0, \infty)$ km; 0 km |
| SSRAT | ✓ | ✓ | $[0, \infty)$; 1 |
| DTMF | ✓ | ✓ | $[0, 1]$; 1 |
| RHD | ✓ | ✓ | $\geq 0$; 0 |

# 4. Results of hyperparameter experiment (based on validation data)

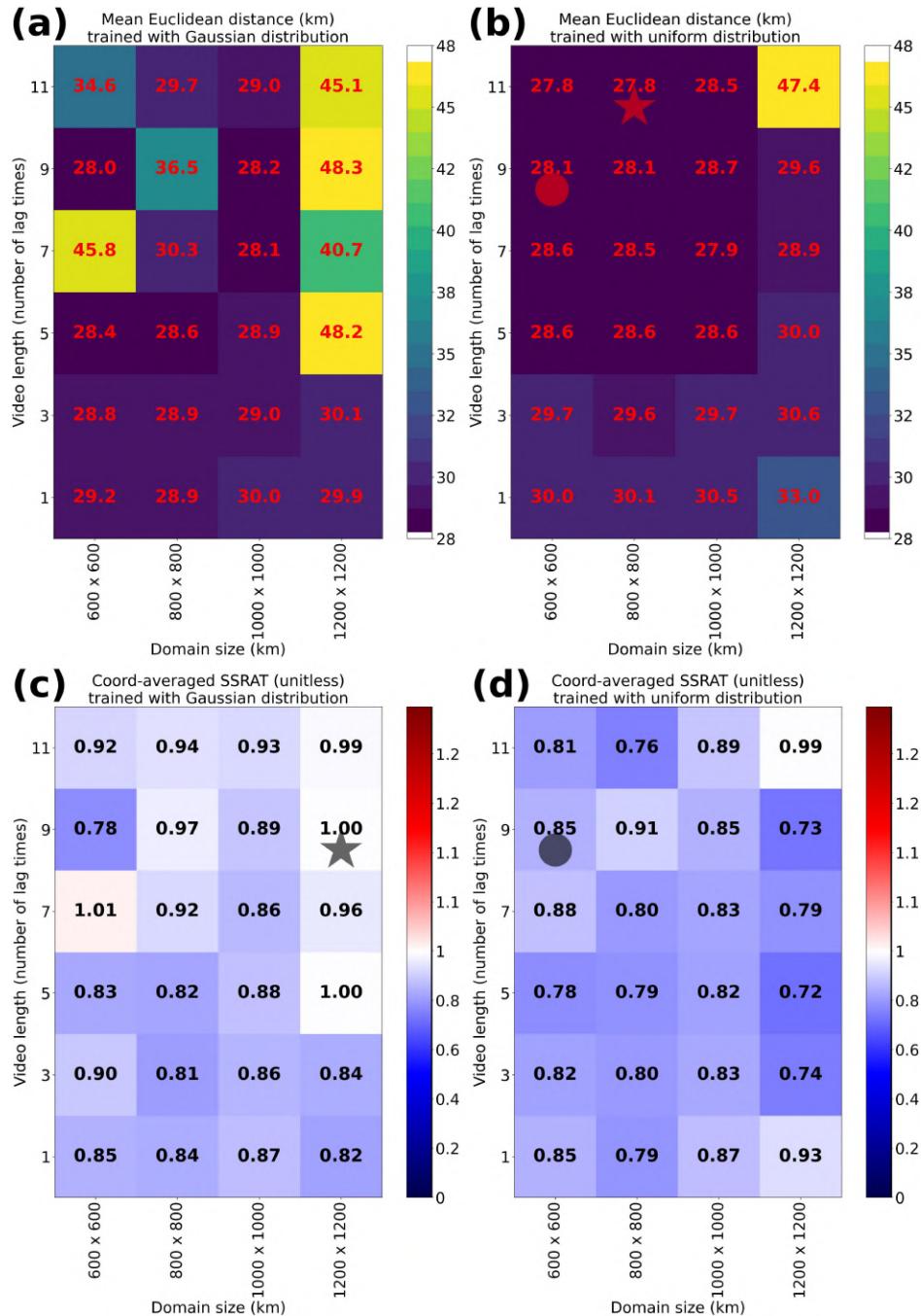

Figure 6: CNN performance on all validation data (both tropical and non-tropical systems), with respect to all three experimental hyperparameters. The circle marks the selected (best overall) model, while the star marks the best model according to the evaluation metric shown in the given panel. [a] Mean Euclidean distance for CNNs trained with Gaussian distribution of first-guess errors. [b] Same but for uniform distribution. [c] SSRAT for CNNs trained with Gaussian distribution of first-guess errors. [d] Same but for uniform distribution.

This section evaluates the CNNs on all systems (both tropical and non-tropical) in the validation data, with the aim of selecting the best CNN. The validation data contain 1141 original TC samples; we create 8 random first guesses for each sample (following the procedure in Section 3b with Gaussian whole-track error), yielding a final sample size of 9128. Results for two evaluation metrics (mean Euclidean distance and SSRAT) are shown in Figure 6; other evaluation metrics are shown in Supplemental Section 2. We summarize the results below:

1. There is a trade-off between the quality of deterministic predictions and ensemble calibration. For example, training with Gaussian first-guess errors and a long video sequence (corresponding to the top half of Figure 6a or 6c) leads to the worst Euclidean errors but the best values of SSREL, SSRAT, and RHD.

2. There is no obvious relationship between model performance and a single hyperparameter. For example, in Figure 6a (corresponding to models trained with Gaussian first-guess errors), model performance generally deteriorates with video length. However, in Figure 6b (corresponding to models trained with uniform first-guess errors), model performance generally improves with video length.

3. There is no clear optimum for domain size. For some metrics the best value is achieved with the largest domain (1200 × 1200 km), while for others the best value is achieved with the smallest domain (600 × 600 km).

4. There is no clear optimum for video length, except that the best value for all metrics is achieved with a video length > 1. This suggests that the temporal evolution of the TC's satellite presentation – not just a single image – is important.

5. Overall, training with a uniform distribution of first-guess errors does not aid model performance.

Based on all ten metrics, we select the CNN with a 600-by-600-km domain size, animation length of 9 lag times (back to 240 min ago), and trained with uniform first-guess errors – marked by circles in Figure 6. The exact scores achieved by this model are listed in Table 9.

Table 9: Results on validation data for selected CNN. All ranks are out of 56. For all evaluation metrics, 1 is the best rank and 56 is the worst.

| Evaluation metric | Exact value | Rank |
|---|---|---|
| Mean Euclidean distance | 28.1 km | 5th |
| Median Euclidean distance | 23.8 km | 5th |
| RMS Euclidean distance | 34.0 km | 7th |
| Mean absolute bias | 1.57 km | 47th |
| Reliability | 26.5 km$^2$ | 31st |
| CRPS | 13.0 km | 5th |
| SSREL | 3.40 km | 22nd |
| SSRAT | 0.85 | 25th |
| DTMF | 1.00 | Tied for 1st |
| RHD | 0.0055 | 30th |

## 5. Evaluation of final ensemble (based on testing data)

Here we evaluate the final ensemble of center-fixing models, henceforth "GeoCenter". The ensemble consists of several CNNs – all using IR imagery with a domain size of 600 × 600 km, an IR animation with 9 lag times, and trained with a uniform distribution of first-guess errors. By experimental design, the ensemble should contain four CNNs, each using a different set of IR wavelengths (Table 6). However, the CNN trained with wavelengths $\{3.9, 6.185, 6.95\}\mu$m achieved poor performance on the validation data in this particular experiment, for reasons that are unclear. (We suspect that this model failed to converge due to random factors, such as the random seed used to initialize model weights.) The inclusion of this CNN substantially degrades the performance of the ensemble on the validation data; thus, the final GeoCenter ensemble contains only the other three CNNs, leading to a total ensemble size of 150 instead of 200. Note that we base this decision only on validation performance, not on testing performance.

The first subsection below evaluates GeoCenter objectively, based on all tropical systems in the testing dataset (for which the "tropical flag" defined in Table 3 is 1), using Gaussian first-guess errors. For analogous results on the complete testing set and on non-tropical systems in the testing set, see Supplemental Section 3. The second subsection presents case studies for a deeper but more subjective understanding of GeoCenter.

*a. Objective evaluation*

Tropical systems in the testing data comprise 854 original samples (82.4% of the testing data); we create 8 random first guesses for each, yielding a final sample size of 6832.

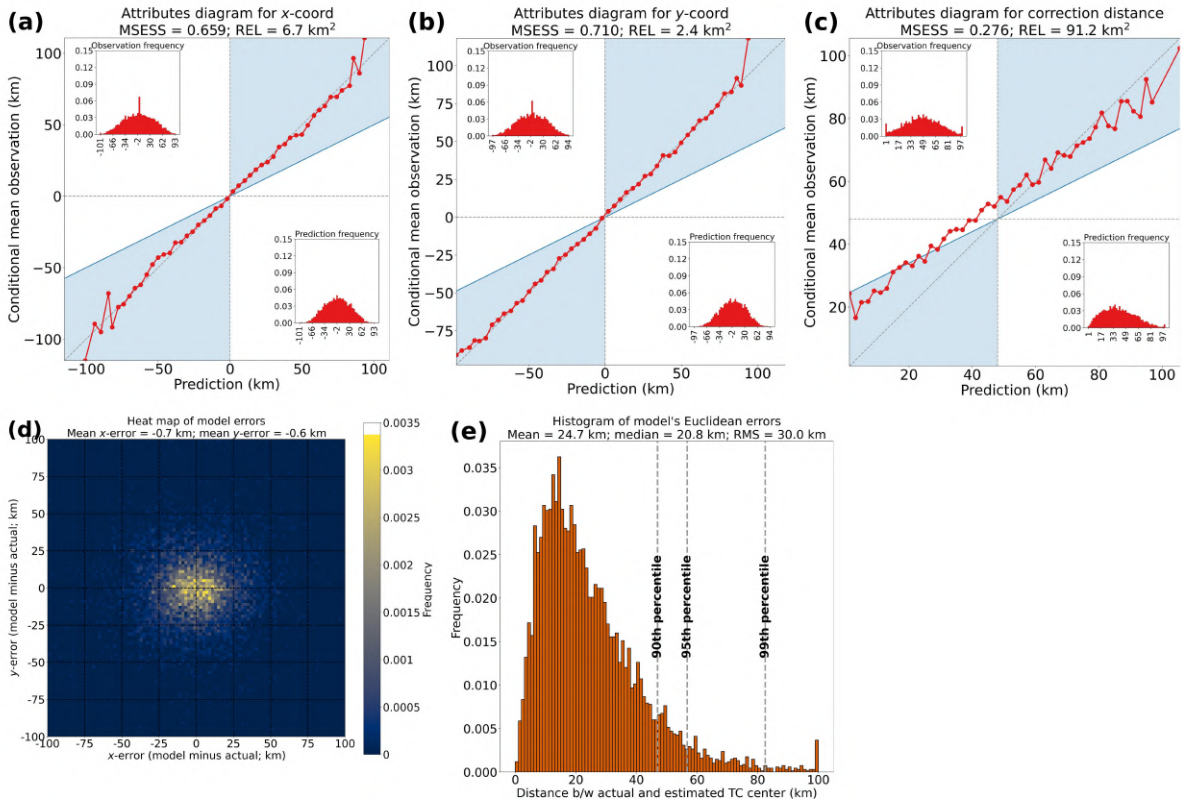

Figure 7: Performance of GeoCenter ensemble mean on tropical systems in testing data. [a] Attributes diagram for image-center-relative *x*-coordinate. The blue shading marks the positive-skill area, where MSE skill score > 0; the dashed 1:1 line is the perfect reliability curve; the red curve is the reliability curve achieved by GeoCenter; and the inset histograms show the distribution of predictions (GeoCenter ensemble means) and observations. [b] Same but for image-center-relative *y*-coordinate. [c] Same but for total correction distance. [d] Heat map of errors. [e] Histogram of Euclidean distance errors.

Figure 7 evaluates the GeoCenter ensemble mean. The attributes diagram for the *x*-coordinate (Figure 7a) shows that, while intermediate predictions are nearly unbiased, extreme predictions are not extreme enough. Specifically, predictions ≲ -80 km are too high (not negative enough), while predictions ≳ +80 km are too low (not positive enough). In other words, when GeoCenter makes an extreme correction to the west/east, it should correct even more to the west/east. Similarly, the attributes diagram for the *y*-coordinate (Figure 7b) shows that predictions ≳ +80 km are not positive enough. In other words, when GeoCenter makes an extreme correction to the north, it should correct even more to the north. The attributes diagram for total correction distance (Figure 7c) shows that GeoCenter has a negative bias for correction distances ≲ 60 km and positive bias for correction distances ≳ 60 km. In other words, when GeoCenter makes a small correction, the correction is systematically too small (by up to 10 km); and when GeoCenter makes a large correction, the correction is systematically too large (by up to 5 km). Despite the shortcomings mentioned, all reliability curves (Figures 7a-c) lie close to the 1:1 line and almost entirely within the positive-skill area, leading to a positive MSE skill score. The error heat map (Figure 7d) shows that GeoCenter has a bias of 0.7 km too far west and 0.6 km too far south; these biases are very small compared with the inherent uncertainty in FBT labels. Also, by comparison to Figure 5a – showing many errors well above 50 km – GeoCenter improves much of the error in the first-guess center fix. The histogram of Euclidean distance errors (Figure 7e) shows that the model has a mean error of 24.7 km and median of 20.8 km. Analogous GeoCenter results for non-tropical systems are 35.4 and 30.7 km, respectively; analogues for the complete testing set (tropical and non-tropical) are 26.6 and 22.2 km, respectively.

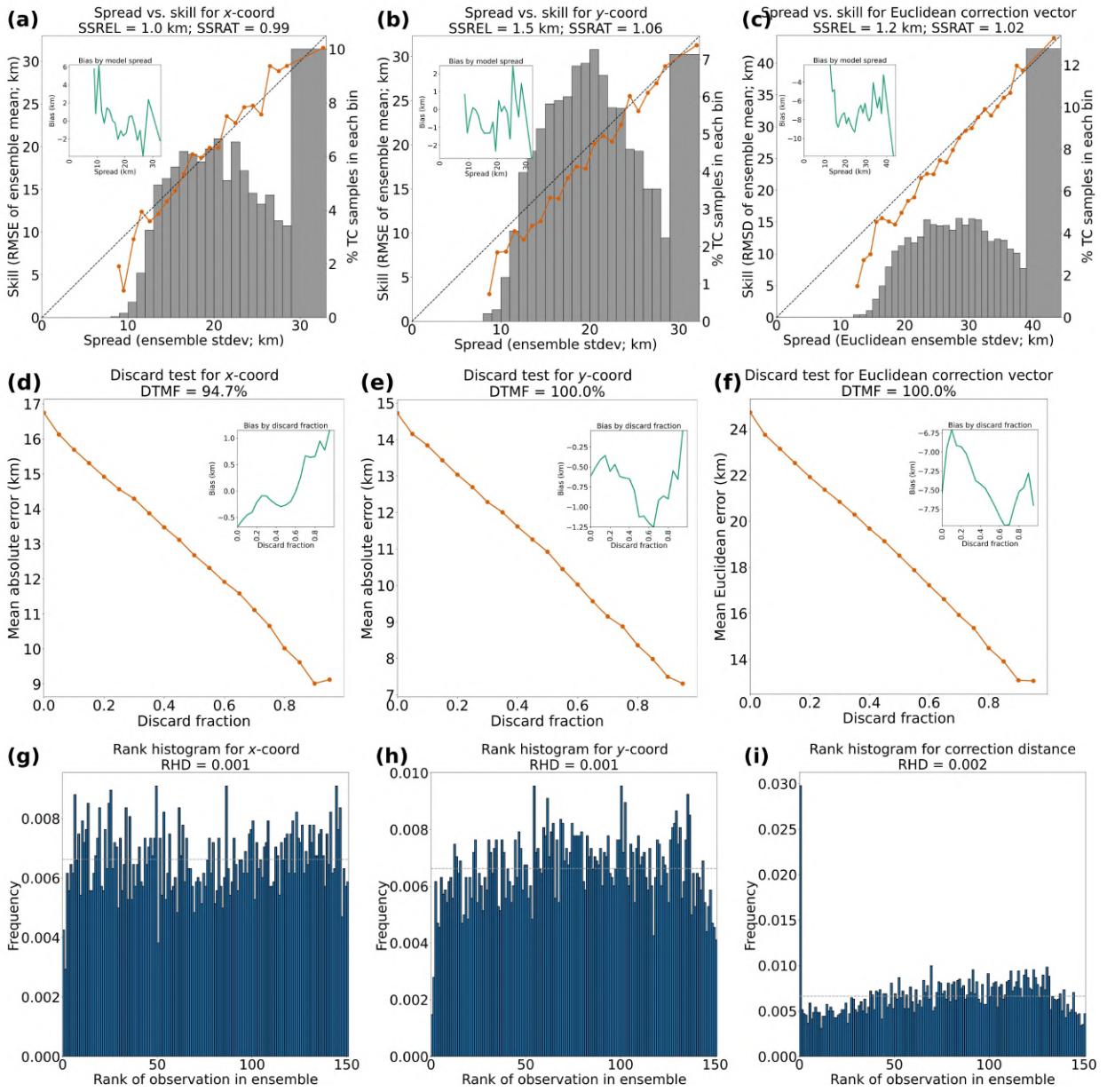

Figure 8: Performance of GeoCenter uncertainty estimates on tropical systems in testing data. [a] Spread-skill plot for image-center-relative *x*-coordinate. The background histogram shows the distribution of GeoCenter spread values; the 1:1 line is the perfect spread-skill curve; the orange curve is the spread-skill curve achieved by GeoCenter; and the inset plot shows bias (mean prediction minus mean target) as a function of spread. [b] Same but for image-center-relative *y*-coordinate. [c] Same but for full correction vector. Values shown here are Euclidean: root mean squared distance (RMSD) instead of RMSE, EESD (Equation 1) instead of scalar standard deviation. [d] Discard test for image-center-relative *x*-coordinate. The inset plot shows bias (mean prediction minus mean target) as a function of discard fraction. For a higher discard fraction (lower uncertainty threshold), the mean uncertainty of the remaining samples is lower. [e] Same but for image-center-relative *y*-coordinate. [f] Same but for full correction vector. [g] Rank histogram for image-center-relative *x*-coordinate. The dashed line is the perfect rank histogram. The horizontal axis is the rank achieved by the observation within the ensemble distribution; 0 = below all ensemble members, and 150 = above all ensemble members. [h] Same but for image-center-relative *y*-coordinate. [i] Same but for total correction distance.

Figure 8 evaluates the full GeoCenter ensemble, including uncertainty estimates. The spread-skill plots (Figures 8a-c) show that GeoCenter is slightly underconfident, producing more spread than it should. Also, GeoCenter is more underconfident when producing small spread values ($\lesssim$ 15 km for $x$, $\lesssim$ 20 km for $y$, $\lesssim$ 25 km for the full vector) than when producing large spread values. However, across all three quantities, the overall SSRAT (see panel titles) ranges from 0.99 to 1.06, very close to the ideal value of 1.0. Discard tests (Figures 8d-f) show that whenever a higher percentage of the most uncertain data samples is thrown out, model error for all three quantities decreases. In an operational setting, this property could be useful as an "ignore" flag. For example, if the maximum acceptable distance error is 22 km – which corresponds to a discard fraction of ~0.2 (see Figure 8f) and EESD of 36.4 km (not shown in Figure 8f) – forecasters could choose to ignore GeoCenter whenever its EESD > 36.4 km. Rank histograms show that the GeoCenter ensemble is almost perfectly calibrated for the *x*- and *y*-coordinates (Figures 8g-h), with a mean deviation of 0.001 from the optimal histogram, which has a uniform frequency of $\frac{1}{151} = 0.0066$. However, rank histograms for the *y*-coordinate and total correction distance (Figures 8h-i) show that GeoCenter is slightly overdispersive, which is consistent with an SSRAT > 1.0 (Figures 8b-c). For example, the overdispersion for *y* manifests primarily as a negative bias, corresponding to the high frequency for ranks of ~55 to 135 in the 150-member ensemble, along with a low frequency for other ranks. In other words, the *y*-coordinate of the true TC center falls within the ranks 55 to 135 – mostly above the ensemble median (75.5) – more often than it should. Lastly, for total correction distance, the overdispersion manifests primarily as a low frequency of ranks $\lesssim$ 35 and $\gtrsim$ 140. In other words, the total correction distance needed falls on the extreme ends of the GeoCenter distribution less often than it should.

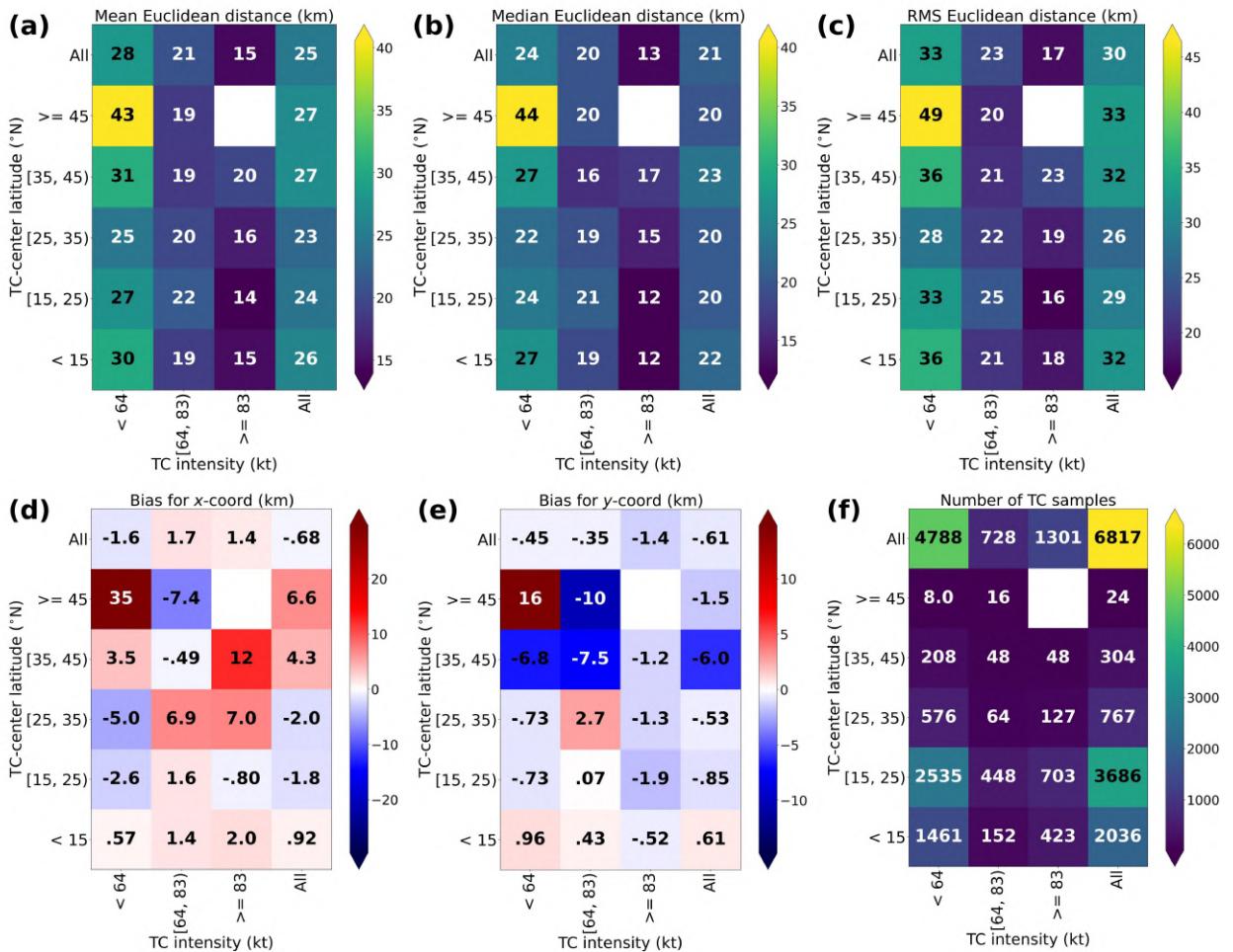

Figure 9: Performance of GeoCenter on tropical systems in testing data, as a function of TC intensity and true TC-center latitude. The three intensity bins correspond to sub-hurricane-strength, category-1 hurricane, and category-2–5 hurricane on the Saffir-Simpson scale. In each panel, the top row averages over all latitudes, so is a function of intensity only; and the rightmost column averages over all intensities, so is a function of latitude only. [a] Mean Euclidean distance error. [b] Median Euclidean distance. [c] RMS Euclidean distance. [d] Bias for *x*-coordinate (GeoCenter minus truth). [e] Bias for *y*-coordinate (GeoCenter minus truth). [f] Number of TC samples in each bin.

Figure 9 breaks down GeoCenter performance by both TC intensity and latitude. For all metrics based on Euclidean distance (Figures 9a-c), we note three patterns. First, error decreases with TC intensity, reaching a maximum for sub-hurricane-strength storms (intensity < 64 kt) and a minimum for category-2–5 hurricanes[6] (intensity ≥ 83 kt). This behaviour is similar to ARCHER-2, which also shows larger error for weak systems. Second, error reaches a minimum at 25-35°N and increases while moving equatorward. This is because low-latitude TCs are often at an early stage in their life cycle and therefore less organized. Third, error also increases while moving poleward from 25-35°N. This is because high-latitude TCs often become extratropical or dissipate; these systems include complicated structures that make center-fixing difficult (Wang et al. 2020). Again, ARCHER-2 has similar behaviour for high-latitude systems.

---

[6]This division of intensities – into three broad categories instead of seven fine categories (depression, storm, categories 1-5) – is partly motivated by small sample size in some of the fine categories. For example, categories 2-5 include only 96 samples in category 5. Furthermore, these 96 samples come from only 12 original samples, each with 8 first guesses.

For bias in the *x*- and *y*-coordinates (Figures 9d-e), we again note three patterns. First, extreme biases generally occur at high latitudes. Second, extreme biases in *x* are mostly positive (too far east), while extreme biases in *y* are mostly negative (too far south). Third, extreme biases at high latitude are typically worse for weaker TCs, although this pattern may be unduly affected by small sample size. For example, the bins at latitude $\geq 45°N$ contain no more than 24 TC samples each (Figure 9f). Furthermore, these 24 samples come from only 3 original samples, each with 8 first guesses.

Table 10 breaks down GeoCenter performance by ocean basin, showing that the EP basin has the lowest error. We believe that this is because [1] EP storms reach high latitudes less often; [2] EP storms recurve less often, leading to smaller acceleration vectors and therefore greater time continuity in the TC location. Also, Table 10 suggests that biases in the overall dataset (Figure 7d) are driven largely by the AL basin.

Table 10: Performance of GeoCenter on tropical systems in testing data, as a function of basin.

|  | Basin | | |
| --- | --- | --- | --- |
| **Evaluation metric** | **AL** | **EP** | **WP** |
| Mean Euclidean distance (km) | 24.9 | 20.7 | 27.3 |
| Median Euclidean distance (km) | 21.0 | 18.4 | 23.0 |
| RMS Euclidean distance (km) | 29.9 | 24.4 | 33.2 |
| Bias in *x*-coordinate (km) | -2.32 | +0.10 | +0.29 |
| Bias in *y*-coordinate (km) | -1.51 | -1.96 | +1.09 |

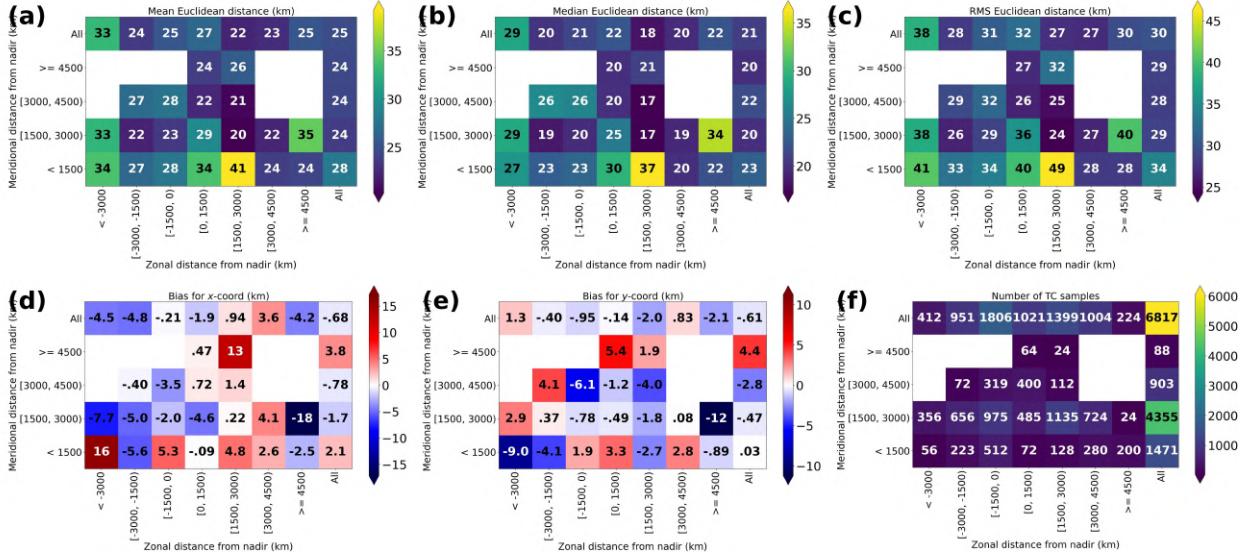

Figure 10: Performance of GeoCenter on tropical systems in testing data, as a function of nadir-relative position. In each panel, the top row averages over meridional distances, so is a function of zonal distance only; and the rightmost column averages over all zonal distances, so is a function of meridional distance only. [a] Mean Euclidean distance error. [b] Median Euclidean distance. [c] RMS Euclidean distance. [d] Bias for *x*-coordinate (GeoCenter minus truth). [e] Bias for *y*-coordinate (GeoCenter minus truth). [f] Number of TC samples in each bin.

Figure 10 breaks down GeoCenter performance by position relative to nadir, *i.e.*, the satellite subpoint. Satellites register off-nadir clouds with a parallax error, which increases with distance from nadir and can be $\mathcal{O}(10$ km$)$. For clouds north/east/west of nadir, the satellite registers cloud position as farther north/east/west than it actually is, respectively. Thus, for TCs east/west of nadir, we expect a positive/negative bias in GeoCenter's *x*-coordinate; and for TCs north of nadir, we expect a positive bias in GeoCenter's *y*-coordinate. For the metrics based on Euclidean distance (Figures 10a-c), we note two patterns. First, error generally increases with the absolute zonal distance from nadir, as can be seen in the top row of each panel. Second, in terms of meridional distance from nadir ($\phi_{\text{nadir}}$), error is actually worst for the smallest off-nadir distances. To explain this pattern, we first note that because all three satellites are located at 0°N, there is a unique relation between $\phi_{\text{nadir}}$ and latitude ($\Phi$). Thus, higher errors at $\phi_{\text{nadir}} < 1500$ km ($\Phi < 13.5$°N) can be explained by the higher errors near the equator noted in Figure 9, associated with disorganized TCs. For GeoCenter's *x*-coordinate, biases east/west of nadir are generally positive/negative (see top row of Figure 10d), as hypothesized above. For GeoCenter's *y*-coordinate, the magnitude of the bias generally increases with $\phi_{\text{nadir}}$ (see rightmost column of Figure 10e).

Although the 3.9-$\mu$m channel in our predictors behaves differently between day and night (Kim and Hong 2019; Kim et al. 2019), the same performance metrics do not differ between day and night by more than 1.53 km (see Supplemental Section 3c).

Figures 11a-b compare GeoCenter with other center-fixing algorithms, focusing on the operational ARCHER-2 and recent deep-learning efforts. For ARCHER-2, we take results from Table 3 of WV16; for the other algorithms, we use Table 1 of Smith and Tuomi (2021, ST21), Table 2 of Wang et al. (2019, W19), Section 3.2 of Yang et al. (2019, Y19), and Table 3 and Figure 8 of Wang and Li (2023, W23). Because these studies filter non-tropical systems out of the dataset to varying degrees[7], we include GeoCenter results on two versions of the testing data: all systems (worst-case estimate) and tropical systems only (best-case estimate). However, the comparisons in Figures 11a-b are still not apples-to-apples for several reasons, the main one being that each study has a different testing dataset. Thus, the purpose of Figures 11a-b is to show that GeoCenter is at least on par with other center-fixing guidance, not to claim conclusively that GeoCenter outperforms any other method. However, GeoCenter has several advantages over other methods that are not conveyed by Figures 11a-b; see Section 1e.

---

[7]Y19 use typhoons only; WV16 exclude systems over land and poleward of 40°, many of which are non-tropical; ST21 exclude systems with a minimum central pressure > 1005 hPa, many of which are non-tropical. W19 and W23 do not mention any filtering of their datasets.

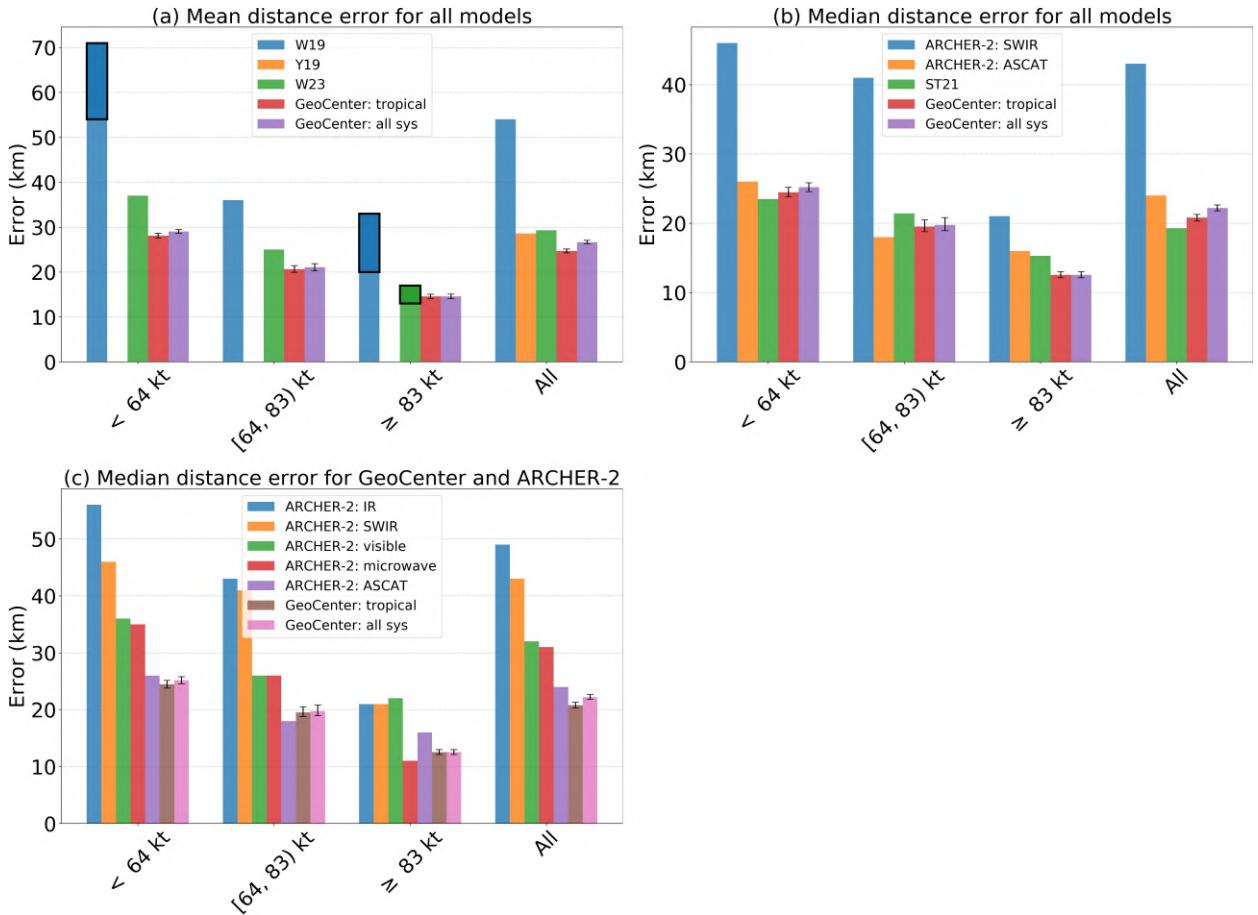

Figure 11: Error comparison.

"ARCHER-2: IR" = ARCHER-2 applied to longwave IR data. "ARCHER-2: SWIR" = ARCHER-2 applied to shortwave IR data, which leads to better ARCHER-2 performance than longwave IR. "ARCHER-2: ASCAT" = ARCHER-2 applied to scatterometer data, which leads to better ARCHER-2 performance than *any* other data type. "GeoCenter: tropical" = GeoCenter on tropical systems only; "GeoCenter: all sys" = GeoCenter on all systems.

ST21 = Smith and Tuomi (2021); W19 = Wang et al. (2019); Y19 = Yang et al. (2019); W23 = Wang and Li (2023).

Error bars for GeoCenter indicate a 95% confidence interval, based on 1000 iterations of bootstrapping. Black-outlined boxes for W19 and W23 (panel a only) indicate a range over multiple intensity categories, because these articles split intensity into smaller bins.

Figure 11c compares GeoCenter only to different versions of ARCHER-2. When considering all storms (rightmost grouping) and sub-hurricane-strength storms (leftmost grouping), GeoCenter performs best. When considering another intensity category, GeoCenter performs second-best, beaten only by "ARCHER-2: microwave" or "ARCHER-2: ASCAT". However, these versions of ARCHER-2 depend on data that are often unavailable, whereas GeoCenter depends only on routinely available geostationary data. Comparing to the versions of ARCHER-2 that use geostationary data (IR, SWIR, or visible), GeoCenter outperforms ARCHER-2 by at least 8 km in every intensity category.

*b. Homogeneous comparison with ARCHER-2*

We perform a homogeneous comparison between GeoCenter and ARCHER-2, using the same TC samples to evaluate both models. Specifically, we use 17 TCs from the 2024 season.[8] We obtain ARCHER-2 estimates from ATCF A-deck files provided by the Cooperative Institute for Meteorological Satellite Studies (CIMSS).[9]

We run GeoCenter for every TC every 10 min, using a short-term forecast as the first guess, henceforth "short track". In other words, rather than the data-augmentation procedure described in Section 3b, we create a first guess using data available in real-time operations. We initialize a new short-track forecast every 6 hours. For every initialization time $t_0$, we create a 36-hour track with valid times $t_0 - 24$ hours, $t_0 - 18$ hours, ..., $t_0 + 12$ hours. For points before $t_0$, we use the CARQ (combined automated response to query) line in the operational ATCF A-deck file. For points after $t_0$, we draw A-deck data from the following priority list (using the $n^{\text{th}}$ source only if the $[n-1]^{\text{th}}$ source is unavailable):

1. The official track forecast (labeled "OFCL" or "OFCI" for the AL/EP basins, "JTWC" or "JTWI" for the WP basin).

2. A model forecast – namely the Trajectory and Beta (TAB) model for the AL/EP basins, or Beta and Advection Model (BAM) for the WP basin (DeMaria et al. 2022).

3. Linear interpolation.

After creating the 36-hour track, we interpolate the data from 6-hour to 1-second time steps, so that short-track forecasts can be matched with an ARCHER-2 or GeoCenter prediction at any time.

We use the following procedure to match ARCHER-2 predictions, GeoCenter predictions, and truth labels for one TC:

---

[8]The randomly selected TCs are AL09-AL15, AL90-91, AL93, EP10-11, EP96, EP99, WP18-19, and WP21. This dataset is small because the analysis is performed on a local CIRA machine, which obviates the need for complicated data-transferring to a supercomputer but limits the computing power available for analysis.

[9]Available at https://tropic.ssec.wisc.edu/real-time/archerOnline/cyclones. For example, the F-deck for AL142024 is available at https://tropic.ssec.wisc.edu/real-time/archerOnline/cyclones/2024_14L/web/archer_fdeck.txt.

1. We match every ARCHER-2 prediction ($\hat{Y}_{A2}$) to the nearest GeoCenter prediction in time ($\hat{Y}_{GC}$), as long as $\hat{Y}_{GC}$ is within 30 min of $\hat{Y}_{A2}$. If there is no GeoCenter prediction within 30 min of $\hat{Y}_{A2}$, then $\hat{Y}_{A2}$ is removed from the dataset.

2. To obtain the truth label for an ARCHER-2 prediction valid at time $t$, we interpolate the FBT data to time $t$.

3. To obtain the truth label for a GeoCenter prediction valid at time $t$, we do the same.

Note that two matched predictions – $\hat{Y}_{A2}$ and $\hat{Y}_{GC}$ – generally have different truth labels, because $\hat{Y}_{A2}$ and $\hat{Y}_{GC}$ generally occur at different times. While GeoCenter predictions are available every 10 min at times divisible by 10 min (0000 UTC, 0010 UTC, 0020 UTC, etc.), ARCHER-2 predictions are typically available at times *not* divisible by 10 min, due to the irregular overpasses of polar-orbiting satellites that provide input data for ARCHER-2.

Before the matching procedure, we had 2170 ARCHER-2 samples and 7847 GeoCenter samples. After the matching procedure, we have 278 TC samples, each with an ARCHER-2 prediction and a GeoCenter prediction. Every ARCHER-2 prediction in this dataset is based on microwave data.[10] The homogeneous error comparison is shown in Figure 12. In terms of coordinate-averaged absolute bias, ARCHER-2 outperforms GeoCenter, and the difference is significant at the 95% level (based on a paired bootstrapping test with 1000 iterations). In terms of the other three metrics, GeoCenter outperforms ARCHER-2, although the difference is not significant at the 95% level. Remaining metrics in Table 8 are not reported here, because reliability requires a larger sample size (due to the use of binning in the calculation procedure, which further subdivides the sample size) and the others (CRPS, SSREL, SSRAT, DTMF, RHD) require more uncertainty information than is provided by ARCHER-2.

Overall, GeoCenter performs competitively with the microwave (best-case) version of ARCHER-2, while providing an updated estimate every 10 min.

---

[10]The F-decks contain only predictions based on microwave data or an unknown source ("UNKN"); we remove the latter from our dataset.

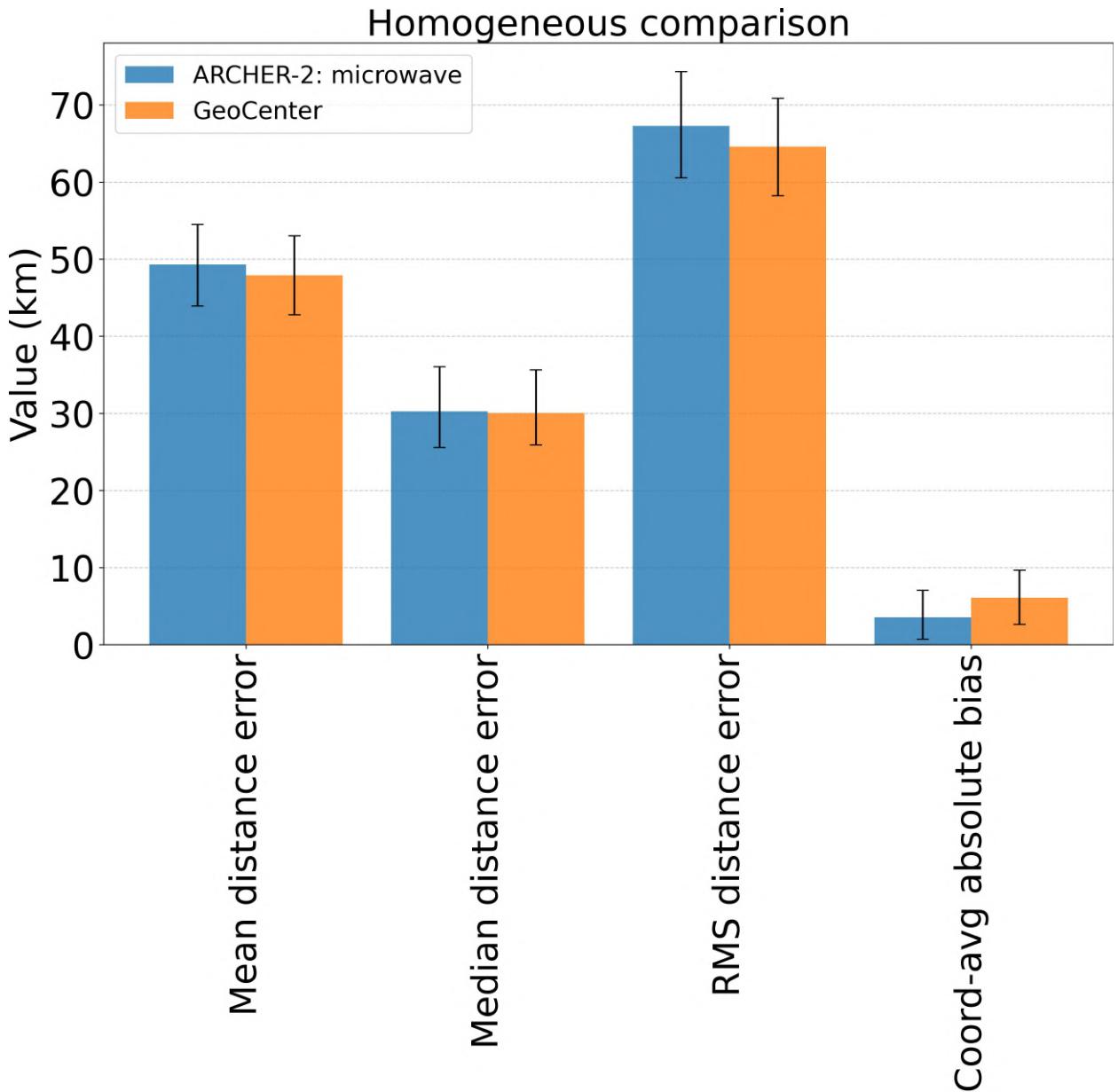

Figure 12: Homogeneous comparison between GeoCenter and the microwave version of ARCHER-2. Error bars indicate a 95% confidence interval, based on 1000 iterations of bootstrapping.

*c. Case studies*

Here we investigate four case studies from the testing data. To save space, every figure in this section shows a 3-D slice of the IR predictors: all channels at the 0-min lag time. Furthermore, every figure shows one random first guess for one TC sample. For every case study, Supplemental

Section 3d shows an analogous figure, containing only a 2-D slice of the IR predictors (the 8.5-$\mu$m channel at the 0-min lag time) but all eight random first guesses. To determine which of the eight first guesses to show here in the main text, we choose a representative example, with GeoCenter error close to the median over all eight first guesses. Additionally, Supplemental Section 3e shows the full track for all TCs used in case studies, according to both FBT and GeoCenter. These figures show that GeoCenter estimates are temporally consistent, without, *e.g.*, erratic jumps between consecutive time steps.

The first case study (Figure 13) is a low-intensity TC (Dujuan; WP012021) at low latitude, identified in Figure 9 as a trouble area for GeoCenter. Dujuan contains several areas of deep convection with no obvious circulation center; also, the true TC center (black star in Figure 13) is in a region of small brightness-temperature gradients. The GeoCenter ensemble (green contours) places the TC center southeast of the image center (red square), but it is actually due south of the image center. The Euclidean distance error is 47.5 km, and the true TC center lies on the third-to-outermost probability contour, indicating that GeoCenter assigns a small probability to the correct location. Supplemental Figure S17 shows that all eight first guesses for this TC sample – not just the one shown in Figure 13 – result in large GeoCenter errors.

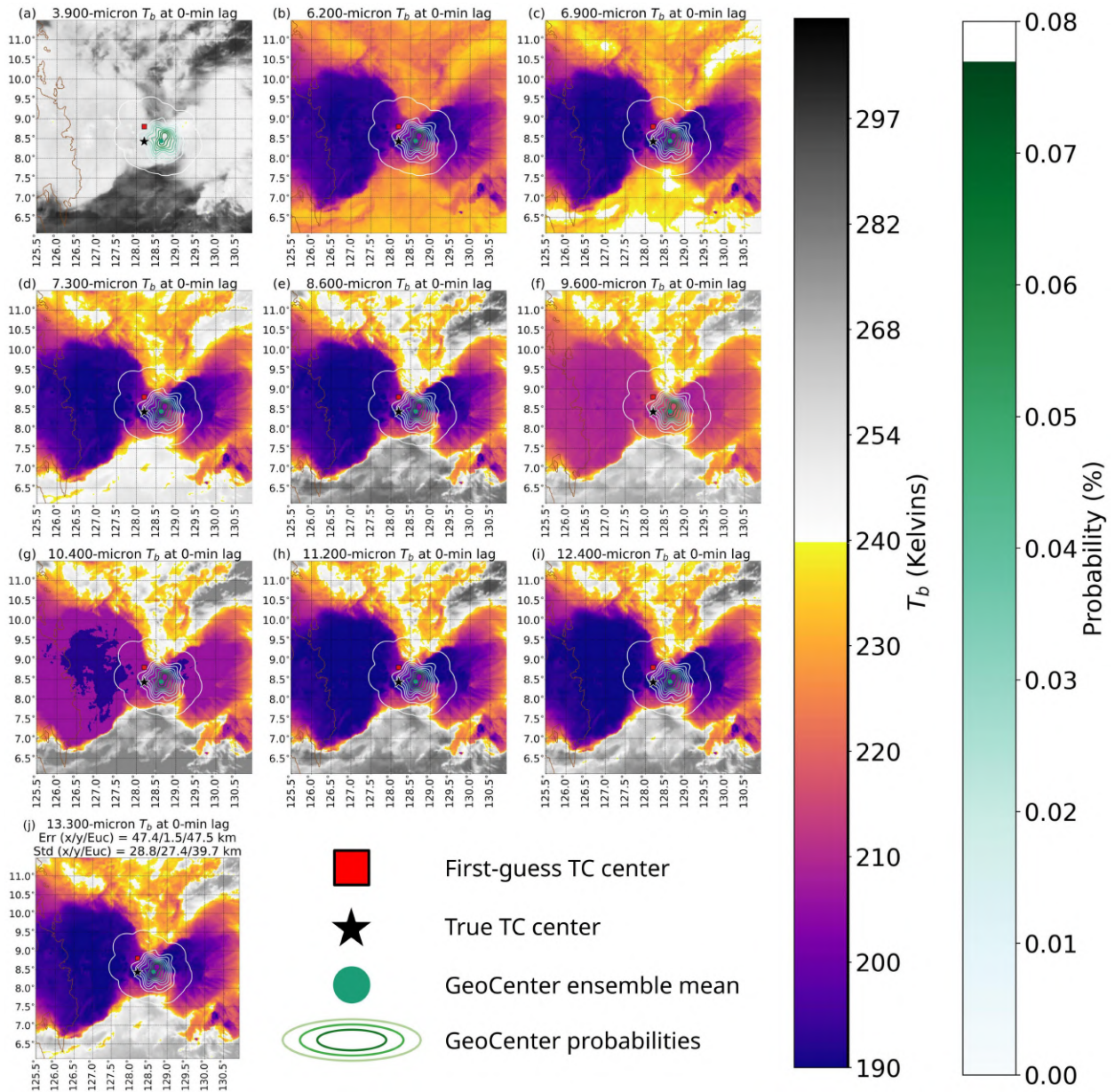

Figure 13: Case study: Tropical Depression Dujuan/Auring (WP012021), with an intensity of 35 kt at 0600 UTC 21 Feb 2021. Each panel shows the brightness temperature at a different IR channel, along with probability contours representing estimated TC-center locations from the 150 members of the GeoCenter ensemble. Darker greens correspond to higher GeoCenter probabilities, and the green circle inside the innermost contour is the GeoCenter ensemble mean. All probabilities are quite low (see colour bar), because they are per-pixel probabilities; there is a large number of pixels ($300^2$); and the sum over all pixels must be 100%. The red square is the image center, and the black star is the true TC center. The title of panel j gives the error of the ensemble mean (*x*-coordinate, *y*-coordinate, and Euclidean) and the standard deviation of the ensemble members, *i.e.*, the spread.

The second case study (Figure 14) is a high-intensity TC (Chanthu; WP192021), identified in Figure 9 as a strong point of GeoCenter. Chanthu has a classic (albeit somewhat overcast) eye signature, and GeoCenter corrects most of the large first-guess error, achieving a Euclidean distance error of 13.8 km. Supplemental Figure S18 shows that this is not a fluke; the average error across the eight first guesses is 15.6 km.

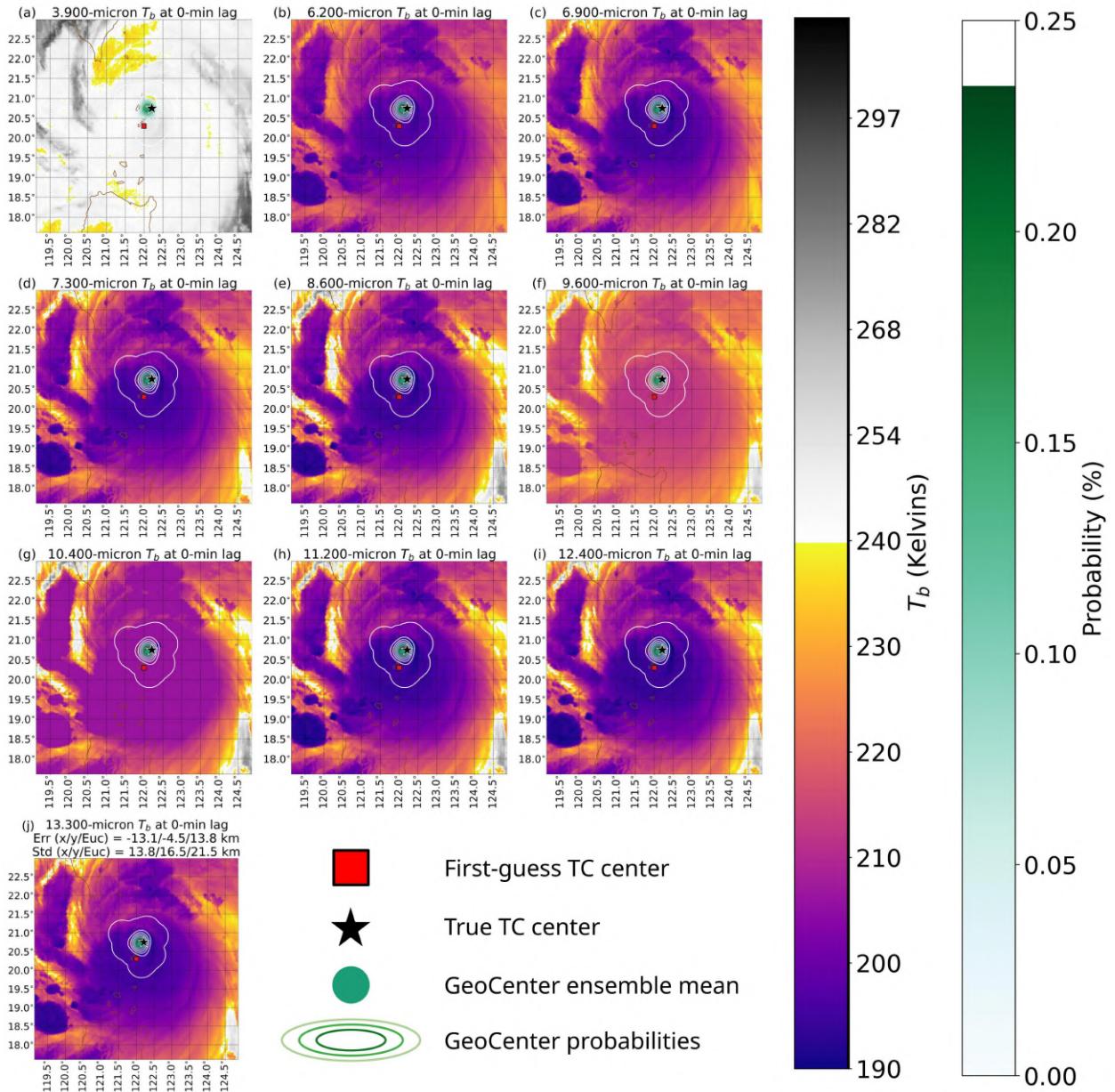

Figure 14: Case study: Super Typhoon Chanthu/Kiko (WP192021), with an intensity of 145 kt at 0000 UTC 11 Sep 2021. Formatting is explained in the caption of Figure 13.

The third case study (Figure 15) is a tropical depression (Fred; AL062021), with aircraft reconnaissance likely making the "true" FBT center highly accurate. There are several blobs of deep convection surrounding a relatively dry central area, which is quite expansive and does not have an obvious elliptical or spiral shape. Despite these issues, GeoCenter is fairly accurate, with a Euclidean distance error of 25.9 km (average of 28.9 km for all first guesses; see Supplemental Figure S19). This case study is just one example of GeoCenter performing well on a weak system with ambiguous structure.

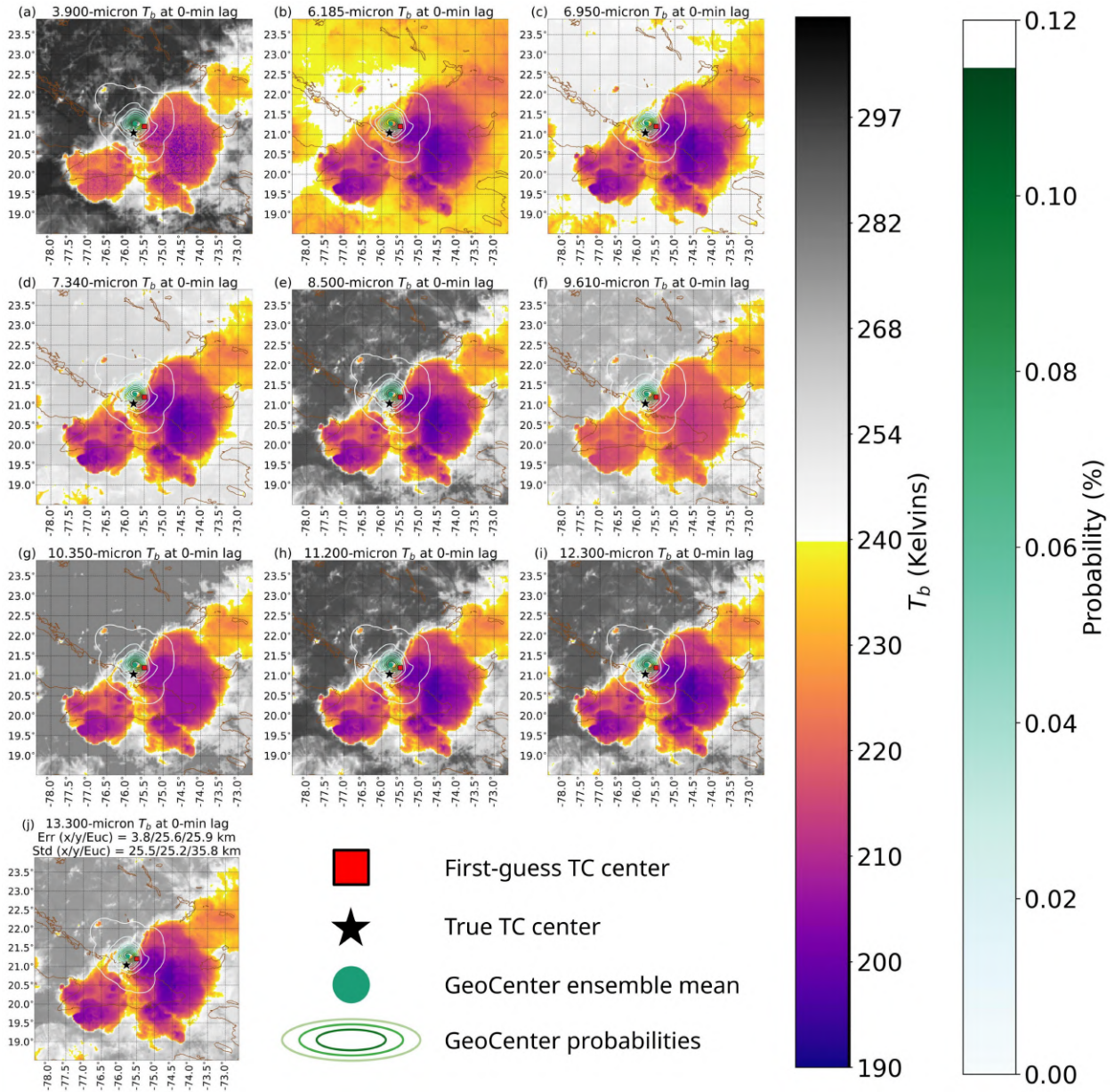

Figure 15: Case study: Tropical Depression Fred (AL062021), with an intensity of 30 kt at 0000 UTC 13 Aug 2021. Formatting is explained in the caption of Figure 13.

The fourth case study (Figure 16) is a storm that underwent extratropical transition four days prior (Odette; AL152021). Despite its highly asymmetric satellite presentation, GeoCenter locates the center of Odette quite accurately, with a Euclidean error of 15.1 km. GeoCenter's success for this case might be attributable[11] to the spiral cloud bands still present south and northeast of the circulation center, which are most visible in panels a, e, and g-i. Supplemental Figure S20 shows that the mean error over all eight first guesses is 16.1 km.

---

[11] A more confident attribution could be possible with explainable artificial intelligence, which is a subject of future research for GeoCenter.

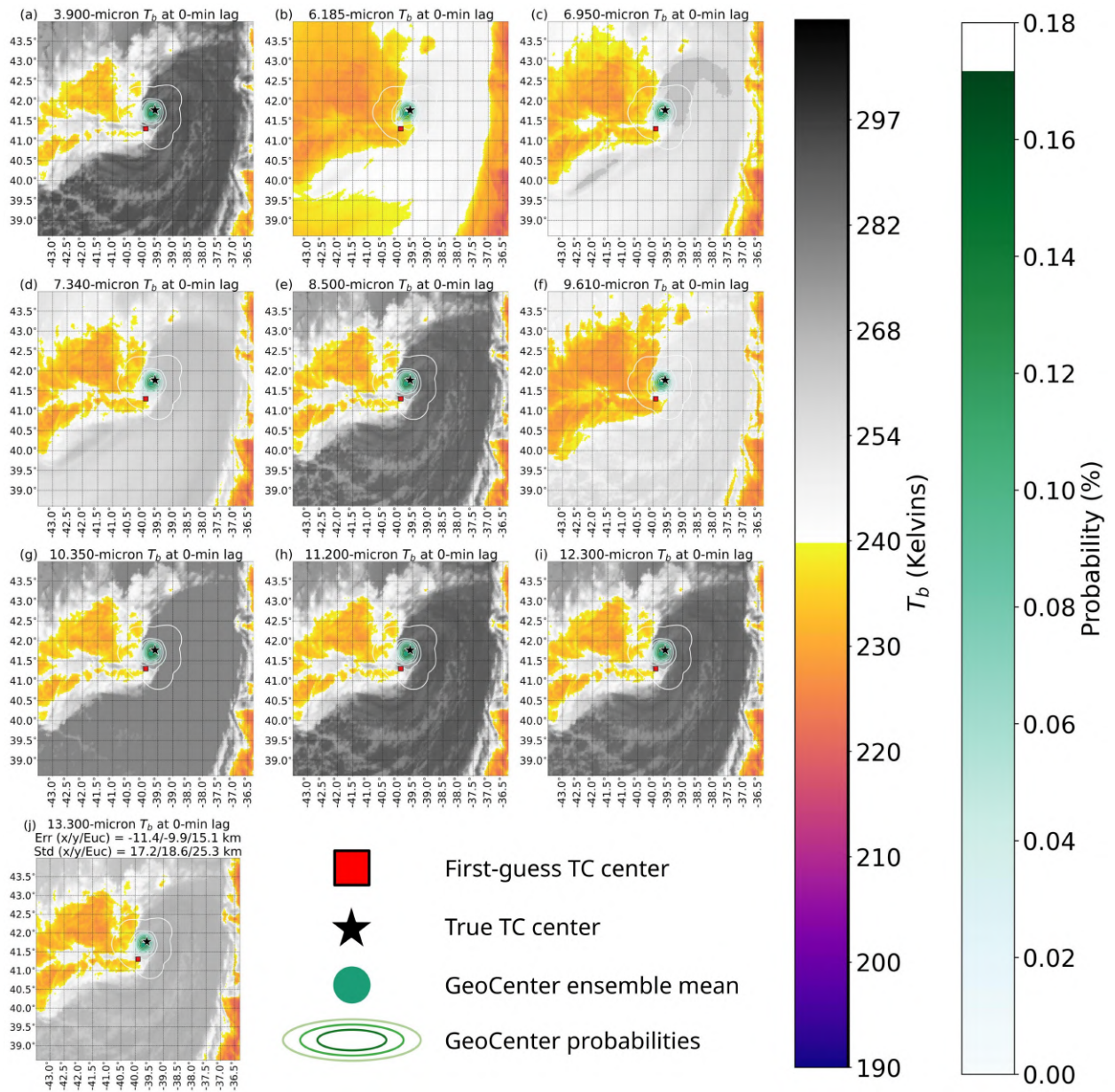

Figure 16: Case study: extratropical remnants of Tropical Storm Odette (AL152021) at 0600 UTC 22 Sep 2021. Formatting is explained in the caption of Figure 13.

The case studies for Fred and Odette – as well as Newton (Supplemental Figure S21), a different kind of remnant system – show that GeoCenter often performs well on complicated systems, not just major hurricanes with cloud-free eyes. However, we do not claim that GeoCenter always handles complicated systems well, as demonstrated by the Dujuan case study (Figure 13).

## 6. Additional analyses (based on testing data)

Additional analyses, relegated to the Supplement for the sake of brevity, are discussed briefly here. In Supplemental Section 4 we compare GeoCenter with and without bias correction, showing that bias correction soundly improves the performance of both the ensemble mean and uncertainty estimates. In Supplemental Section 5 we compare GeoCenter with and without ATCF predictors, showing that although the ATCF predictors lead to a notable improvement in performance, GeoCenter-no-ATCF is still competitive with the IR/SWIR versions of ARCHER-2. Hence, GeoCenter-no-ATCF could be used as a backup model in cases where ATCF data are unavailable. In Supplemental Section 6 we rank the importance of IR channels, showing that the best individual wavelength appears to be 8.5 or 9.61 $\mu$m. However, no 1-channel CNN performs as well as the 3-channel CNNs making up the final GeoCenter ensemble. Finally, in Supplemental Section 7 we analyze the sensitivity of GeoCenter to the magnitude of the first-guess error. We show that although GeoCenter error with increases with first-guess error, GeoCenter remains competitive with the IR/SWIR versions of ARCHER-2 at first-guess errors up to 120 km.

## 7. Summary and future work

Center-fixing is a critical first step in the TC-forecasting process, with current and future estimates of TC properties being highly sensitive to this initial location estimate. Operational centres often rely on the subjective Dvorak technique, with ARCHER-2 being the only objective method used in operations. ARCHER-2's best-case performance is achieved when using microwave or scatterometer data, which are typically available twice a day or less for a given TC. ARCHER-2 performs much worse when using routinely available IR data – collected by geostationary satellites during both day and night, at 10–15-min resolution and < 10-min latency, across all TC basins. Preliminary results show that GeoCenter (the algorithm developed herein), which uses only routinely available IR data, performs similarly to the best-case ARCHER-2 (with microwave/scatterometer data) and greatly outperforms the worst-case ARCHER-2 (with IR data).

GeoCenter is a deep-learning algorithm, specifically an ensemble of CNNs, trained with multi-spectral animations of IR data. Our hyperparameter experiment shows that CNNs perform better when trained with an animation (time series) than a single image. The predictors include nine IR channels at nine lag times – 0, 30, ..., 240 minutes before the valid time $t_0$ – centered at a first-guess location which is randomly offset from the true TC center at $t_0$. GeoCenter's task is to correct this random offset, which averages 48 km and can exceed 100 km, and find the true center. GeoCenter is trained on data from three ocean basins – the AL, EP, and WP – representing the vast majority of northern-hemisphere TCs. GeoCenter is also trained with data across the full lifetime of every TC, which includes post-/sub-/extra-tropical systems. These properties – combined with GeoCenter being trained on routinely available IR data – make GeoCenter a potentially operational tool with wide applicability. Furthermore, GeoCenter performs skillful uncertainty quantification, providing a well calibrated ensemble of 150 possible TC centers.

The objective evaluation of GeoCenter – on an independent testing set – focuses mainly on tropical systems, which comprise the majority (82.4%) of the dataset. GeoCenter achieves a small bias – 0.7 km to the west and 0.6 km to the south – with mean/median/RMS Euclidean errors of 24.7/20.8/30.0 km. GeoCenter errors vary strongly with TC intensity (similar to ARCHER-2), reaching a maximum of 28.1/24.5/33.4 km for sub-hurricane-strength TCs and a minimum of 14.6/12.5/17.3 km for category-2–5 hurricanes. Errors also vary somewhat with ocean basin, reaching a maximum of 27.3/23.0/33.2 km for the WP and a minimum of 20.7/18.4/24.4 km for the EP. GeoCenter provides skillful estimates of its own uncertainty, with a spread-skill ratio of 0.99 to 1.06 (close to the ideal value of 1.0), a monotonic decrease shown in discard tests (indicating that model uncertainty is perfectly rank-correlated with model error), and nearly flat rank histograms. The following scenarios are still challenging for GeoCenter: (1) extremely large offsets between the first-guess and true TC center, which are not corrected enough by GeoCenter; (2) weak TCs near the equator, which are often developing systems without a well defined center; (3) high-latitude TCs, which are often interacting with mid-latitude weather systems; and (4) TCs far from the satellite subpoint, *i.e.*, with a large parallax shift. Nonetheless, GeoCenter performs competitively with ARCHER-2 microwave-based estimates and other center-fixing algorithms, which are known to have many of the same challenges.

We provide several case studies, which (1) show how GeoCenter output might be visualized in operations; (2) demonstrate successes and failures of GeoCenter. Most importantly, we show that GeoCenter performs well not just for strong hurricanes, but for a wide variety of systems. GeoCenter often performs well on TCs with a central dense overcast, sheared TCs undergoing extratropical transition, extratropical systems, and remnant lows.

Future work includes the following items. First, we will incorporate daytime visible imagery and nighttime synthetic visible imagery (ProxyVis; Chirokova et al. 2023) into GeoCenter, which may further improve its performance. Second, we will investigate whether enforcing a notion of temporal consistency – *i.e.*, a reasonable movement of the estimated TC center over time – can improve GeoCenter accuracy. Third, we will create an operational prototype that provides estimates for real-time TCs. We will run GeoCenter whenever new IR data become available (every 10 min). For example, to center-fix a storm at 1200 UTC, the IR predictor times will be {0800, 0830, 0900, …, 1200} UTC. If the next valid time is 1210 UTC, the new IR predictor times will be {0810, 0840, 0910, …, 1210} UTC. Fourth, we will expand GeoCenter by training on southern-hemisphere storms and other geostationary satellites that have similar IR bands to the GOES ABI and Himawari AHI – namely the Geo-Kompsat-2A AMI and Meteosat Third Generation FCI. Fifth, we are considering developing a special version of GeoCenter for the Meteosat Second Generation imager, which does not have similar IR bands but is the only GEO satellite providing data over the Indian Ocean. The fourth and fifth items would make GeoCenter truly global by expanding its coverage to all TC basins. Lastly, it is possible that a global mosaic of satellite imagery could improve GeoCenter's performance, especially for TCs far off nadir.

*Acknowledgments.* This work was supported by NOAA Award Number NA19OAR4320073. All computing was performed on NOAA's Hera supercomputer.

*Data availability statement.* A-deck files and final best tracks for the North Atlantic (AL), eastern North Pacific (EP), and central North Pacific (CP) basins are available from the NHC at `https://ftp.nhc.noaa.gov/atcf/`. Final best tracks for the western North Pacific (WP), Indian Ocean (IO), and southern hemisphere (SH) are available from the JWTC at `https://www.metoc.navy.mil/jtwc/jtwc.html?best-tracks`. A-decks for these basins are not publicly available; they were used by permission from the JTWC for this research, under the condition that they are not redistributed. IR satellite imagery from the GOES-16/17/18 ABI (Himawari-8/9 AHI) is available from NOAA at `https://registry.opendata.aws/noaa-goes/` (https://registry.opendata.aws/noaa-himawari/). All code used for this study, including pre-processing of input data and training/evaluation of CNN models, is contained in a Python library called `ml4tccf`, available at `https://doi.org/10.5281/zenodo.15116854`. The trained models comprising GeoCenter (four CNNs and four isotonic-regression models), as well files needed for pre-processing data (normalization parameters based on the training data), are available at `https://zenodo.org/records/15116855`.

# APPENDIX A

### Quality control of IR satellite data

Satellite data sometimes contain missing or corrupt values. Thus, we perform quality control, using the following procedure for every 2-D image (definition: one channel for one TC sample).

1. Remove brightness temperatures outside the acceptable range. Specifically, we replace any value outside $[170, 330]$ K with NaN to indicate a missing value. These limits approximately represent the coldest cloud top and warmest surface temperature observed on Earth.

2. Count the number of pixels with missing (NaN) values, $N_{\text{missing}}$, noting that these values may appear in the original image or be created by step 1. If $N_{\text{missing}} > 6250$, proceed to step 3. Otherwise, proceed to step 4.

3. Flag the entire TC sample as invalid and remove it from the dataset, so that it cannot be used for training or evaluation. By subjective analysis, images with more than 6250 missing pixels (0.4% of the image) often cannot be made to look plausible by interpolation.

4. Fill the missing values by interpolation. Specifically, each NaN is replaced with its nearest real-valued neighbour on the grid.

Figure A1 shows a randomly selected TC sample affected by the loop-heat-pipe issue on the GOES-17 satellite. By visual inspection, this TC sample contains seven channels[12] with too much corrupted information to be recovered by interpolation. Our quality-control algorithm correctly flags the TC sample as invalid and removes it from the dataset.

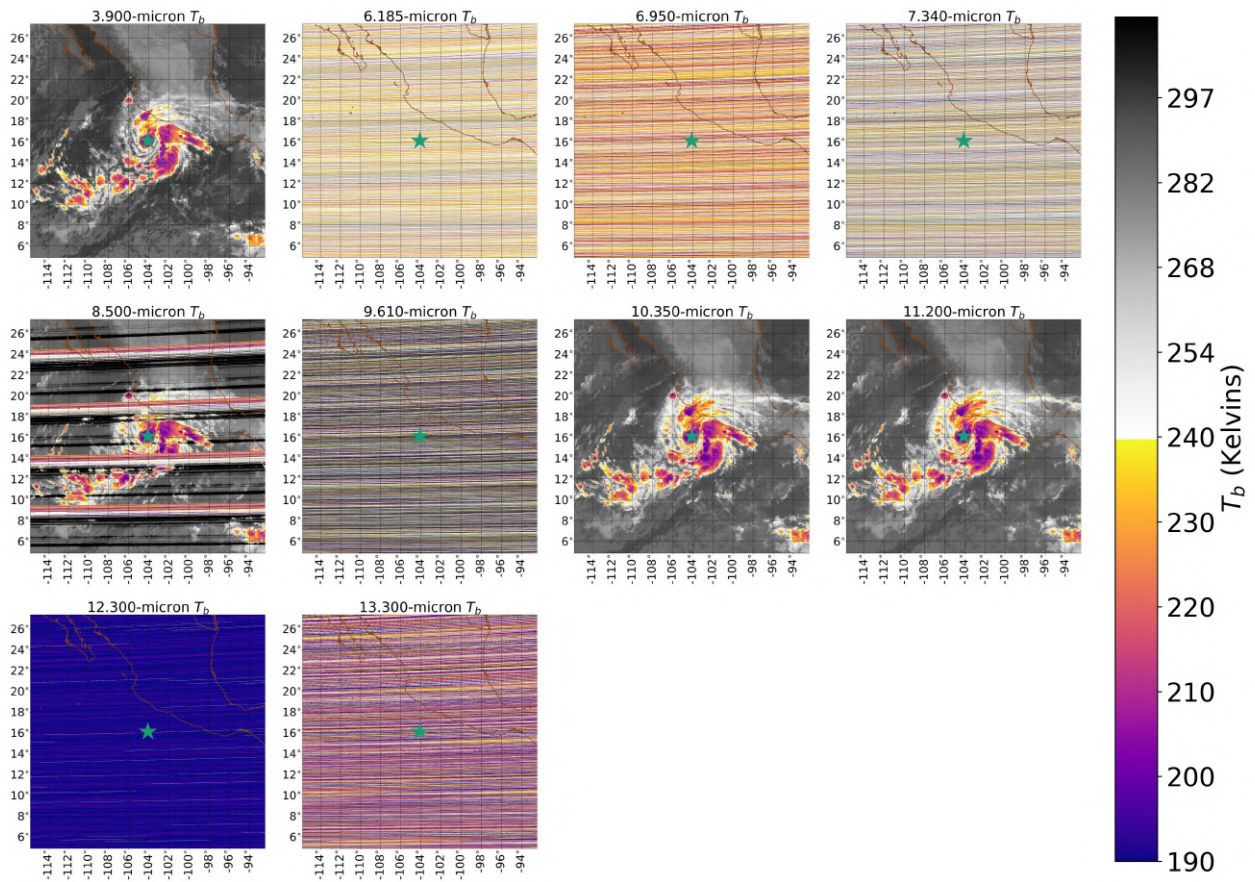

Figure A1: Raw (not quality-controlled) IR satellite imagery valid at 1200 UTC 21 Oct 2022 for Hurricane Roslyn (EP192022). Each panel shows brightness temperature at one spectral channel (colour fill) and the FBT center (green star).

---

[12]These are band 8 (6.185 $\mu$m), band 9 (6.95 $\mu$m), band 10 (7.34 $\mu$m), band 11 (8.5 $\mu$m), band 12 (9.61 $\mu$m), band 15 (12.3 $\mu$m), and band 16 (13.3 $\mu$m).

## APPENDIX B

## CNN details

For a thorough background on CNNs tailored to environmental scientists, see Lagerquist et al. (2020). This section describes our adaptation of the temporal U-net architecture developed by Chiu et al. (2020, henceforth C20), shown in Figure 3.

We modify the C20 architecture in three crucial ways. First, because our outputs are scalars rather than a gridded image, we replace the U-net backbone with a CNN backbone. The U-net backbone includes decoder layers to convert the features detected by convolutional layers into an output image, while the CNN backbone includes fully connected layers (orange arrows in Figure 3) to convert said features into scalar outputs. Second, we use time-distributed convolution, where the weights in each convolutional filter are shared across lag times (Muriga et al. 2023; Nigam and Srivastava 2023). For example, in the column marked $\mathcal{A}$ ($\mathcal{B}$) in Figure 3, each convolutional layer contains 20 (40) filters with the same weights across the nine lag times. Thus, each filter is applied identically to all lag times, detecting the same spatial features at each lag time. The approach of C20, on the other hand, is to use a different set of filters for each lag time. Our time-distributed approach reduces the number of CNN parameters to learn, which reduces computational demand and the risk of overfitting. Also, our time-distributed approach is physically motivated, as features that help locate the TC center (*e.g.*, a clear eye, spiral rain bands, etc.) do not change in *general* appearance over time. Thus, the same convolutional filters should be able to detect these features at different lag times. Said features do change in *specific* appearance (*e.g.*, intensity and location within the image) over time; these temporal changes are captured by the temporal convolution shown in Figure 3 (called the "forecasting module" in C20). Third, our CNNs perform uncertainty quantification, as discussed in Section 3c.

## APPENDIX C

## Details on uncertainty quantification

For a thorough background on machine-learned uncertainty quantification (ML-UQ) tailored to environmental scientists, see Haynes et al. (2023, henceforth H23). To capture aleatory uncertainty, we use the CRPS method, which performed best for this purpose in the experiments of H23.

Whereas the traditional deterministic approach is to provide one estimate that minimizes the mean squared error, the CRPS approach is to provide an ensemble of estimates that minimizes the CRPS. In other words, the CRPS is the loss function:

$$\text{CRPS} = \frac{1}{N} \sum_{i=1}^{N} |o_{\text{true}} - o_{\text{est}}^i| - \frac{1}{2} \frac{1}{N^2} \sum_{i=1}^{N} \sum_{j=1}^{N} |o_{\text{est}}^i - o_{\text{est}}^j|, \tag{C1}$$

where $N = 50$ is the ensemble size; $o_{\text{true}}$ is the true value; and $o_{\text{est}}^k$ is the estimate from the $k^{\text{th}}$ ensemble member. The CRPS is negatively oriented with an optimal value of 0.0. The first term is the classic mean absolute error (MAE) for the ensemble average, and the second term is the mean absolute pairwise difference (MAPD) among ensemble members. We apply Equation C1 separately to both coordinates of the translation vector (Section 3b) – namely $\Delta r$ and $\Delta c$ – and take the average of the two values as our CNN loss.

To capture epistemic uncertainty, we use the MME method. Specifically, we train four CNNs, each with a different set of IR channels, as discussed in Section 3c. Although we introduce different predictor sets out of necessity, this brings an additional advantage, namely that it increases ensemble diversity. Training each CNN in the ensemble with different predictors makes the CNNs more independent of each other, *i.e.*, their error sources and failure modes are less likely to be aligned. See Bourel et al. (2024) and references therein for a discussion of ensemble diversity.

## APPENDIX D

### Details on isotonic regression

Bias correction is a common application for isotonic regression; the model being corrected (here, a CNN) is called the "base model". For each target variable $z$, isotonic regression creates a mapping (or "lookup table") $z_i \rightarrow z_i'$, where $z_i$ is the $i^{\text{th}}$ anchor point for base-model predictions and $z_i'$ is the bias-corrected value. For example, if the CNN has a positive bias when predicting $z = 10$, isotonic regression might "ratchet down" these predictions via the mapping $z = 10 \rightarrow z' = 9.5$. For $z$-values that fall between two anchor points, isotonic regression uses linear interpolation to achieve the bias correction. During training, the mapping is adjusted to minimize mean squared error, subject to the isotonic constraint: if $z_k > z_j$, then $z_k' > z_j'$. In other words, isotonic regression cannot change the rank order of predictions.

Isotonic regression is a univariate method, with one input variable $z$ and one output variable $z'$. Thus, we train four separate isotonic-regression models for each CNN, letting zonal correction distance be $\Delta c$ and meridional correction distance be $\Delta r$:

1a. One to correct the ensemble-mean $\Delta c$. Here, the input variable is the CNN ensemble mean $\Delta c$, and the output variable is the correct $\Delta c$.

1b. Same as above but for $\Delta r$.

2a. One to correct the ensemble spread for $\Delta c$. Here, the input variable is the CNN ensemble variance for $\Delta c$, and the output variable is the squared error of the CNN ensemble mean for $\Delta c$. Thus, the goal is to make CNN variance = mean squared error (MSE), or equivalently, to make CNN standard deviation = RMSE. In other words, the goal is to achieve a spread-skill ratio of 1.0.

2b. Same as above but for $\Delta r$.

The effects of isotonic regression are illustrated in Figure D1.

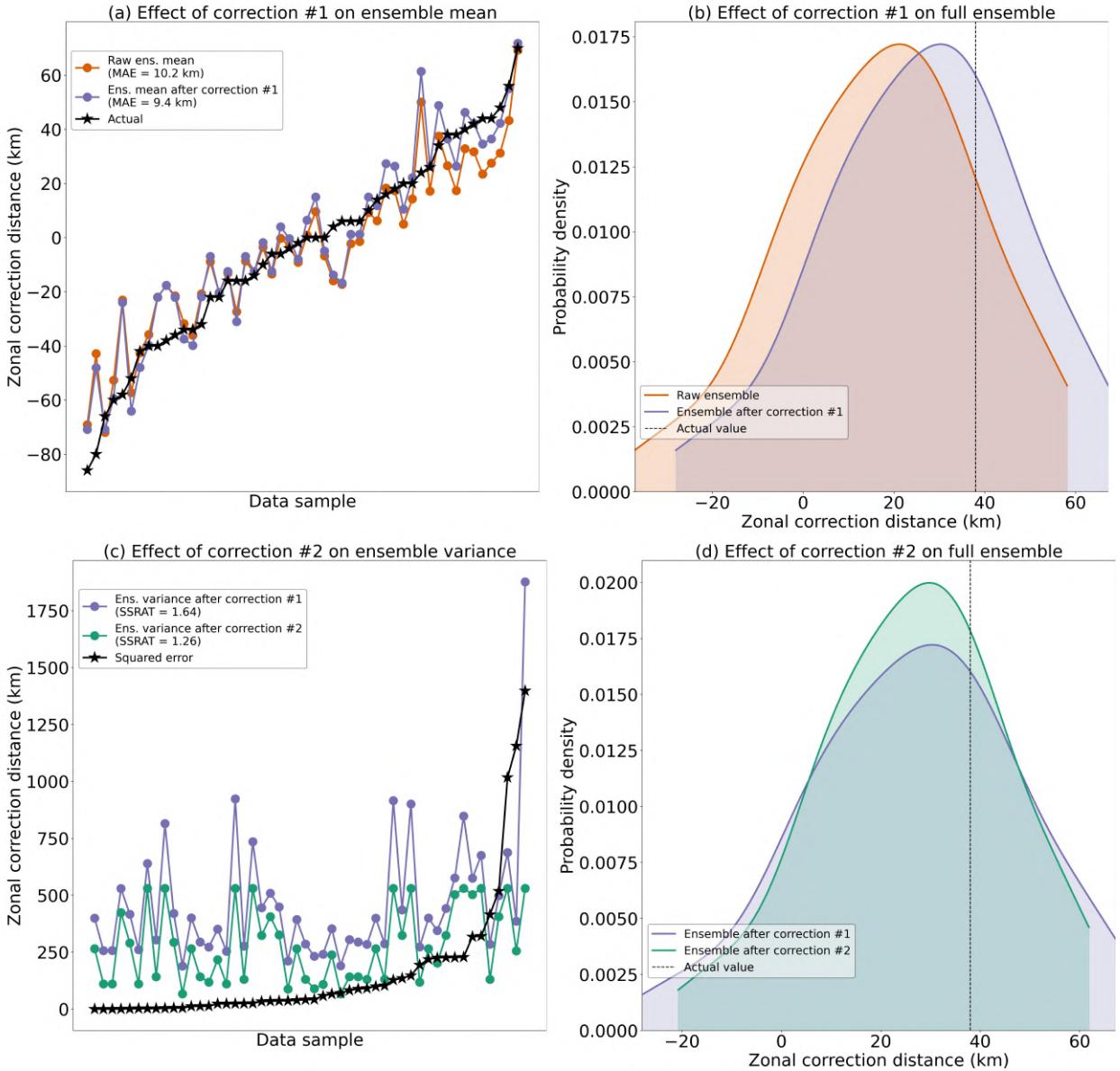

Figure D1: Demonstration of isotonic regression (iso-reg), applied here to the zonal correction distance. Correction #1 affects only the ensemble mean, and correction #2 affects only the ensemble spread. [a] Effects of correction #1 on the ensemble mean for 50 randomly selected TC samples in the testing data. For actual values (black), "zonal correction distance" = ($x$-coordinate of true center) minus ($x$-coordinate of first-guess center). For the predictions, "zonal correction distance" = ($x$-coordinate of predicted center) minus ($x$-coordinate of first-guess center). Note that iso-reg preserves the rank order of predictions. [b] Effects of correction #1 on the full ensemble for one randomly selected TC sample in the testing data. [c] Effects of correction #2 on the ensemble spread for the same 50 TC samples as in panel a. [d] Effects of correction #2 on the full ensemble for the same TC sample as in panel b. To convert the ensemble into a smooth probability curve, we use the Scott (2015) method for kernel-density estimation, as implemented in `gaussian_kde` in the scipy library (Virtanen et al. 2020).

# Uncertainty-aware center-fixing of tropical cyclones using convolutional neural networks with infrared satellite imagery

Supplemental material

This is the third version (second revision) of an article under review in the AMS journal *Weather and Forecasting*.

## 1. CNN hyperparameters

Some details of our CNNs are discussed in the main text, including the architecture (Main Section 3a), training procedure (Main Section 3b), uncertainty quantification (Main Section 3c), and experimental hyperparameters (Main Section 3e). However, many hyperparameter choices are not justified in the main text (Table S1), and some hyperparameters are not mentioned there at all (Tables S2-S3).

Table S1: Fixed hyperparameters, part 1: convolutional layers (CL), pooling layers, and fully connected layers (FCL). The first FCL is the orange arrow (Figure 3 of main text) connecting boxes labeled "1129 features" and "232 features". The number of weights in an FCL connecting shallower layer $\mathcal{S}$ to deeper layer $\mathcal{D}$ is $N_\mathcal{S}(N_\mathcal{D}+1)$, where the $N$ are the number of features in the two layers.

| Hyperparameter | Value chosen | Justification |
| --- | --- | --- |
| Size of convolutional filters | $3 \times 3$ pixels | Smallest filter size possible (other than $1 \times 1$, which cannot detect spatial patterns at all). Allows the first CLs (black arrows and left column of purple arrows in Main Figure 3) to learn features at the finest scale (where 1 pixel = 2 km), while pooling allows deeper CLs to learn features at coarser scales (where 1 pixel > 2 km). |
| Number of feature maps by CL | 10 for first CLs (black arrows in Main Figure 3), increasing to 70 for last CLs | A common choice in the ML literature (*e.g.*, Simonyan and Zisserman 2014; Wimmers et al. 2019; Sha et al. 2020) is to double the number of feature maps after each pooling; in Main Figure 3 this would lead to a progression from $10 \to 20 \to 40 \to 80 \to 160 \to 320 \to 640$ feature maps, rather than $10 \to 20 \to 30 \to 40 \to 50 \to 60 \to 70$. The first NN would be too large for our memory resources and therefore impossible to train. |
| Number of features by FCL | Exponential decrease | A superlinear decrease is needed to prevent the weight count in the FCLs from exploding. For example, in Main Figure 3, the first FCL has $(1129+1) \times 232$ = 0.26 million weights. With a linear decrease from 1129 to 100 features, this layer would have $(1129+1) \times 786$ = 0.89 million weights, nearly tripling the weight count in the entire NN from 0.44 million to 1.06 million. Thus, we implement an exponential decrease from the number of concatenated features (1129 in Main Figure 3) to the number of output variables (2; an *x*- and *y*-coordinate). We then increase the feature count for the last layer to 100, allowing for ensemble prediction. |
| Number of pooling layers | Whatever leads to a final image size of 4-7 pixels per dimension. In Main Figure 3, this means 6 pooling layers. | From prior ML experience, the coarsest grid should have ~4 pixels in each direction. More pooling (*e.g.*, the next grid in Main Figure 3 would be $2 \times 2$) destroys too much spatial information; less pooling (*e.g.*, the previous grid in Main Figure 3 is $8 \times 8$) prevents the NN from leveraging the largest-scale features. |
| Type of pooling filter (mean or maximum) | Maximum | Maximum-pooling adds more non-linearity to the NN than the alternative (mean-pooling), which is a linear operation. Also, maximum-pooling better emphasizes sharp gradients, such as those associated with deep convection (cold cloud tops) around the TC center (typically warmer cloud tops). |

Table S2: Fixed hyperparameters, part 2: other layers (activation, batch normalization) and regularization (dropout, $L_2$ penalty). "AF" = activation function. "Hidden layers" include every convolutional and fully connected layer except the output layer. We use a different regularization method for CLs (the $L_2$ penalty) vs. FCLs (dropout), because dropout was designed specifically for FCLs (Hinton et al. 2012). We use dropout only for hidden FCLs (all except the output layer), because from prior ML experience, dropout on the output layer yields extremely poor predictions.

| Hyperparameter | Value chosen | Justification |
| --- | --- | --- |
| AF for hidden layers | Leaky ReLU (Maas et al. 2013) with slope of 0.2 | Defined as $f(x) = \max(x, 0.2x)$, where $x$ is a feature value. This leaves positive values untouched and multiplies negative values by 0.2. Leaky ReLU combines the best properties of strict ReLU and the identity function (Chapter 4 of Lagerquist 2020). We choose a slope of 0.2 based on prior experience. |
| AF for output layer | None | Common AFs clip the data to a range of $0\ldots\infty$ (*e.g.*, ReLU), $0\ldots 1$ (*e.g.*, sigmoid and softmax), or $-1\ldots +1$ (*e.g.*, tanh). The outputs are estimates of $\Delta r$ and $\Delta c$ – the row and column offsets between the image center and true TC center – which both can be large positive or large negative numbers. Thus, we do not want to clip the output range with an AF. |
| Batch normalization (BN; Ioffe and Szegedy 2015) | On for hidden layers, off for output layer | BN alleviates the vanishing-gradient problem often experienced by NNs with many layers (Chapter 4 of Lagerquist 2020). BN transforms data to a standard normal distribution – where nearly all values range from $[-3, +3]$ – making it inappropriate for the output layer. |
| Strength of $L_2$ regularization | $10^{-6}$ | Chosen from an earlier hyperparameter experiment (not shown), where we varied $L_2$ strength from $10^{-7}$ to $10^{-5}$. |
| Dropout rate | 0.5 | Chosen from the same earlier hyperparameter experiment, where we varied dropout rate from 0.125 to 0.5. |

Table S3: Fixed hyperparameters, part 3: training procedure.

| Hyperparameter | Value chosen | Justification |
| --- | --- | --- |
| Number of epochs | 1000 | In each epoch, 36 training batches and 12 validation batches – with 20 samples per batch (see Section 3b of main text) – are presented to the NN. |
| Optimizer | AdamW (Loshchilov and Hutter 2017) | More sophisticated version of stochastic gradient descent (Section 8.3.1 of Goodfellow et al. 2016). |
| Learning rate | Start with 0.001, reduce by 5% upon 10-epoch plateau | A starting value of 0.001 is the default for the AdamW optimizer in Keras; a "10-epoch plateau" means that validation loss has not reached a new minimum in the last 10 epochs. While developing GeoCenter, we found that a 10-epoch waiting period and 5% reduction factor lead to better model performance than our previous preference, namely a 20-epoch waiting period and 40% reduction factor. |

## 2. Further validation results for hyperparameter experiment

Here we expand on the results in Section 4 of the main text, which shows two evaluation metrics: mean Euclidean distance and SSRAT. Figures S1-S8 show the other eight metrics.

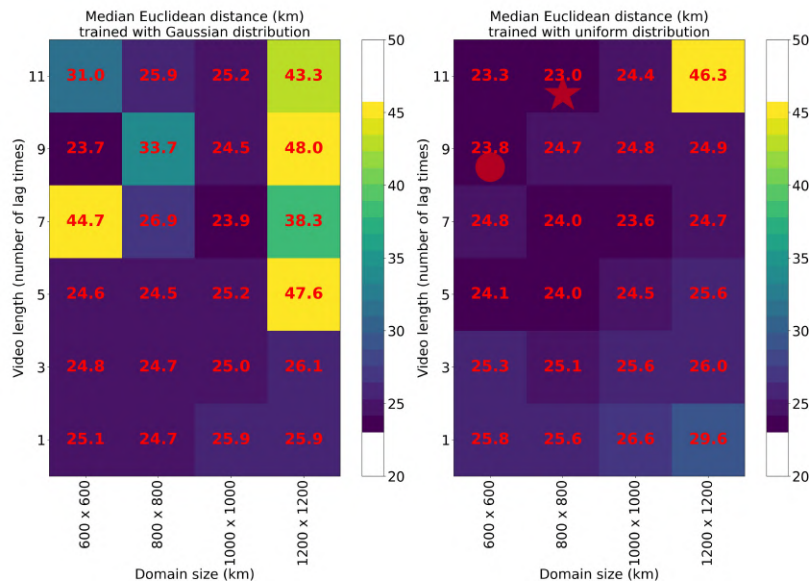

Figure S1: Median Euclidean distance on all validation data (both tropical and non-tropical systems), with respect to all three experimental hyperparameters. The circle marks the selected (best overall) model, while the star marks the best model according to median Euclidean distance.

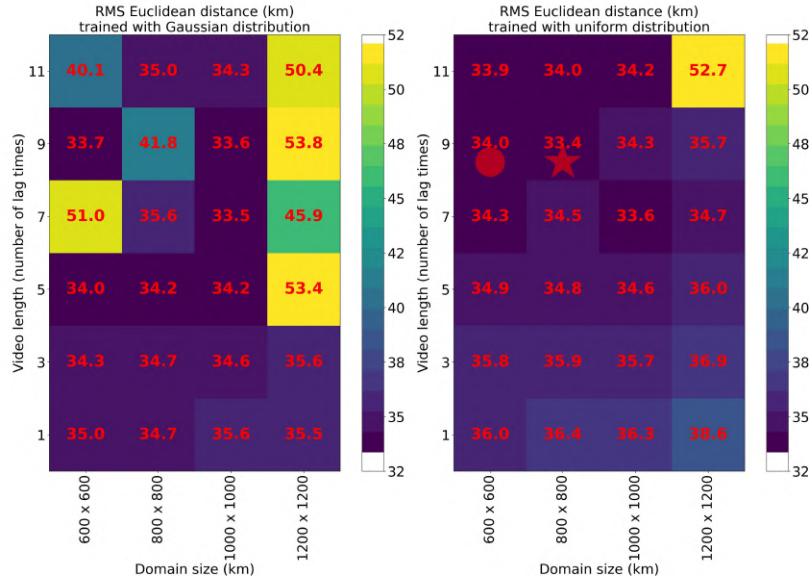

Figure S2: RMS Euclidean distance on all validation data (both tropical and non-tropical systems), with respect to all three experimental hyperparameters. The circle marks the selected (best overall) model, while the star marks the best model according to RMS Euclidean distance.

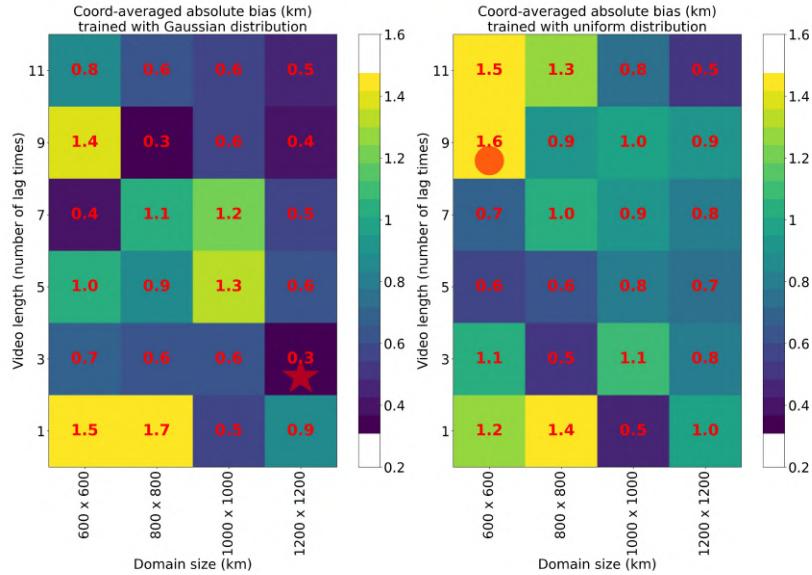

Figure S3: Coordinate-averaged absolute bias on all validation data (both tropical and non-tropical systems), with respect to all three experimental hyperparameters. The circle marks the selected (best overall) model, while the star marks the best model according to coordinate-averaged absolute bias.

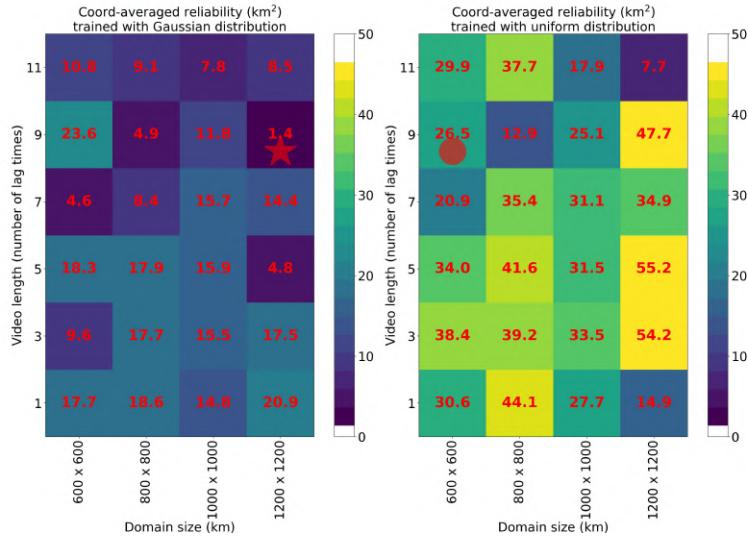

Figure S4: Coordinate-averaged reliability on all validation data (both tropical and non-tropical systems), with respect to all three experimental hyperparameters. The circle marks the selected (best overall) model, while the star marks the best model according to coordinate-averaged reliability.

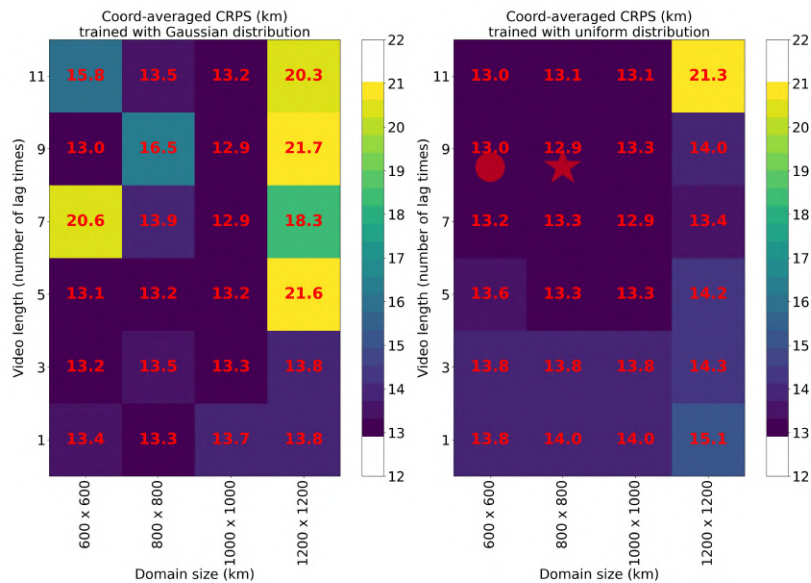

Figure S5: Coordinate-averaged CRPS on all validation data (both tropical and non-tropical systems), with respect to all three experimental hyperparameters. The circle marks the selected (best overall) model, while the star marks the best model according to coordinate-averaged CRPS.

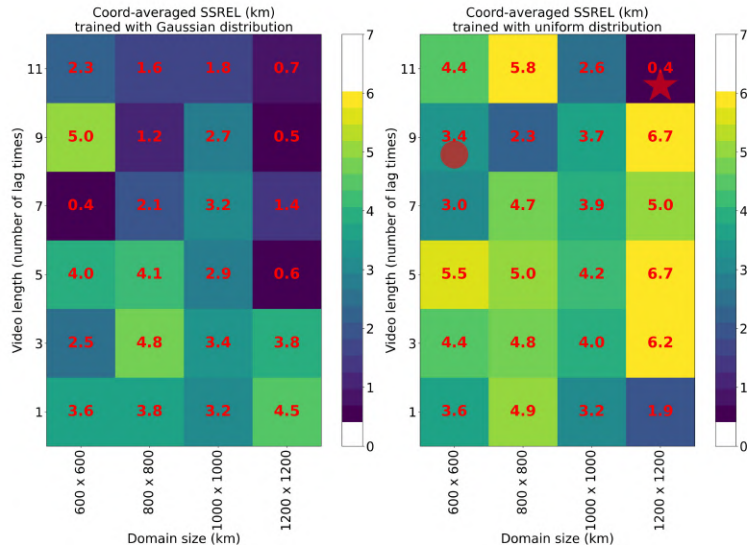

Figure S6: Coordinate-averaged SSREL on all validation data (both tropical and non-tropical systems), with respect to all three experimental hyperparameters. The circle marks the selected (best overall) model, while the star marks the best model according to coordinate-averaged SSREL.

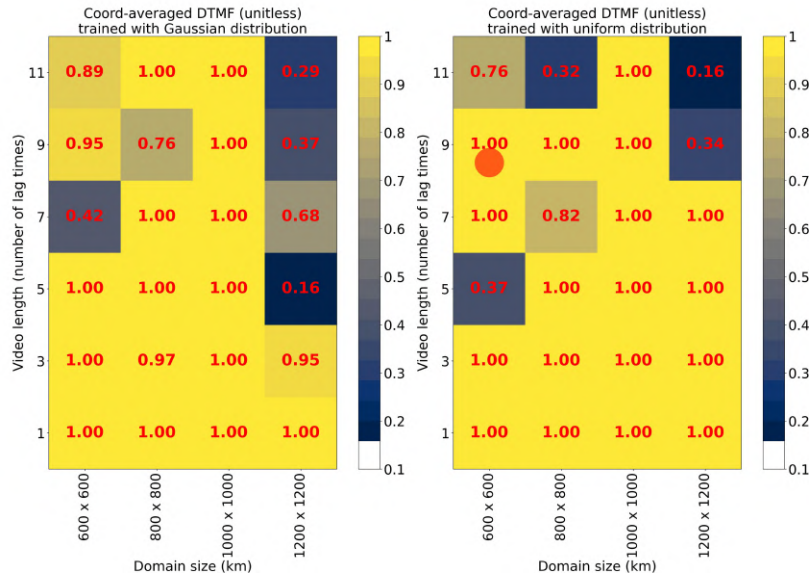

Figure S7: Coordinate-averaged DTMF on all validation data (both tropical and non-tropical systems), with respect to all three experimental hyperparameters. The circle marks the selected (best overall) model. There is no marker for the best coordinate-averaged DTMF, because most models achieve the optimal DTMF of 1.0.

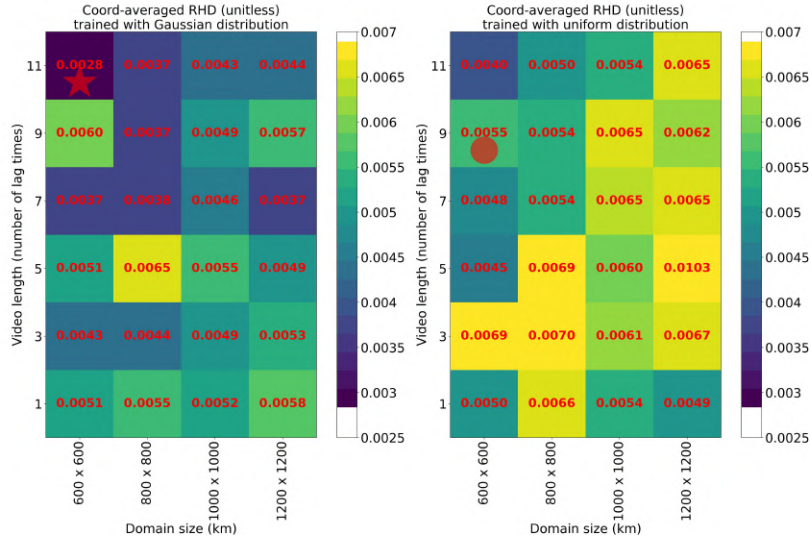

Figure S8: Coordinate-averaged RHD on all validation data (both tropical and non-tropical systems), with respect to all three experimental hyperparameters. The circle marks the selected (best overall) model, while the star marks the best model according to coordinate-averaged RHD.

## 3. Further testing results for final ensemble

Section 5 of the main text presents an objective evaluation of the final ensemble – henceforth, GeoCenter – along with several case studies. That evaluation is based only on tropical systems in the testing data. The following subsections present an objective evaluation for *non*-tropical systems and for *all* systems, then expand upon the case studies.

*a. Objective evaluation for non-tropical systems*

Figures S9-S12 are analogous to Figures 7-10 in the main text but for *non*-tropical systems in the testing data. Tropical systems are defined in Table 3 of the main text; non-tropical systems encompass the rest of the dataset.

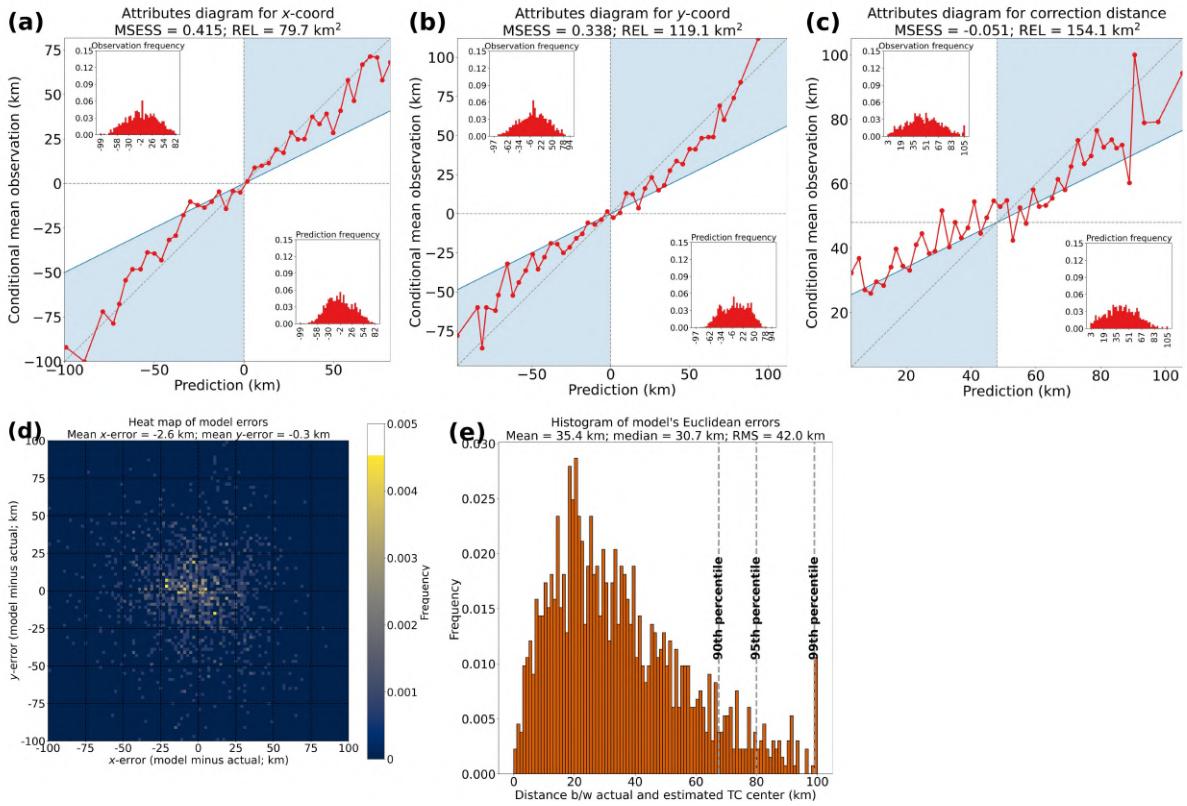

Figure S9: Performance of GeoCenter ensemble mean on non-tropical systems in testing data. Formatting is explained in the caption of Figure 7 in the main text.

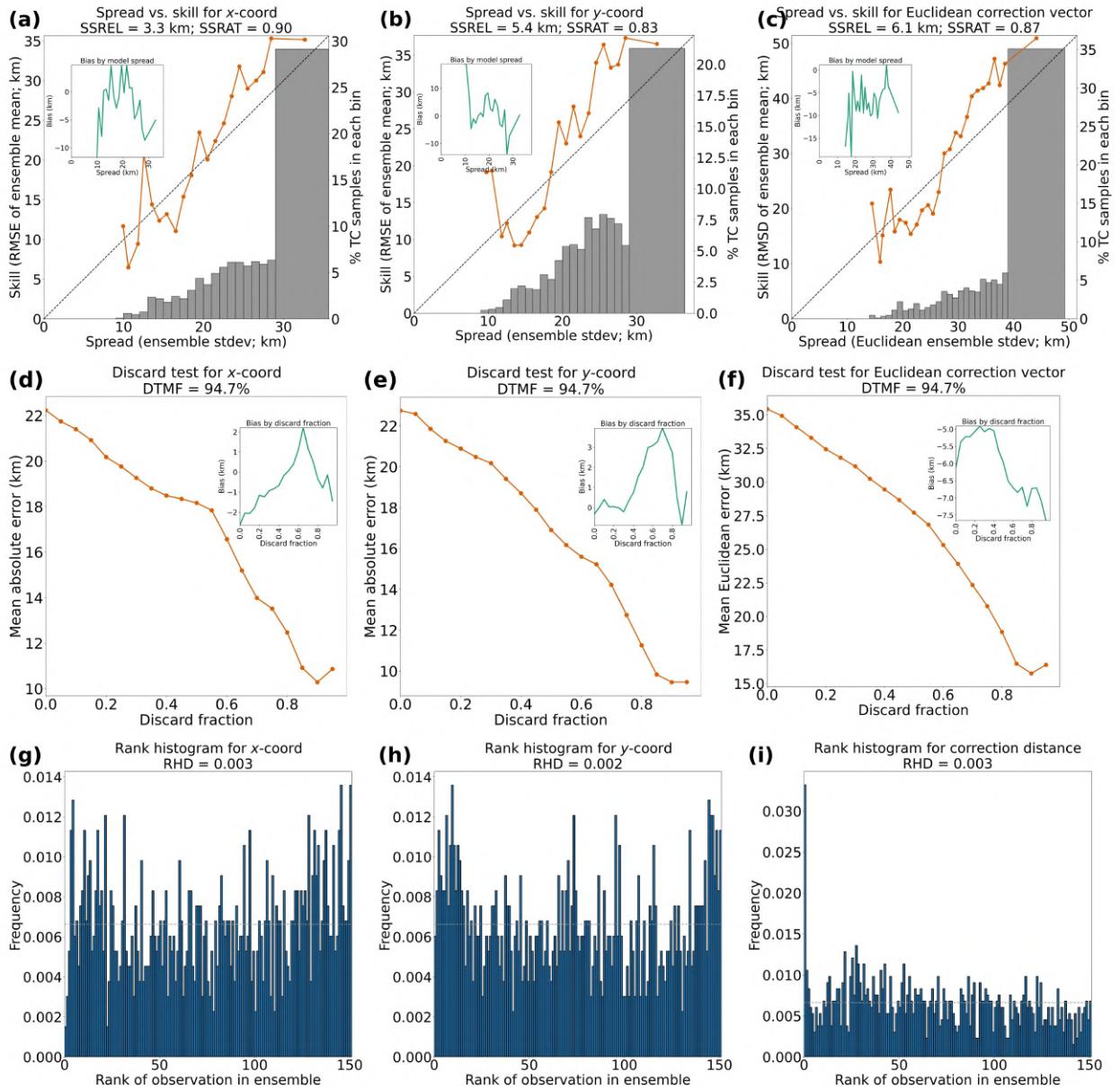

Figure S10: Performance of GeoCenter uncertainty estimates on non-tropical systems in testing data. Formatting is explained in the caption of Figure 8 in the main text.

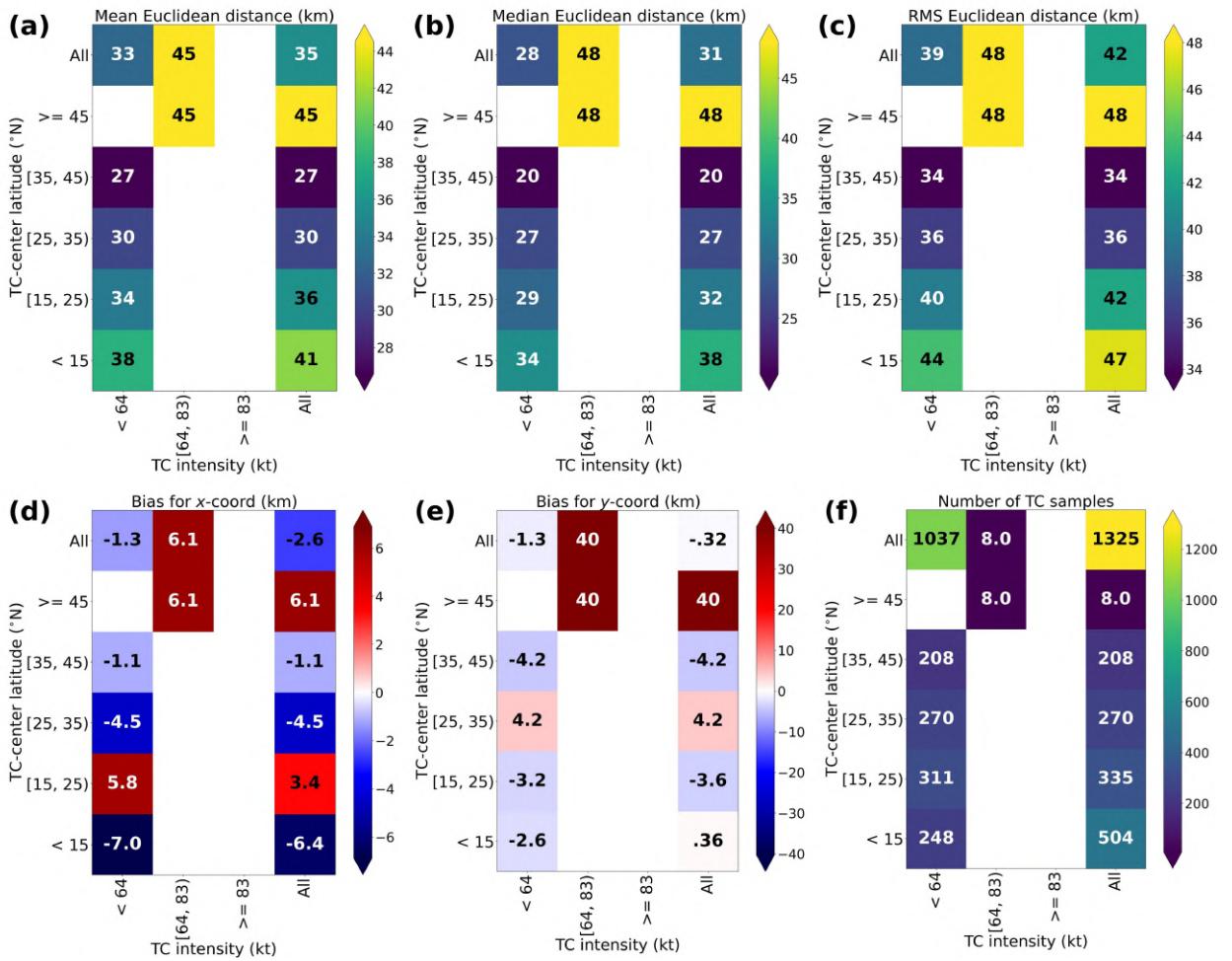

Figure S11: Performance of GeoCenter on non-tropical systems in testing data, as a function of TC intensity and true TC-center latitude (from FBT). Formatting is explained in Figure 9 of the main text.

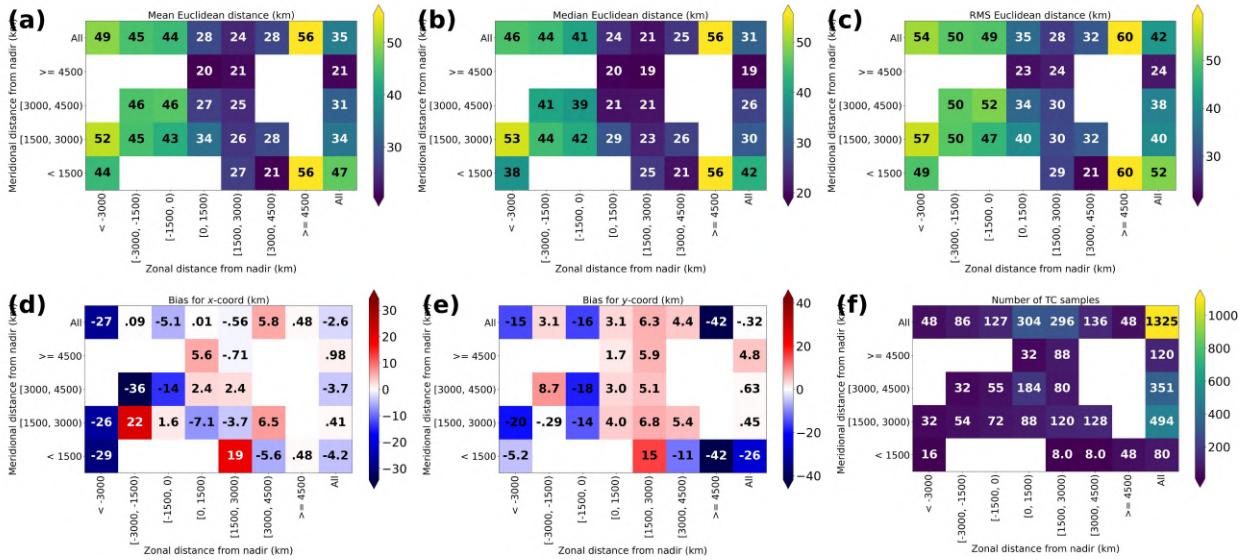

Figure S12: Performance of GeoCenter on non-tropical systems in testing data, as a function of nadir-relative position (FBT center). Formatting is explained in Figure 10 of the main text.

## b. Objective evaluation for all systems

Figures S13-S16 are analogous to Figures 6-10 in the main text but for *all* systems, rather than only tropical systems, in the testing data. Note that the two sets of figures are quite similar, as the testing dataset is dominated by tropical systems (6832 samples vs. 1328).

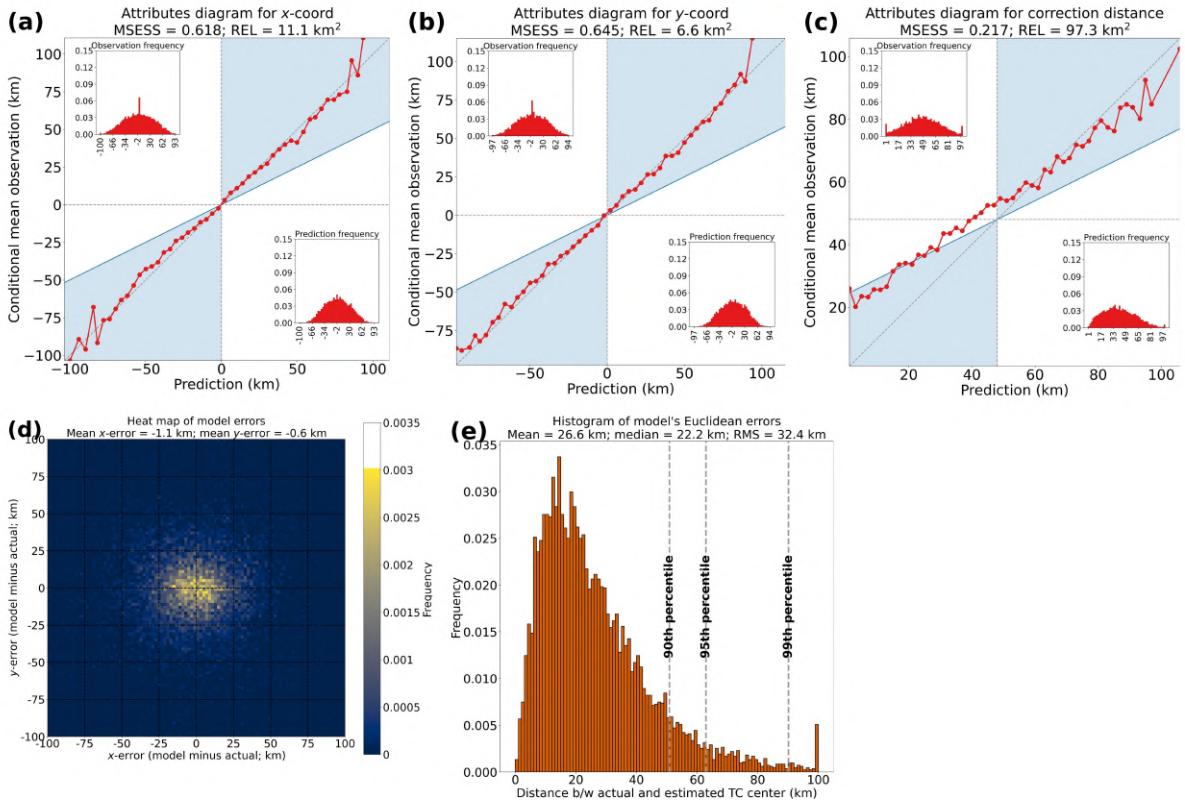

Figure S13: Performance of GeoCenter ensemble mean on all systems in testing data. Formatting is explained in the caption of Figure 7 in the main text.

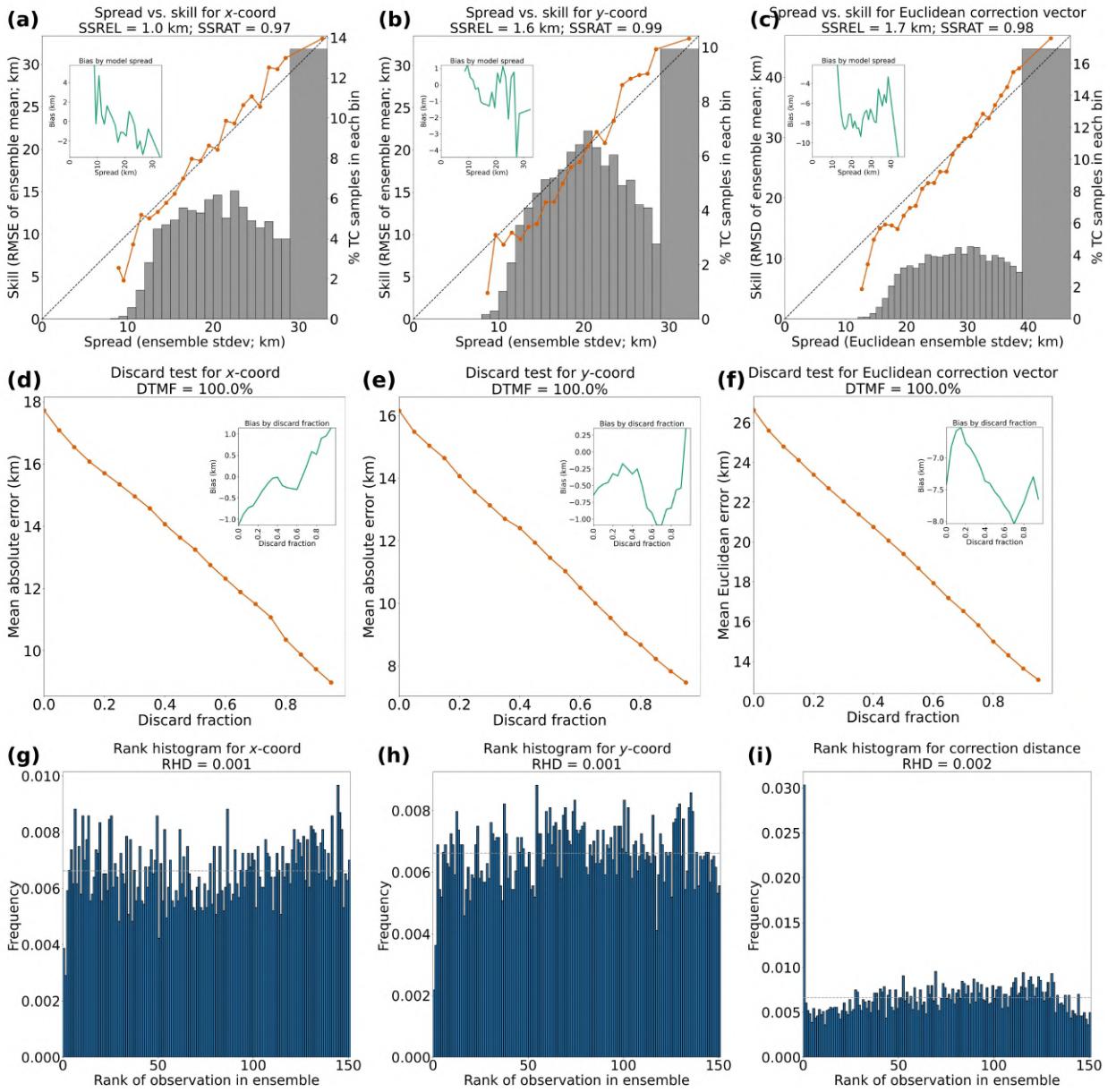

Figure S14: Performance of GeoCenter uncertainty estimates on all systems in testing data. Formatting is explained in the caption of Figure 8 in the main text.

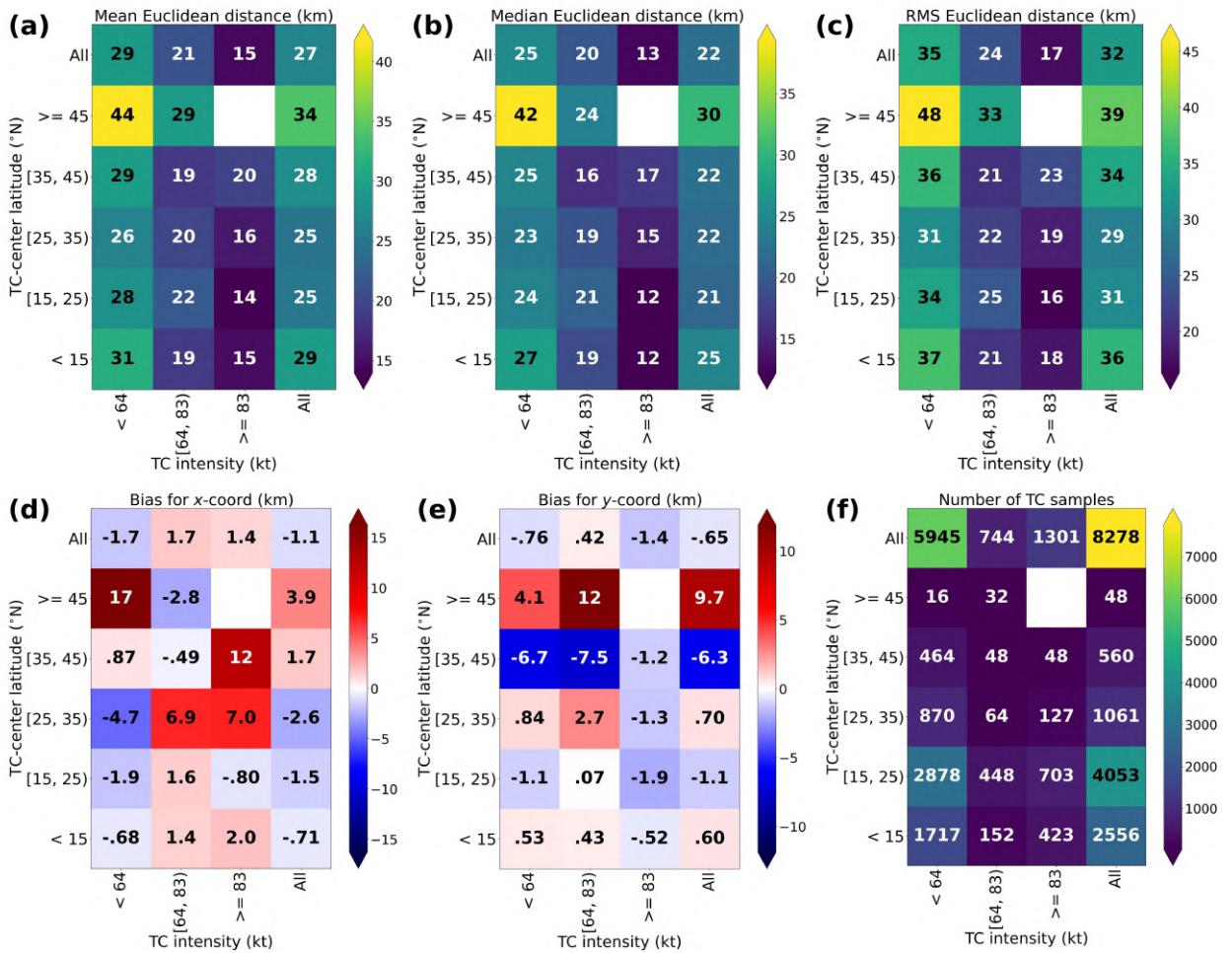

Figure S15: Performance of GeoCenter on all systems in testing data, as a function of TC intensity and true TC-center latitude (from FBT). Formatting is explained in Figure 9 of the main text.

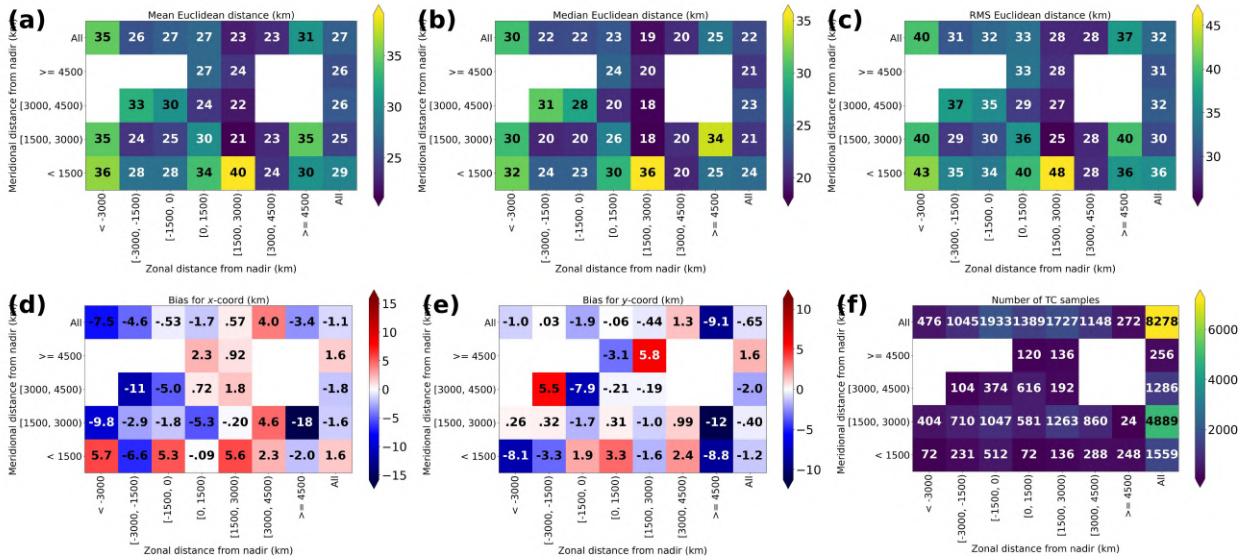

Figure S16: Performance of GeoCenter on all systems in testing data, as a function of nadir-relative position (FBT center). Formatting is explained in Figure 10 of the main text.

*c. Objective evaluation for day vs. night*

Table S4 breaks down GeoCenter performance by daytime (when $\alpha > 0°$, $\alpha$ being the solar altitude angle at the TC center) vs. nighttime (when $\alpha < 0°$). We perform this analysis partly due to concerns about the inclusion of the 3.9-$\mu$m channel in our predictors, which behaves differently during the day, when it receives energy reflected from the Sun, versus during the night, when it does not (Kim and Hong 2019; Kim et al. 2019). However, GeoCenter does not appear to be severely affected by this difference in behaviour; its daytime and nighttime performance are very similar, with no metric differing by more than 1.53 km.

Table S4: Performance of GeoCenter on tropical systems in testing data, divided by time of day.

|  | Time | |
| --- | --- | --- |
| **Evaluation metric** | **Day** | **Night** |
| Mean Euclidean distance (km) | 24.9 | 24.6 |
| Median Euclidean distance (km) | 21.0 | 20.4 |
| RMS Euclidean distance (km) | 30.0 | 29.9 |
| Bias in $x$-coordinate (km) | -1.09 | -0.21 |
| Bias in $y$-coordinate (km) | -1.32 | +0.21 |

*d. Further details on case studies*

For every case study in the main text (Figures 13-16), only one random first guess is shown, along with all IR predictors at the 0-min lag time. Here we show all eight first guesses for every case study, along with a smaller subset of IR predictors: just band 11 (wavelength = 8.5 $\mu$m for GOES, 8.6 $\mu$m for Himawari) at the 0-min lag time. In this way, Figures S17-S20 are analogous to Figures 13-16 of the main text.

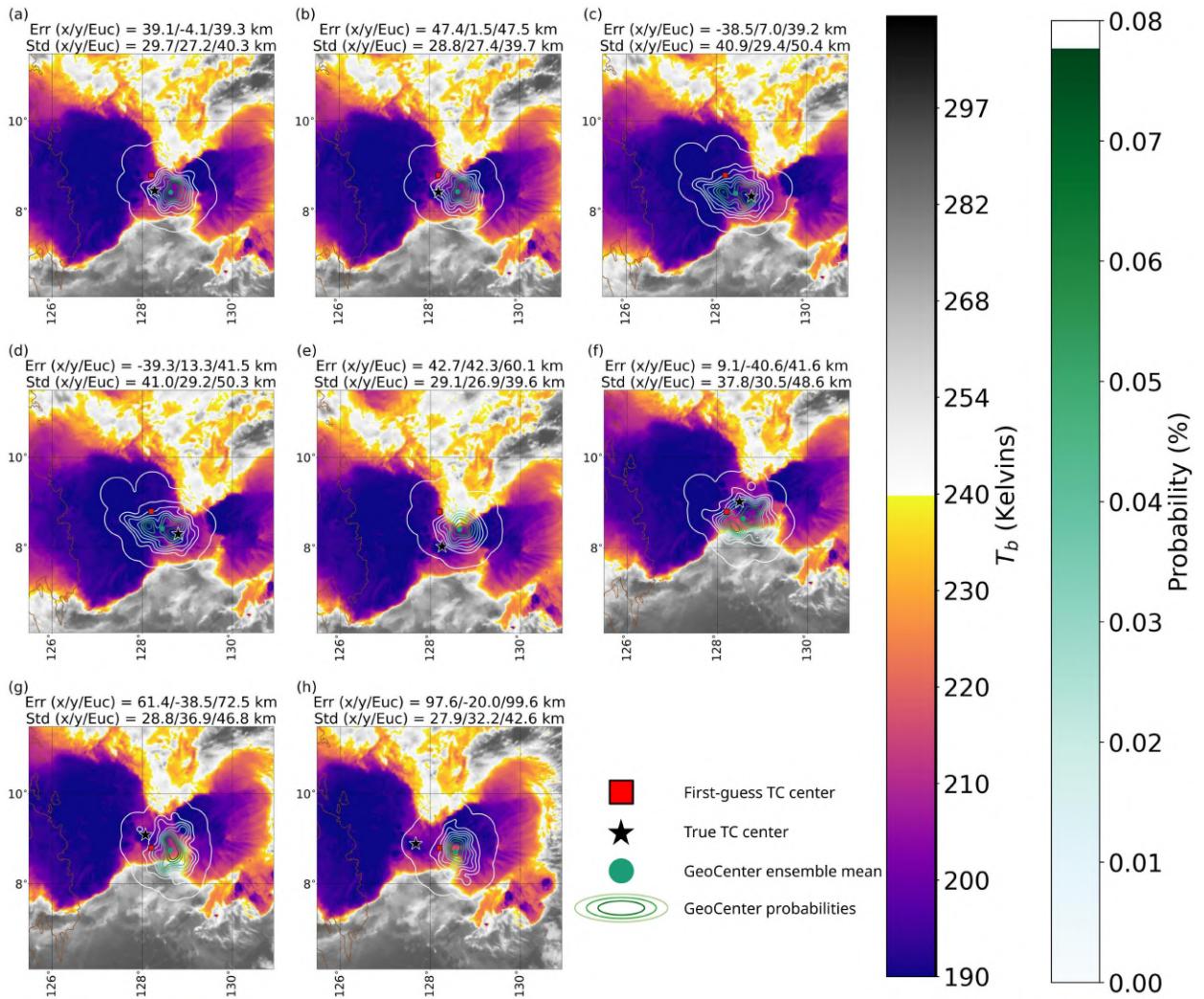

Figure S17: Case study: Tropical Depression Dujuan/Auring (WP012021), with an intensity of 35 kt at 0600 UTC 21 Feb 2021. Each panel corresponds to one of the eight first guesses applied to the original TC sample. Each panel shows one 2-D slice of the IR predictors – band 11 (wavelength = 8.6 $\mu$m) at the 0-min lag time – along with probability contours representing estimated TC-center locations from the 150 members of the GeoCenter ensemble. Darker greens correspond to higher GeoCenter probabilities, and the green circle inside the innermost contour is the GeoCenter ensemble mean. The red square is the image center, and the black star is the true TC center. The title of each panel gives the error of the ensemble mean ($x$-coordinate, $y$-coordinate, and Euclidean) and the standard deviation of the ensemble members, *i.e.*, the spread.

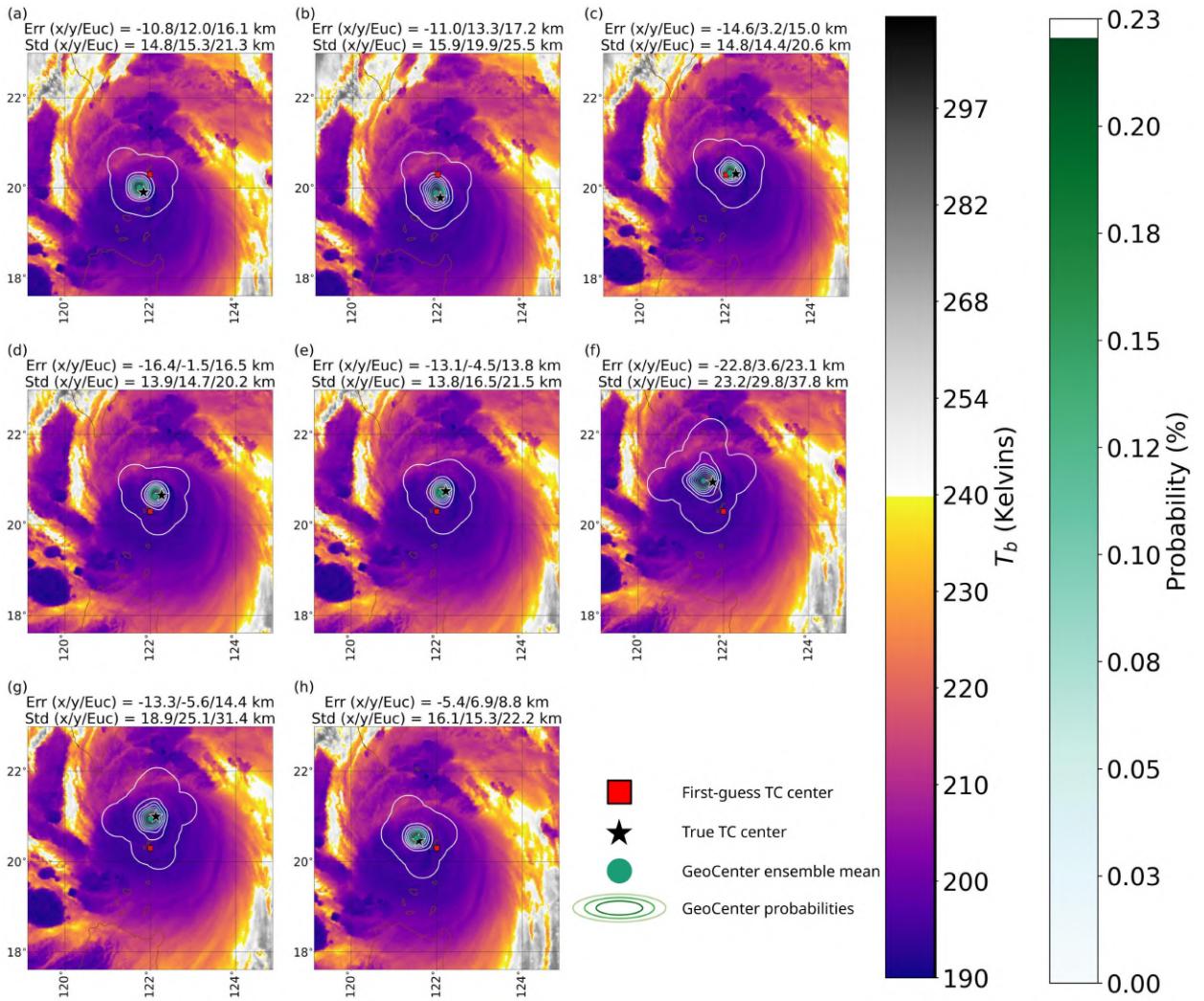

Figure S18: Case study: Super Typhoon Chanthu/Kiko (WP192021), with an intensity of 145 kt at 0000 UTC 11 Sep 2021. Formatting is explained in the caption of Figure S17.

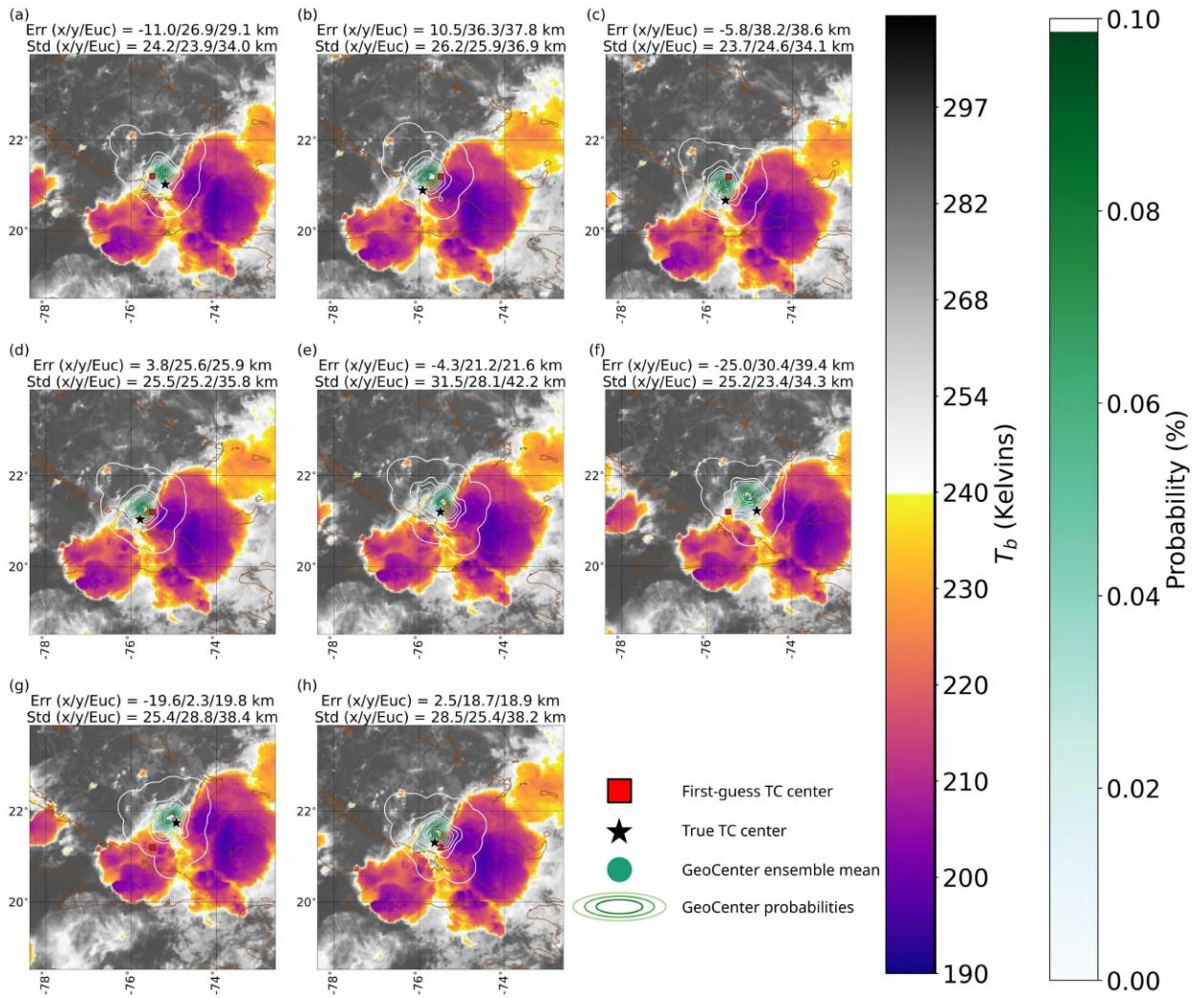

Figure S19: Case study: Tropical Depression Fred (AL062021), with an intensity of 30 kt at 0000 UTC 13 Aug 2021. Formatting is explained in the caption of Figure S17.

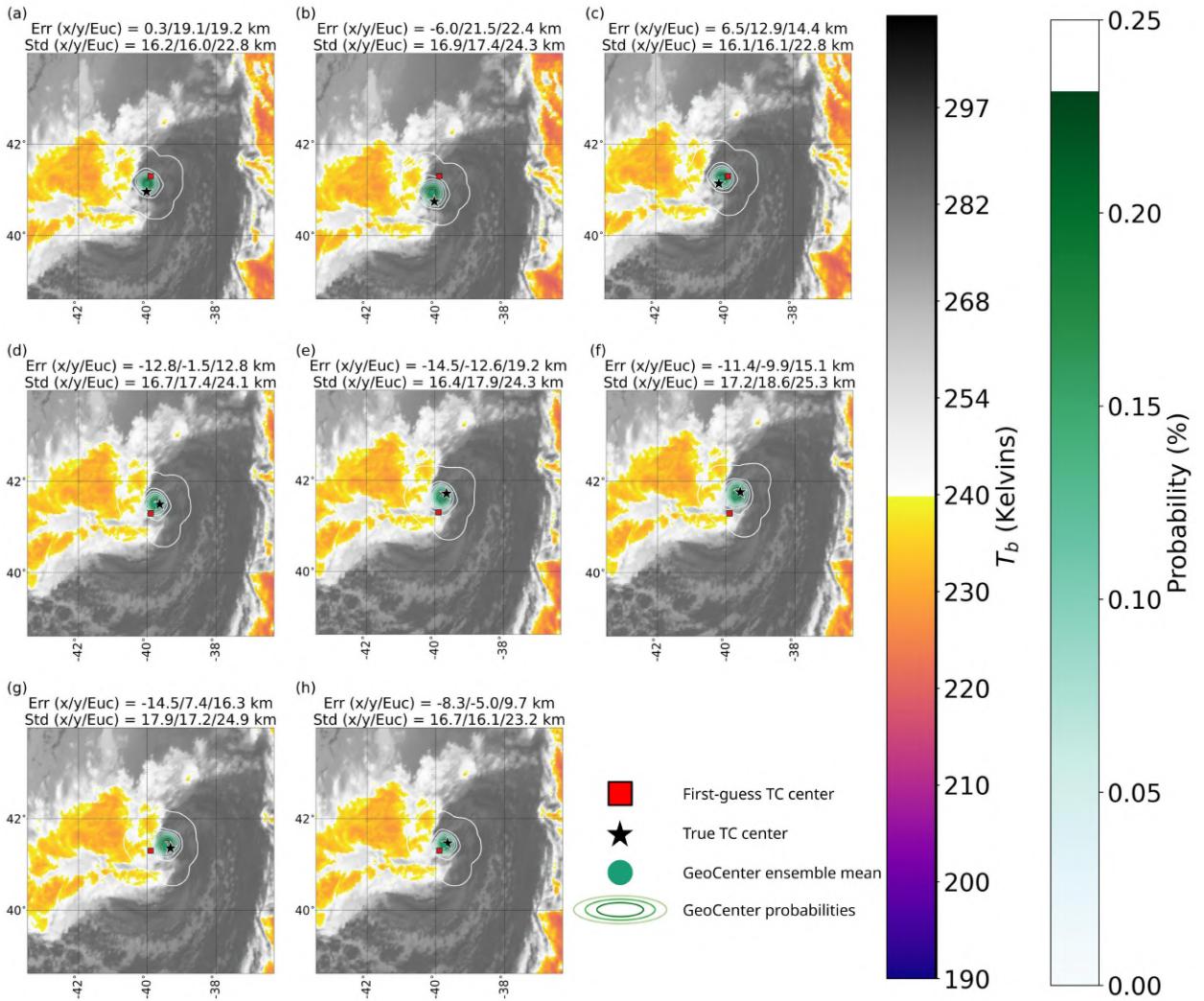

Figure S20: Case study: extratropical remnants of Tropical Storm Odette (AL152021) at 0600 UTC 22 Sep 2021. Formatting is explained in the caption of Figure S17.

The fifth and last case study (Figure S21; not shown in the main text) is a storm that transitioned to a remnant low three days prior (Newton; EP152022). Although the system contains several small areas of deep convection with no obvious center, it still contains remnants of a spiral cloud band around 16°N, most visible at 3.9, 8.5, and 11.2 $\mu$m. Similar to the Odette case, GeoCenter locates the center of Newton quite effectively, with a Euclidean error of 21.1 km. Figure S22 shows that the mean error over all eight first guesses is 23.6 km.

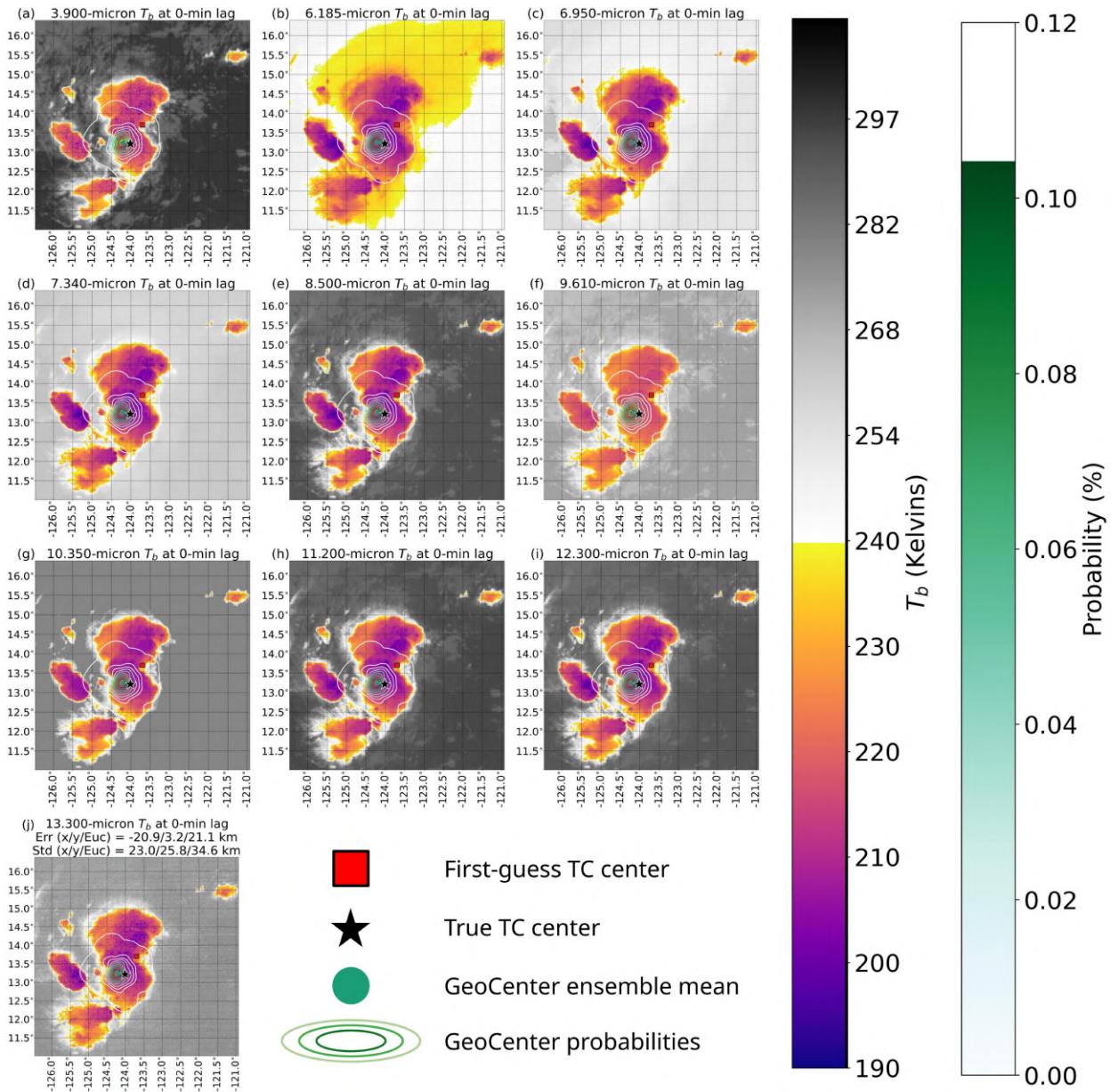

Figure S21: Case study: remnant low of Tropical Storm Newton (EP152022) at 1200 UTC 28 Sep 2022. As for case studies shown in the main text, each panel shows the brightness temperature at a different IR channel, but all panels show the same random first guess.

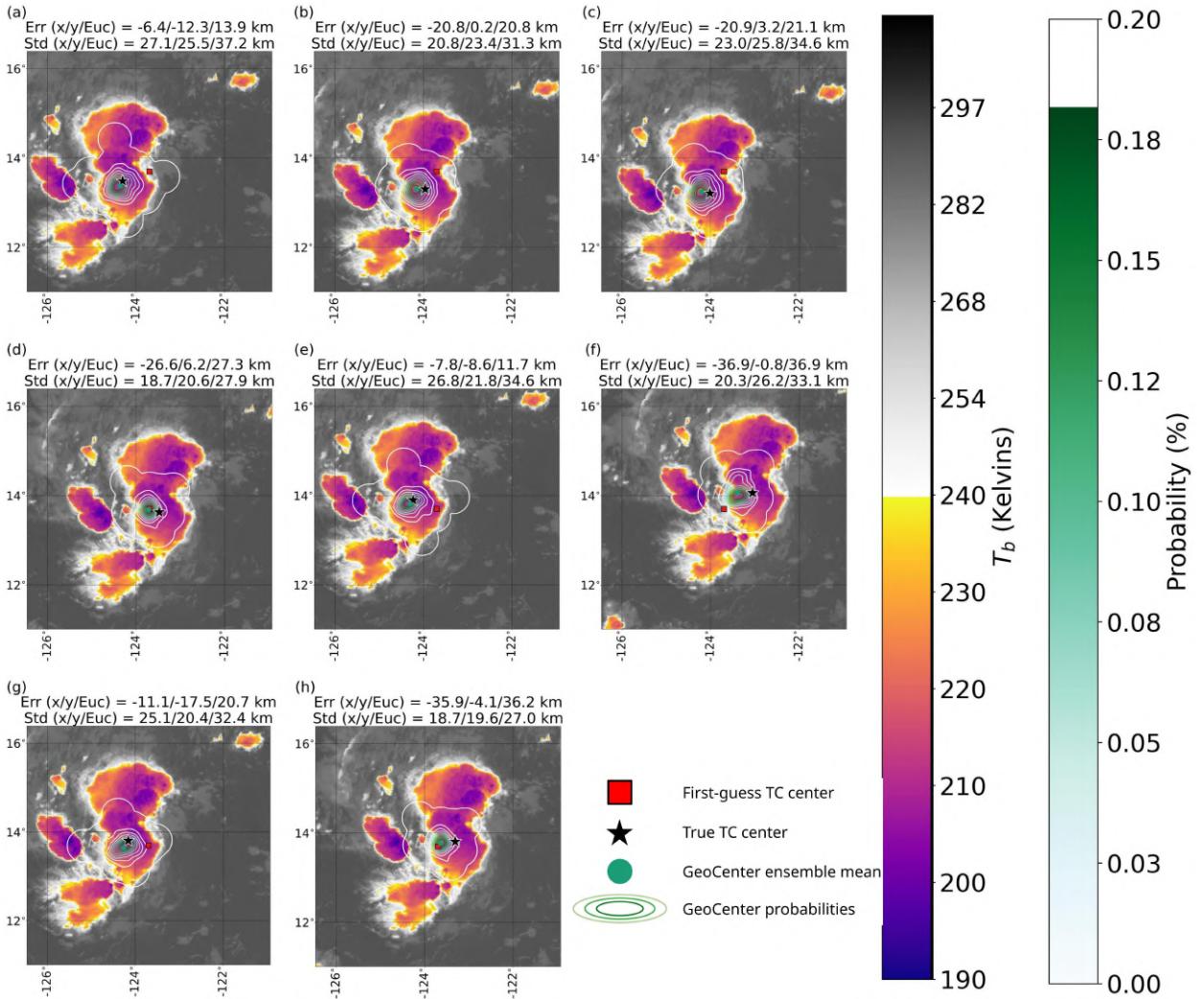

Figure S22: Case study: remnant low of Tropical Storm Newton (EP152022) at 1200 UTC 28 Sep 2022. As for Figures S17-S20, each panel shows a different random first guess.

*e. Full tracks for case studies*

Figures S23-S27 show the full track for every TC used in the case studies, according to both FBT and GeoCenter. Each GeoCenter estimate shown in these figures is a mean over 1200 estimates, generated from 8 random first-guess centers × 150 ensemble members. The title of each figure shows the mean and median Euclidean errors over the track, based only on GeoCenter estimates at synoptic times, *i.e.*, those with a simultaneous FBT center. The "sample size" in the title is the number of such synoptic-time comparisons. Note that GeoCenter estimates are sometimes missing for a day (see AL062021, AL152021, and EP152022), due to missing satellite data in our archive.

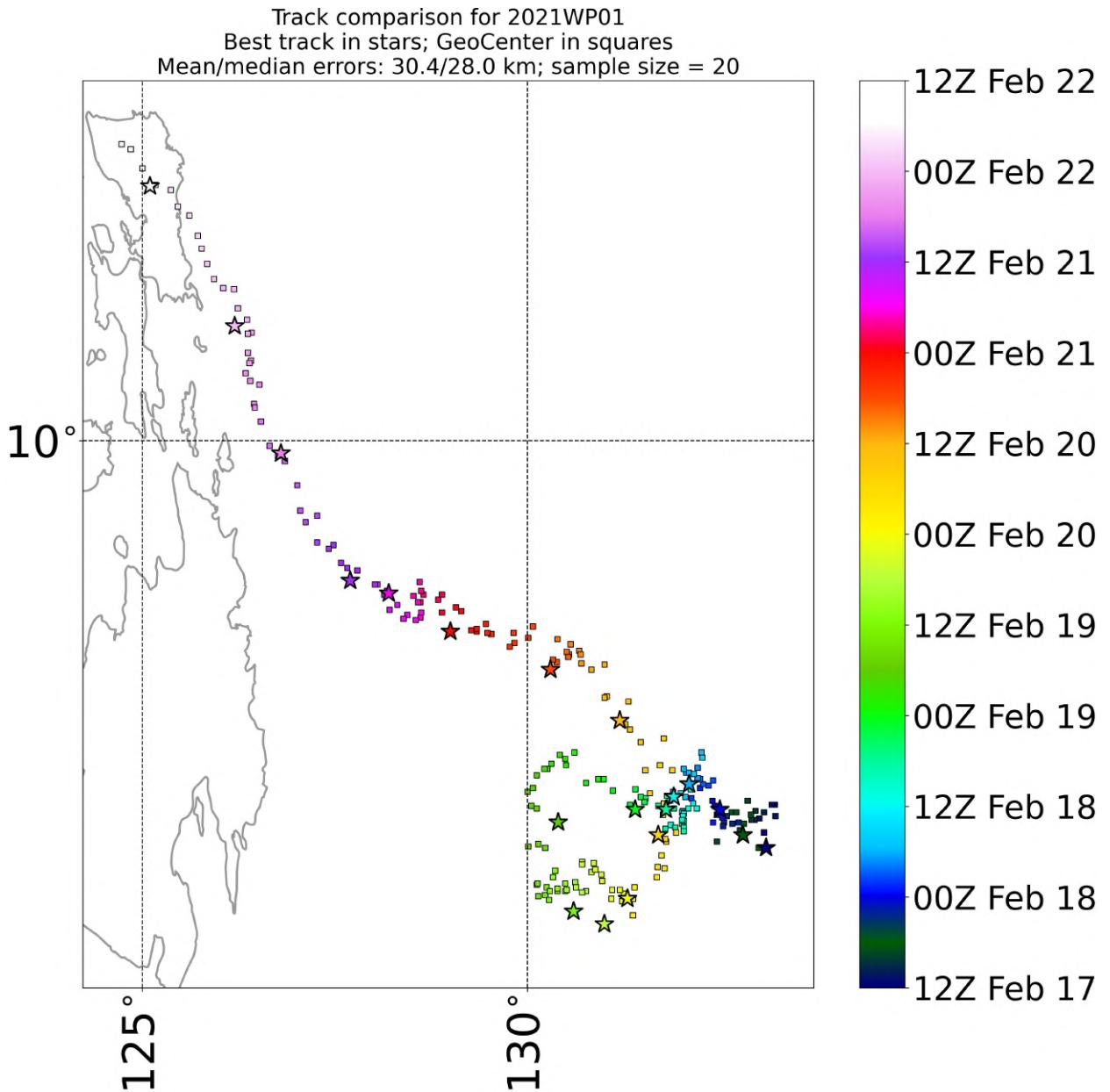

Figure S23: Full track for Tropical Storm Dujuan/Auring (WP012021), according to both the final best tracks (FBT; stars) and GeoCenter (squares). Both FBT and GeoCenter estimates are colour-coded by time. Error statistics in the title are based only on GeoCenter estimates at synoptic times, which can be compared with a simultaneous FBT center. The "sample size" in the title is the number of such synoptic-time comparisons.

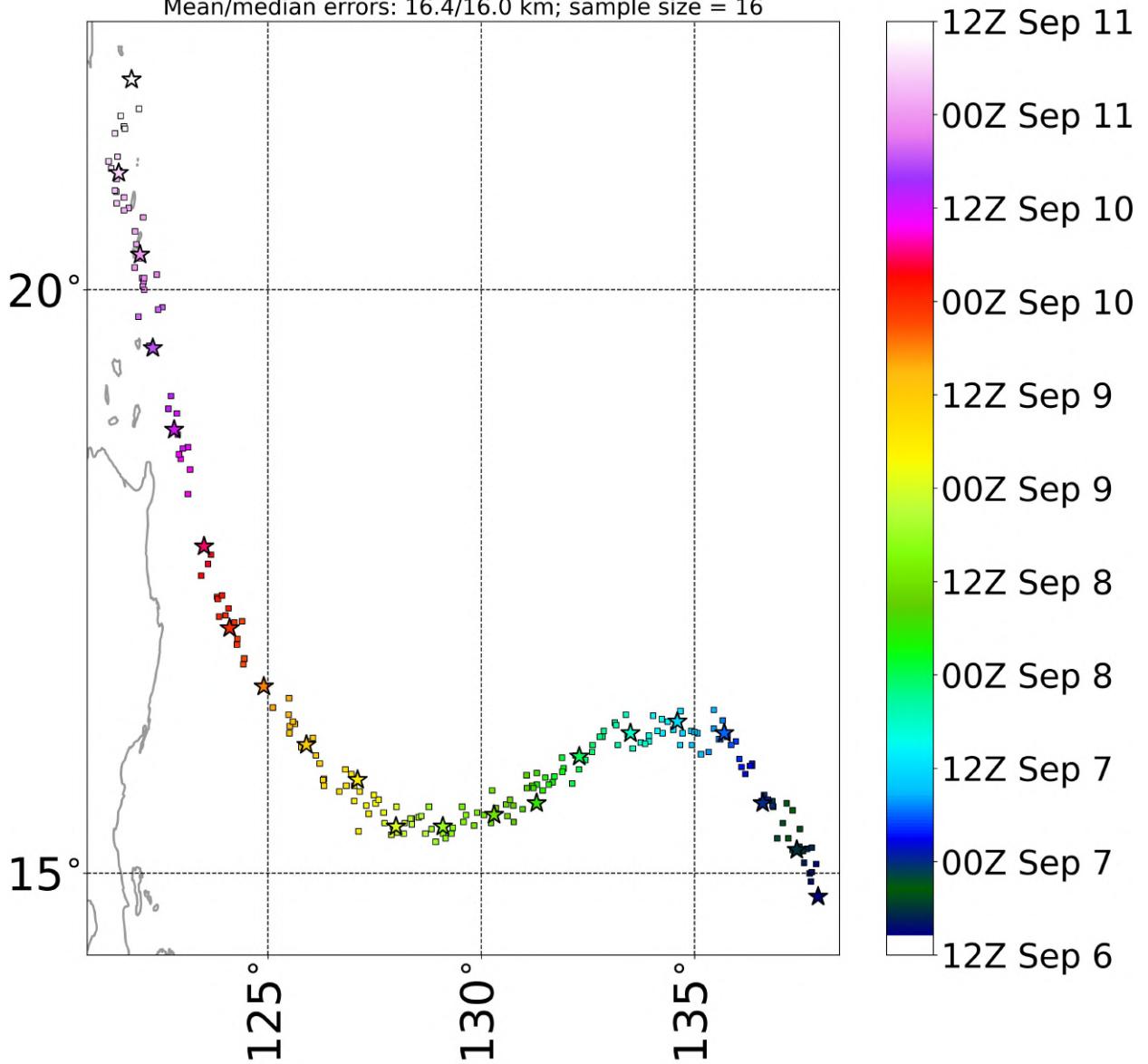

Figure S24: Full track for Typhoon Chanthu/Kiko (WP192021), according to both the final best tracks (FBT; stars) and GeoCenter (squares). Formatting is explained in the caption of Figure S23.

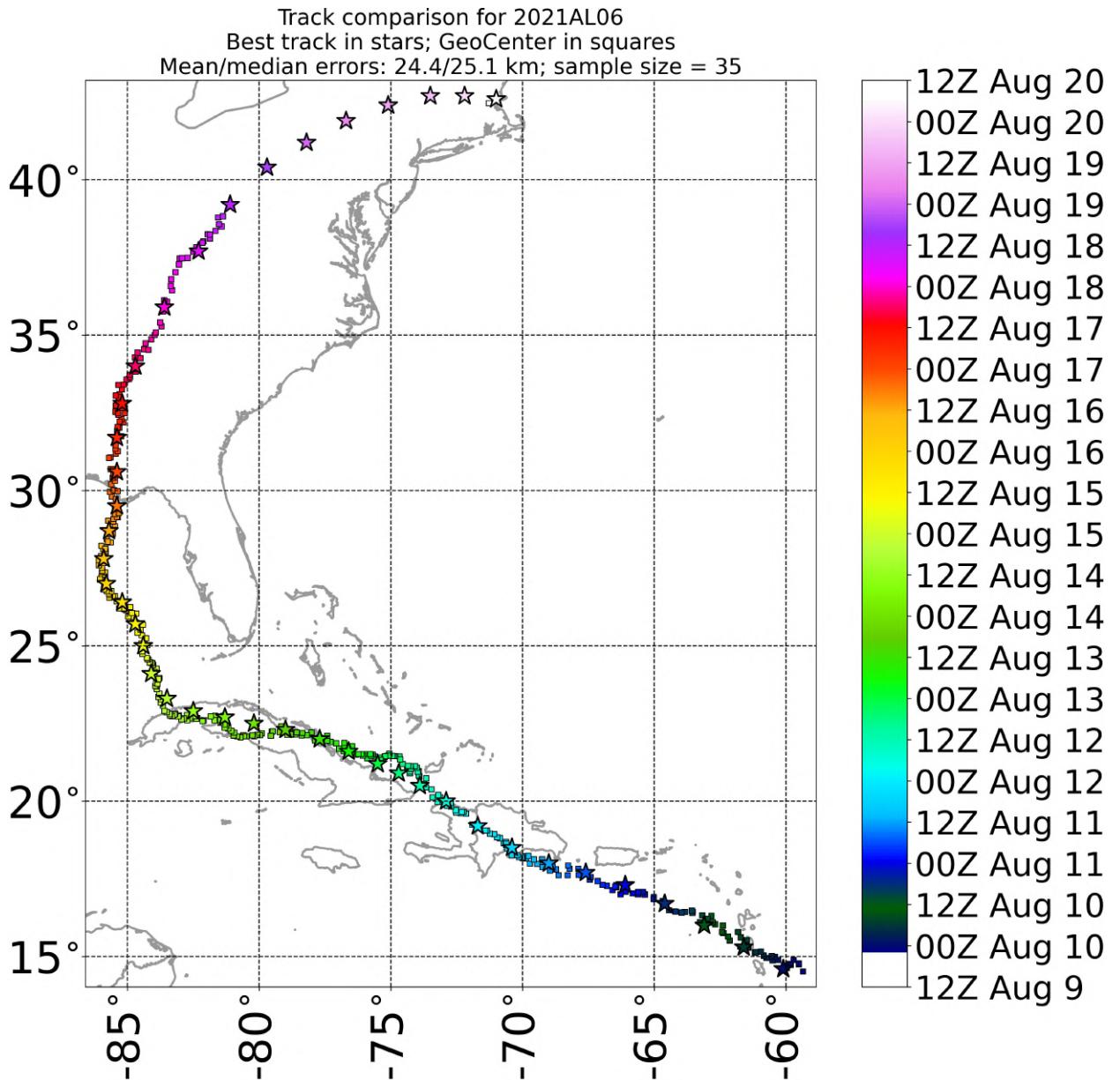

Figure S25: Full track for Tropical Storm Fred (AL062021), according to both the final best tracks (FBT; stars) and GeoCenter (squares). Formatting is explained in the caption of Figure S23.

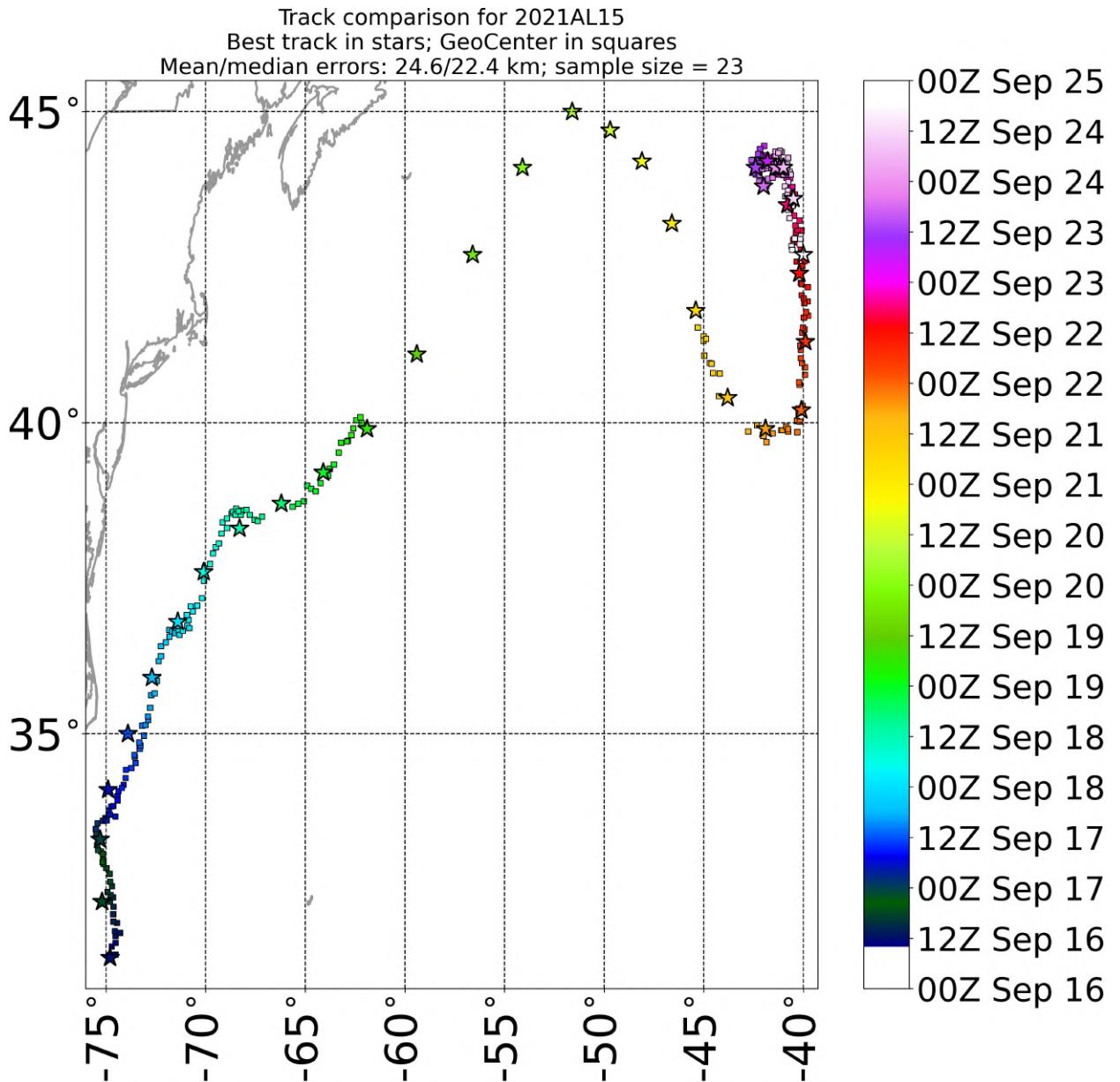

Figure S26: Full track for Tropical Storm Odette (AL152021), according to both the final best tracks (FBT; stars) and GeoCenter (squares). Formatting is explained in the caption of Figure S23.

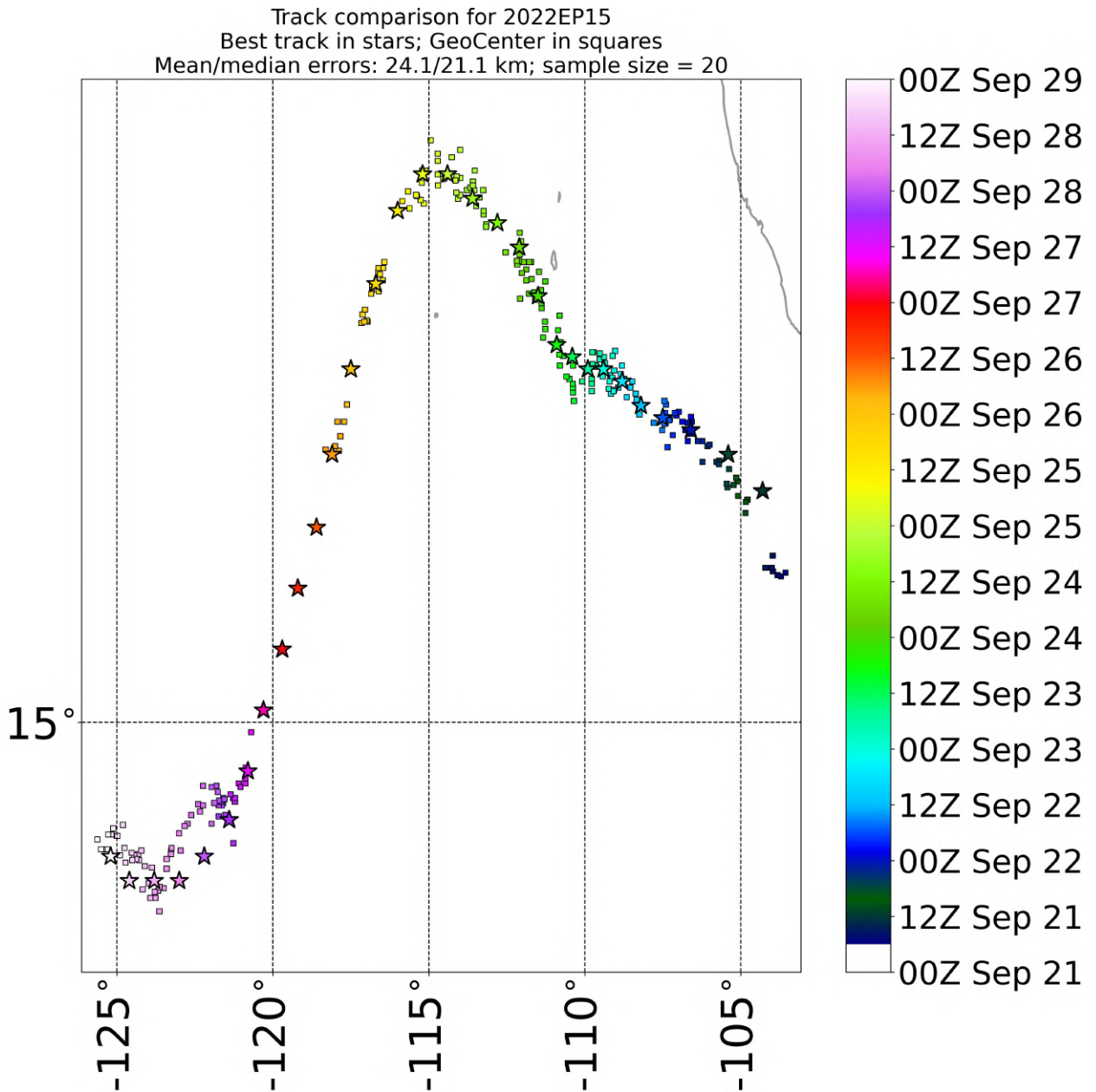

Figure S27: Full track for Tropical Storm Newton (EP152022), according to both the final best tracks (FBT; stars) and GeoCenter (squares). Formatting is explained in the caption of Figure S23.

## 4. Comparison between bias-corrected and uncorrected GeoCenter

Section 5a of the main text evaluates the final GeoCenter ensemble, which uses isotonic regression for bias correction, on tropical systems in the testing data. This section is analogous but evaluates the GeoCenter ensemble without bias correction, henceforth GeoCenter-no-bc. Figure S28 evaluates the GeoCenter-no-bc ensemble mean; Figure 7 of the main text is the analogue for GeoCenter.

Bias correction improves every metric shown in these figures – except for bias in the *x*-coordinate ("mean *x*-error" in panel d), which is -0.7 km for GeoCenter and +0.2 km for GeoCenter-no-bc, and bias in the *y*-coordinate ("mean *y*-error" in panel d), which is -0.6 km for GeoCenter and +0.3 km for GeoCenter-no-bc.

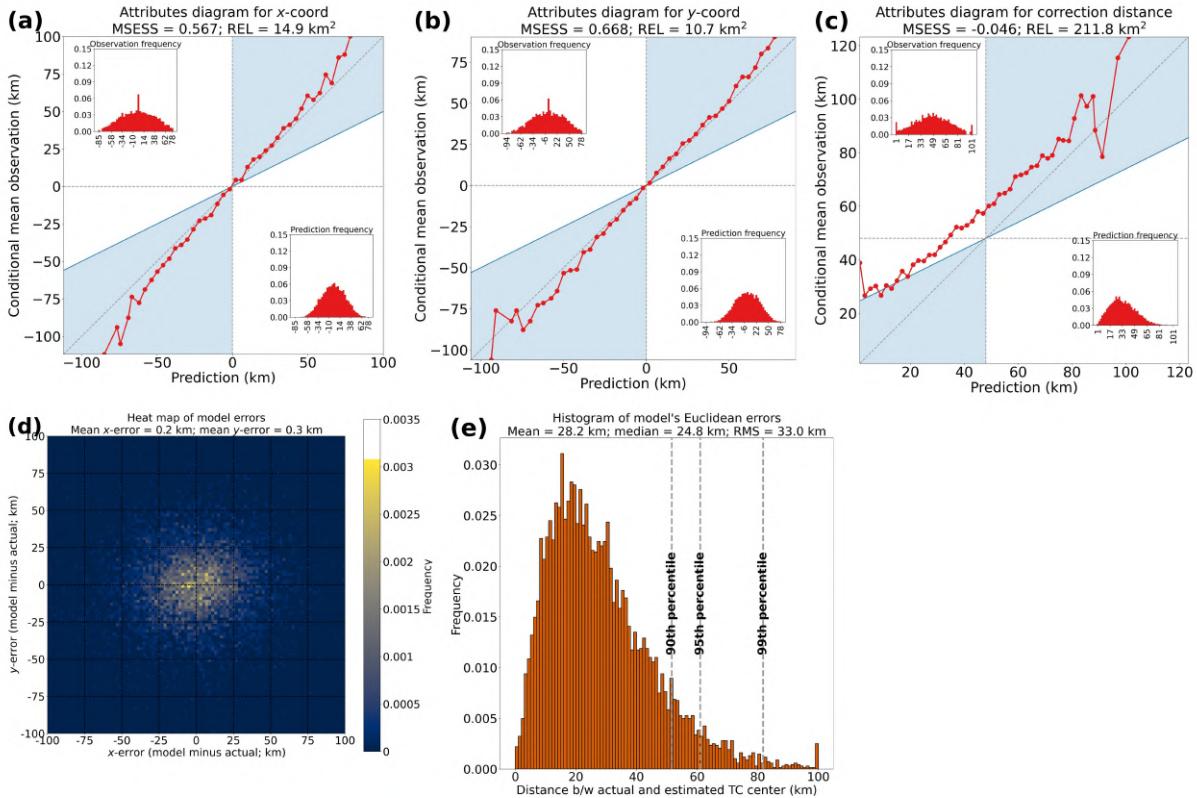

Figure S28: Performance of GeoCenter-no-bc ensemble mean on tropical systems in testing data. Formatting is explained in the caption of Figure 7 in the main text.

Figure S28 evaluates GeoCenter-no-bc uncertainty estimates; Figure 8 of the main text is the analogue for GeoCenter. Again, bias correction improves every metric shown in these figures, with the exception of DTMF for the *y*-coordinate and full vector, which is already 100% (the optimal value) without bias correction. Overall, bias correction soundly improves the performance of both the ensemble mean and uncertainty estimates.

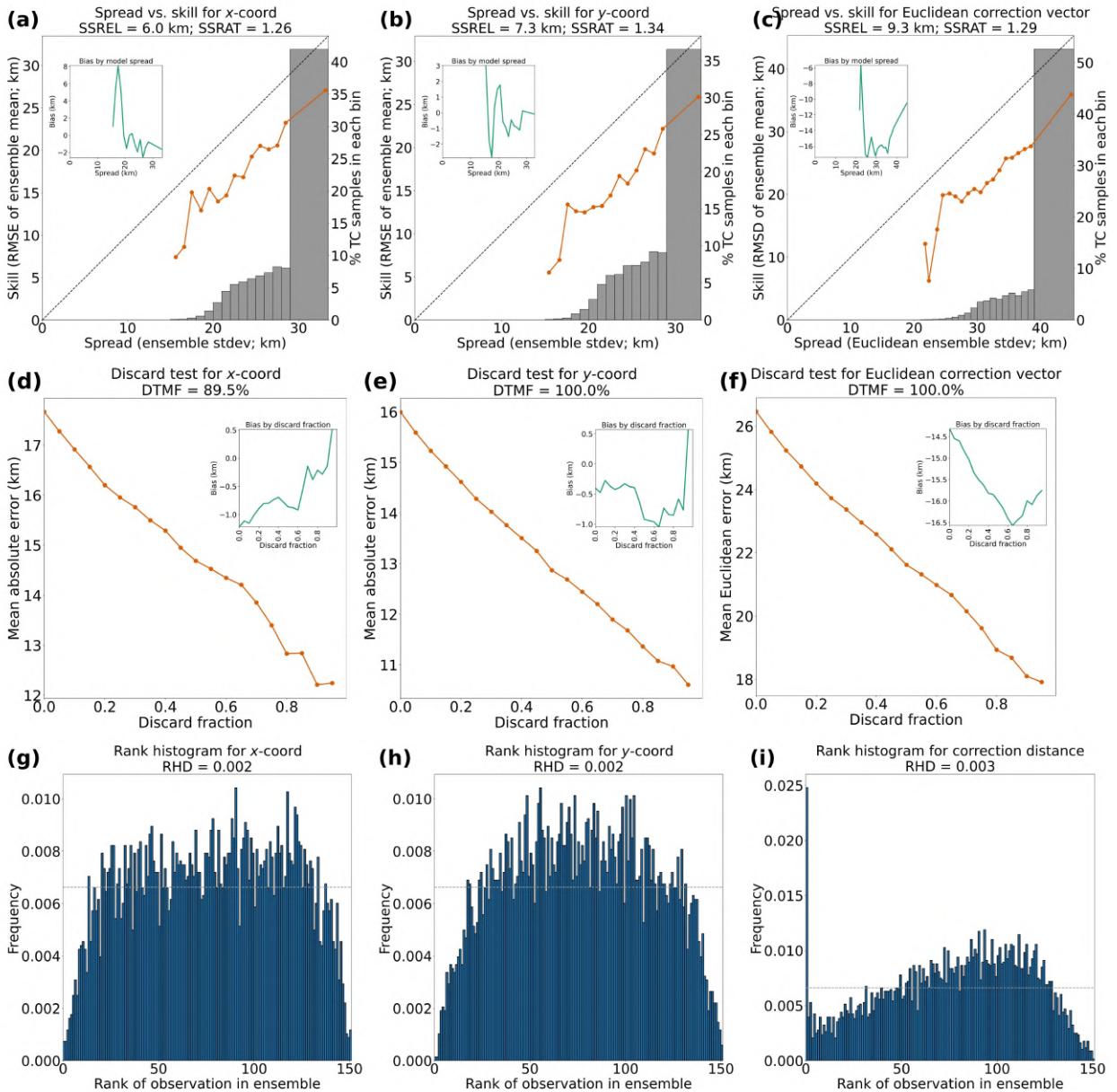

Figure S29: Performance of GeoCenter-no-bc uncertainty estimates on tropical systems in testing data. Formatting is explained in the caption of Figure 8 in the main text.

## 5. Comparison between GeoCenter with and without ATCF predictors

Section 5a of the main text evaluates the final GeoCenter ensemble, which uses nine ATCF predictors (Table 3 of main text), on tropical systems in the testing data. This section evaluates an analogous GeoCenter ensemble, henceforth GeoCenter-no-ATCF, with two differences. First, GeoCenter-no-ATCF is trained without ATCF predictors; thus, its predictors include only IR imagery. The second difference – which arose incidentally and not by experimental design – is that GeoCenter-no-ATCF uses slightly different wavelengths. For GeoCenter we found that a 3-member ensemble – omitting the CNN trained with wavelengths $\{3.9, 6.185, 6.95\}\mu$m – performs better than the 4-member ensemble on validation data. Similarly, for GeoCenter-no-ATCF, we found that a 3-member ensemble – omitting the CNN trained with wavelengths $\{8.5, 9.61, 12.3\}\mu$m – performs better than the 4-member ensemble on validation data. Both of these decisions are based only on the validation data and made before any evaluation on testing data.

Figures S30-S33 are analogous to Figures 7-10 in the main text but for GeoCenter-no-ATCF. Note that GeoCenter outperforms GeoCenter-no-ATCF on nearly every metric shown in these figures, *i.e.*, the ATCF predictors lead to better performance. However, GeoCenter-no-ATCF still soundly outperforms ARCHER-2 when only IR data are available. For example, GeoCenter-no-ATCF achieves a median Euclidean error of 24.8 km (Figure S30e), compared to 43-49 km for ARCHER-2 applied to IR data (Figure 11c of main text).

The superior performance of GeoCenter with ATCF predictors means that these predictors add value. We speculate that the additional information in these predictors (Table 3 of main text) allows the CNNs to selectively focus on, or look for, certain features. For example, the CNNs might have learned that strong tropical systems (with tropical flag = 1, high intensity, and low pressure) typically have a well defined eye, while extratropical systems at high latitude (with extratropical flag = 1) often have a sheared pattern and typically do not have a well defined eye.

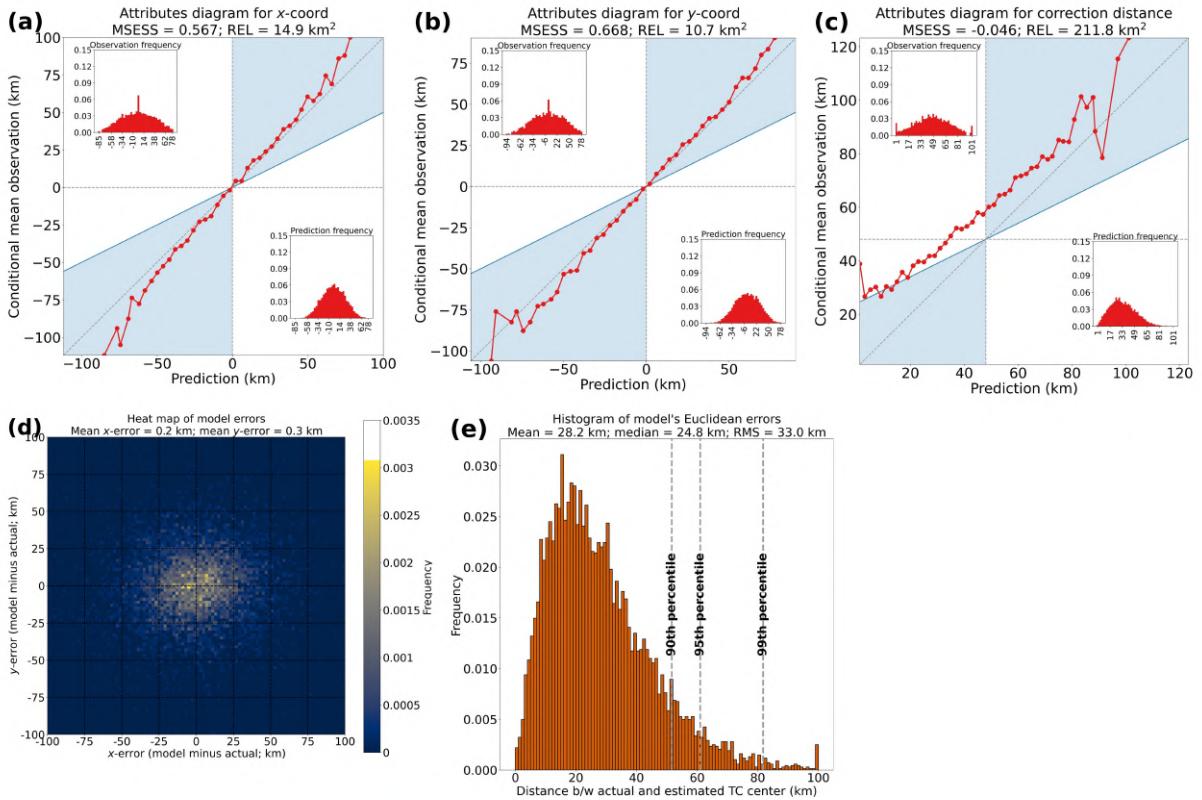

Figure S30: Performance of GeoCenter-no-ATCF ensemble mean on tropical systems in testing data. Formatting is explained in the caption of Figure 7 in the main text.

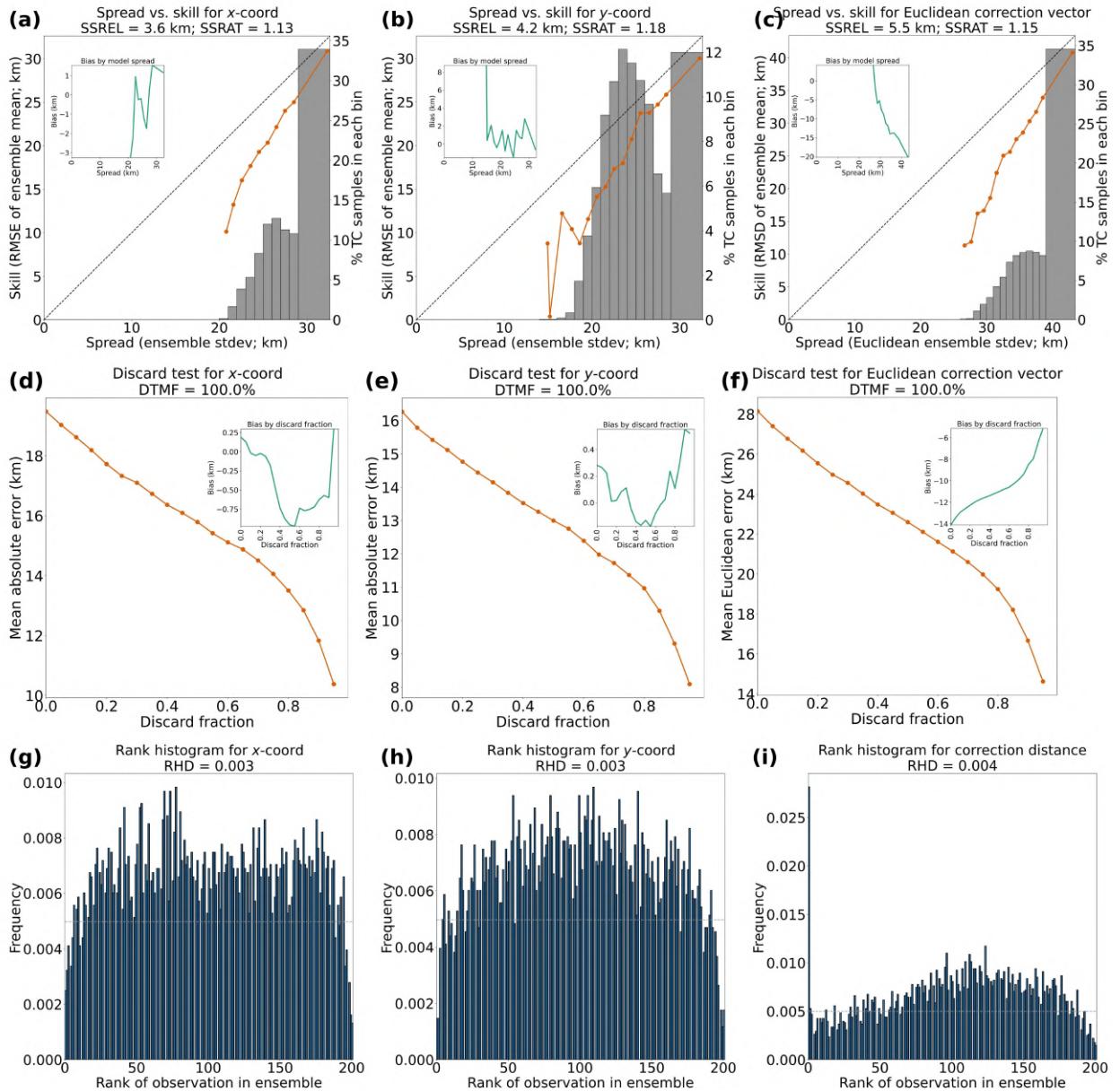

Figure S31: Performance of GeoCenter-no-ATCF uncertainty estimates on tropical systems in testing data. Formatting is explained in the caption of Figure 8 in the main text.

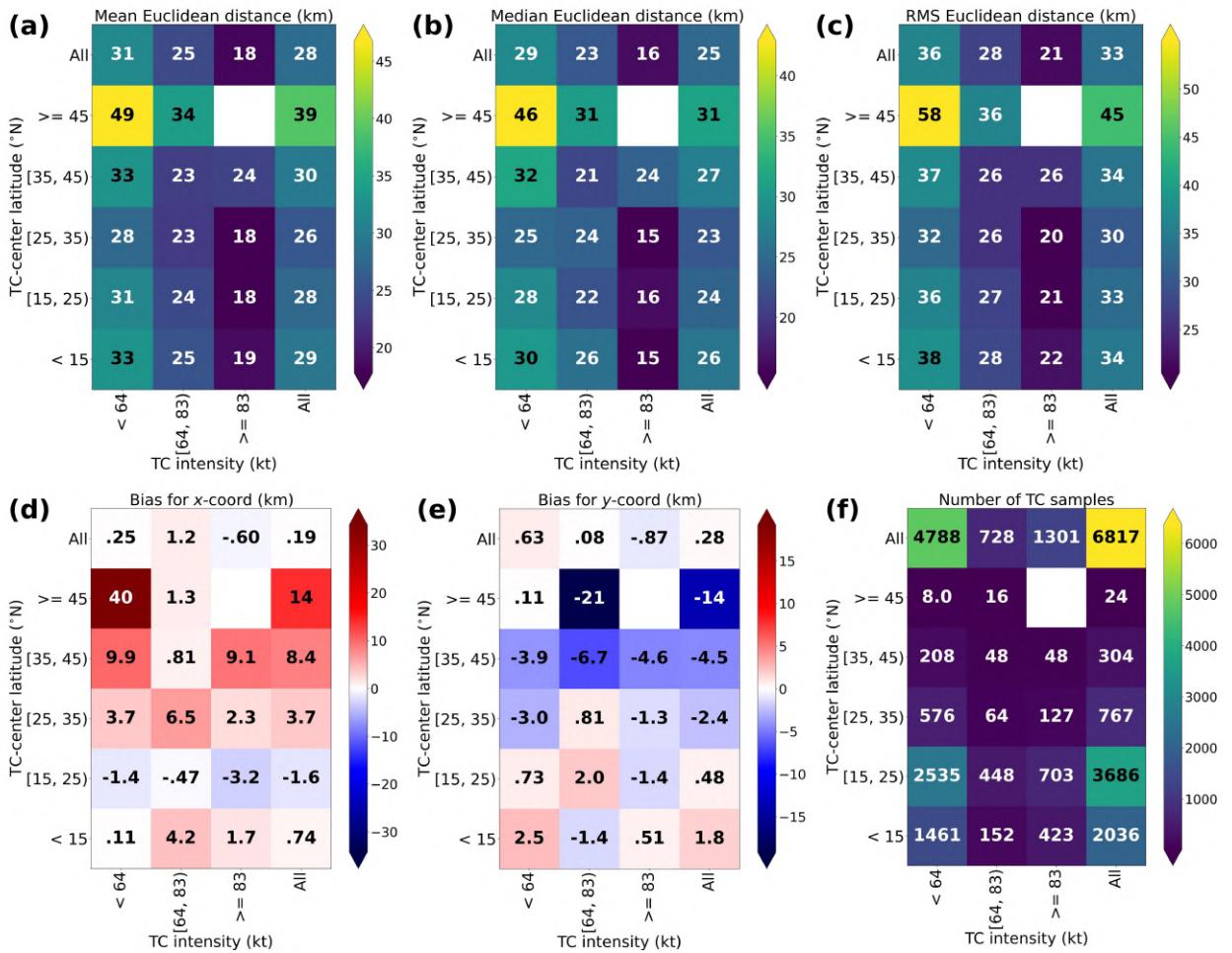

Figure S32: Performance of GeoCenter-no-ATCF on tropical systems in testing data, as a function of TC intensity and true TC-center latitude (from FBT). Formatting is explained in Figure 9 of the main text.

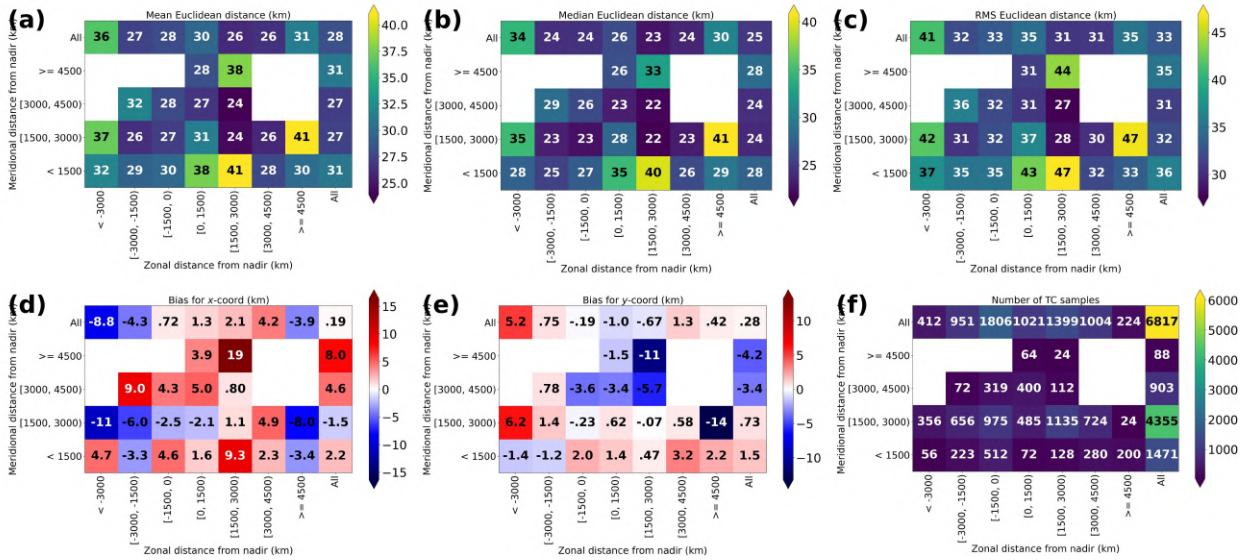

Figure S33: Performance of GeoCenter-no-ATCF on tropical systems in testing data, as a function of nadir-relative position (FBT center). Formatting is explained in Figure 10 of the main text.

## 6. Importance-ranking of individual IR channels

To quantify the importance of individual IR channels, we train 10 different CNNs, each with one channel in the predictors. Other than IR channels, these models use the same hyperparameters as those in the final GeoCenter ensemble: a 600-by-600-km domain, 9 lag times, and trained with a uniform distribution of first-guess errors. The single-channel models are also bias-corrected with isotonic regression. Figure S34 shows evaluation metrics, based on all systems (tropical and non-tropical) in the testing data, for the single-channel models. The best individual wavelengths appear to be 8.5 and 9.61 $\mu$m, while the worst single wavelength appears to be 3.9 $\mu$m.

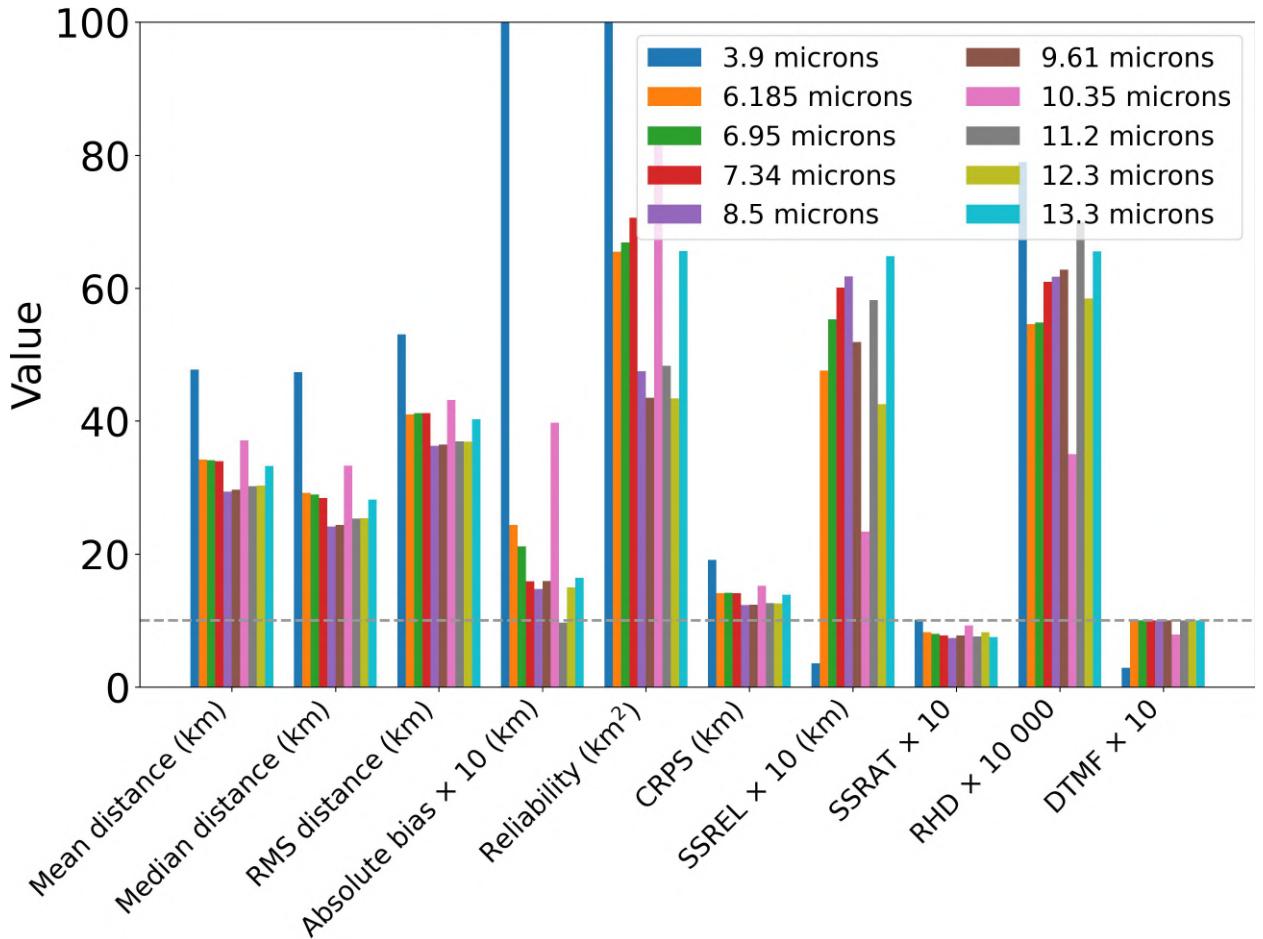

Figure S34: Performance of single-channel GeoCenter models on all systems (tropical and non-tropical) in testing data. For SSRAT, the optimal value is 1.0, indicated by the horizontal dashed line. DTMF is positively oriented, with an optimal value of 1.0. All other metrics are negatively oriented, with an optimal value of 0.0.

## 7. Sensitivity of GeoCenter to magnitude of first-guess error

This section investigates the sensitivity of the final GeoCenter ensemble to the magnitude of the first-guess error, specifically the whole-track error. This is the Euclidean distance between the image center at lag time = 0 min and the true TC center. For every offset distance in $\{2, 4, 6, \ldots, 120\}$ km, we apply the following steps:

1. Create an ancillary testing set. For 17 TCs in the testing set, apply 8 random first guesses to every TC sample. We follow the procedure from Section 3b of the main text, except that here the whole-track error is constant, rather than drawn from a distribution. The *direction* for whole-track error is still random, and the track-shape error is also still random, with distance drawn from the distribution $\max\{0 \text{ km}, \mathcal{N}(10 \text{ km}, 5 \text{ km})\}$.

2. Apply GeoCenter to this ancillary testing set and compute the errors.

Step 1 is the same first-guess procedure used to train GeoCenter, except with a constant distance for the whole-track error. We use only 17 TCs in the testing set – randomly selected – due to limitations on computing time.

The resulting sensitivity curve is shown in Figure S35. As long as the first-guess error is below 55 km, both the mean and median GeoCenter errors remain below 20 km. As first-guess error increases to 120 km, the mean and median GeoCenter errors increase to 40 and 37 km, respectively. Thus, although GeoCenter is sensitive to the magnitude of the first-guess error, even with an extreme first-guess error of 120 km, GeoCenter outperforms the IR version of ARCHER-2 (Figure 11c of main text).

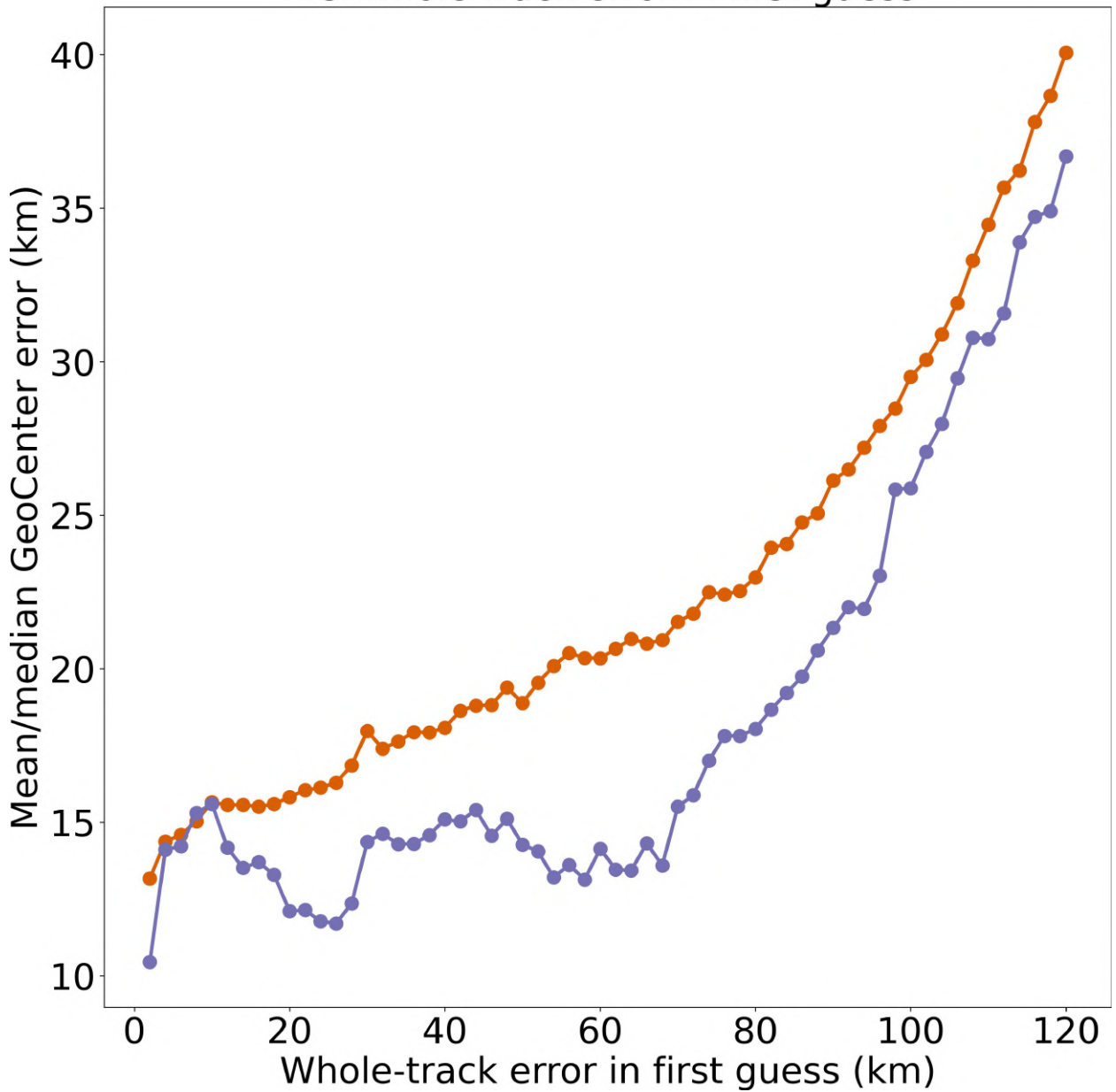

Figure S35: Sensitivity of GeoCenter to magnitude of first-guess error, based on all time steps (tropical and non-tropical) from 17 randomly selected TCs in the testing set.